# Resolving dynamics and function of transient states in single enzyme molecules

**Authors:** Hugo Sanabria[1,2,#,*], Dmitro Rodnin[1,#], Katherina Hemmen[1,#], Thomas Peulen[1], Suren Felekyan[1], Mark R Fleissner[3,7], Mykola Dimura[1,5], Felix Koberling[4], Ralf Kühnemuth[1], Wayne Hubbell[3], Holger Gohlke[5,6], Claus A.M. Seidel[1*]

#: contributed equally
*: corresponding authors

## Affiliations

[1]Institut für Physikalische Chemie, Lehrstuhl für Molekulare Physikalische Chemie, Heinrich-Heine-Universität, Düsseldorf, Germany.
[2]Department of Physics and Astronomy, Clemson University, Clemson, South Carolina, U.S.A.
[3]Jules Stein Eye Institute and Department of Chemistry and Biochemistry, University of California, Los Angeles, U.S.A.
[4]PicoQuant GmbH, Berlin, Germany.
[5]Institut für Pharmazeutische und Medizinische Chemie, Heinrich-Heine-Universität, Düsseldorf, Germany.
[6]John von Neumann Institute for Computing (NIC), Jülich Supercomputing Centre (JSC) & Institute for Complex Systems - Structural Biochemistry (ICS 6), Forschungszentrum Jülich GmbH, 52425 Jülich, Germany
[7]Present address: Avanir Pharmaceuticals Inc., Aliso Viejo, California, U. S. A.

*Correspondence to: cseidel@hhu.de and hsanabr@clemson.edu

## Abstract

We used a hybrid fluorescence spectroscopic toolkit to monitor T4 Lysozyme (T4L) in action. By unraveling the kinetic and dynamic interplay of the conformational states, we sought to elucidate the dynamic structural biology of T4L. In particular, by combining single-molecule and ensemble multiparameter fluorescence detection, EPR spectroscopy, mutagenesis, and FRET-positioning and screening, we characterized three short-lived conformational states within the conformational landscape of the T4L over the ns-ms timescale. The use of 33 FRET-derived distance sets, to screen known T4L structures, revealed that T4L in solution mainly adopts the known *open* and *closed* states in exchange at 4 µs. A newly found *minor* state, undisclosed by at present more than 500 crystal structures of T4L and sampled at 230 µs, may be actively involved in the product release step in catalysis. The presented fluorescence spectroscopic toolkit is anticipated to accelerate the development of dynamic structural biology.

## 1. Introduction

Enzymes adopt distinct conformational states during catalysis [1,2], where transiently populated ("excited") states are often of critical importance in the enzymatic cycle. These states are short-lived and therefore "hidden" to many experimental techniques. Classical structural biology methods often struggle to fully capture enzymes during catalytic action because the conformational rearrangements often span several decades in time (ns-ms) [3,4,5,6,7]. Hence, there is an urgent need to develop experimental and analysis methods to overcome this challenge.

Recently, we demonstrated by synthetic experiments that a new analysis toolkit ("FRET on rails") combined with molecular simulations can resolve short-lived conformational states of proteins [34]. Here, we apply and extend the fluorescence analysis toolkit, developed for dynamic structural biology, to interrogate the catalytic cycle of an enzyme [8]. In particular, the analysis (*1*) captures an excited, short-lived state and (*2*) identifies its potential relevance in the enzyme's catalytic cycle. The presented approach may serve as a blueprint for future enzymological studies with the well-established single-molecule Multiparameter Fluorescence Detection (MFD) experiments in that it enables detecting hidden states by the unique time-resolution (picoseconds) and sensitivity (single-molecule) of fluorescence.

We use lysozyme (T4L) of the bacteriophage T4 as development platform and probe its conformational dynamics and structural features. Structurally, T4L [9] consists of two interrelated subdomains, the N-terminal subdomain (NTsD) and the C-terminal subdomain (CTsD), differing in their folding behavior and stability [10]. A long α-helix (helix c) links the two subdomains (Figure 1A). To date, more than 500 structural models of T4L are available within the Protein Data Bank (PDB). In this ensemble, T4L adopts several opening angles corresponding to a classic hinge-bending motion of the NTsD with respect to the CTsD. The enzymatic function of T4L is to cleave the glycosidic bond between *N*-acetylmuramic acid and *N*-acetylglucosamine of the saccharides of the bacterial cell wall [11].

T4L in solution is thought to adopt conformations that are *open* to various degrees, and a covalent adduct of the protein and its processed enzymatic product can crystallize in a *closed* conformation [12,13] [11]. Therefore, T4L is thought to follow a classical Michaelis-Menten mechanism (MMm) characterized as a two-state system (Figure 1B). Here, an *open* and *closed* conformational state fulfils *unique* functions of substrate binding and substrate cleavage, respectively [14]. In a MMm, the product dissociates stochastically from the enzyme. For other enzymes, e.g. the Horseradish peroxidase[15], an "active" product release state was identified. Recent experimental findings for T4L suggest the involvement of more than two states in catalysis [16] [17], where the turnover rate was estimated between 10-50 ms [17-20], while the conformational dynamics fell within the ns to sub-ms range [4,12,17-26]. Such complex cases, with distinct interconverting conformational states, open additional reaction paths to yield disperse kinetics[27].

For a full description of an enzymological cycle, the number of enzymatic states, their connectivity, the conformational structures of the states, and the states' chemical function have to be unraveled. Technically, we achieve these objectives by a hybrid approach combining classic biochemical methods (mutagenesis & HPLC), probe-based spectroscopy, and molecular simulations. Förster Resonance Energy Transfer (FRET) and Electron Paramagnetic Resonance (EPR) spectroscopy probe distances between bioorthogonally introduced probes through dipolar coupling. In FRET spectroscopy, the coupling is measured between a donor (D) and acceptor (A) fluorophore.

In confocal MFD single-molecule FRET (smFRET) experiments, freely diffusing molecules are repeatedly excited by a pulsed light source, and the emitted fluorescence photon is detected with picosecond time-resolution by Time-Correlated Single Photon Counting (TCSPC) for several milliseconds per molecule (diffusion time, $t_{diff}$) [28] (Figure 1C). smFRET experiments are ideal to study kinetics because no sophisticated strategies are necessary to synchronize molecules prior to the analysis. Consequently, it is possible to probe reliably protein kinetics over seven decades in time (sub ns-ms).

Distinct features of photon streams are highlighted by different representations (Figure 1D). (*1*) A MFD-histogram is particularly valuable to reveal the number of states, identify dynamics, and to inform on state connectivities. A MFD-histogram is generated by analyzing two complementary FRET-indicators, the average FRET-efficiency, $E$, and the fluorescence-averaged donor lifetime in the presence of acceptor, $\langle\tau_{D(A)}\rangle_F$, for individual single-molecule events [28-30]. (*2*) Filtered fluorescence correlation spectroscopy (fFCS) quantifies exchange dynamics among the states by determining relaxation times [31 32]. (*3*) The analysis of fluorescence decays reveals populations of states and equilibrium distance distributions. (*4*) Finally, these experimental distances can be translated to structural models by molecular simulations [66,34].

Following the concepts established for synthetic data for Hybrid-FRET for structural Dynamics [33] or FRET on Rails (Figure 1E), we start out with simultaneously monitoring the dynamics and structural features of T4L by a systematic design of a FRET network for T4L (Section 2.1). We use a combination of MFD to resolve the conformational states stable at the ns timescales (Section 2.1). Next, we use fFCS and Monte Carlo simulations to resolve the connectivity of the conformational states (Section 2.2). In section 2.3, we perform a statistical analysis to further justify the existence of a hidden conformational state of T4L that was clearly identified in Section 2.1. Subsequently, we use three distinct distance sets to screen an ensemble of structural models to compare our identified states to models in the Protein Data Bank (PDB) (Section 2.4). The identified hidden/excited conformational state is corroborated by other analytical tools such as chromatography and Electron Paramagnetic Resonance (EPR) spectroscopy. Finally, we derived an experimental energy landscape of T4L's enzymatic cleavage cycle, based on shifting equilibria, by mutating key active site residues that mimic functional enzyme states at various steps during substrate hydrolysis. (Section 2.5). Overall, our results demonstrate the potential of fluorescence spectroscopy to go beyond traditional experimental methods for obtaining a dynamic structural picture of enzymes in action.

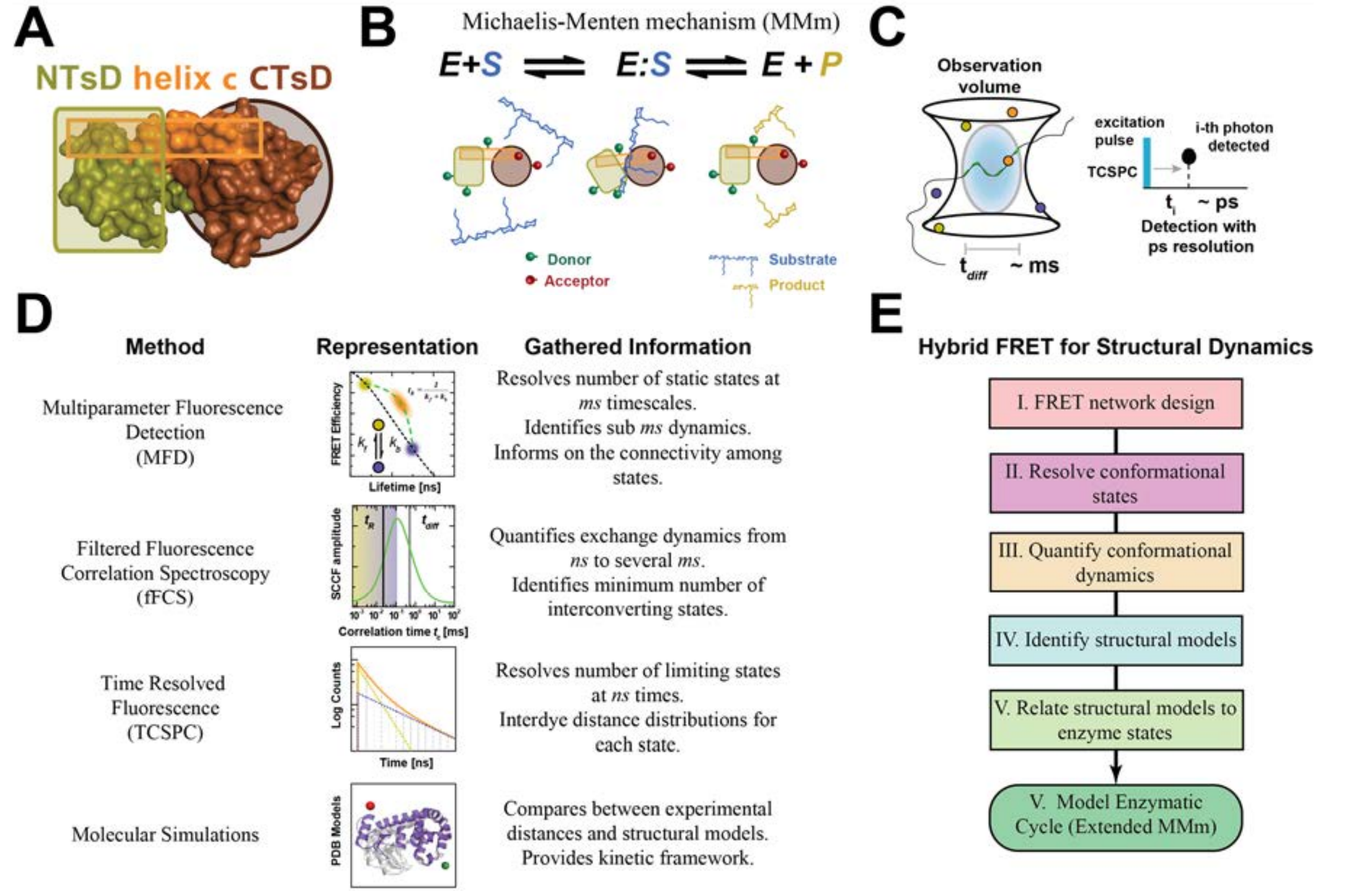

**Figure 1 Schematic of the high precision FRET and fluctuation analysis toolkit. (A)** Subdomain architecture of Phage T4 Lysozyme (T4L). **(B)** Upon cleavage of its substrate peptidoglycan (blue), the N-terminal lobe (green) and the C-terminal lobe (brown) of T4L exhibit a change of closure, which can be observed via the induced change in the FRET indicators [11,12] **(C)** In MFD-experiments of freely diffusing single molecules, the emitted fluorescent photons ("bursts") are detected with ps resolution (with respective to the exciting laser pulse) during the diffusion (on the ms time scale) of the molecule through the observation volume (diffusion time, $t_{diff}$). **(D)** The labeled T4L molecules were studied with different experimental methods. In single-molecule Multiparameter Fluorescence Detection (MFD) experiments, the fluorescence bursts - averaged over ms – are analyzed e.g. with respect to their fluorescence lifetime $\langle\tau_{D(A)}\rangle_F$

or FRET-efficiency $E$ and displayed in multidimensional frequency histograms (2D MFD-histogram). Molecules that adopt a stable conformation during burst duration are located along the static FRET-line (black) (Supplementary Information 1.2). Assuming that the two limiting states (yellow and blue) exchange on timescales faster than ms with exchange rate constants $k_f$ and $k_b$, we find only a single population (orange) shifted towards a longer fluorescence lifetime that is located on the dynamic FRET-line (green) connecting these two limiting states. Thus, FRET-lines serve as a visual guide to interpret 2D MFD-histograms, with deviations from the static FRET-line being indicative for the dynamic averaging and dynamics at the sub-ms and ms timescales. Filtered fluorescence correlation spectroscopy (fFCS) computes the species-specific cross-correlation function (*sCCF*, Supplementary Information 1.4). The observed anti-correlation reveals a characteristic relaxation time $t_R$ related to the inverse of the sum of the exchange rate constants, $k_f$ and $k_b$. In eTCSPC, the distribution of the fluorescent photons is detected with respect to an excitation pulse with ps resolution to reveal populations stable on the ns timescale (Supplementary Information 1.5-1.7). Finally, in molecular simulations, the experimental results are compared to available structural models. **(E)** Flowchart for the hybrid FRET toolkit for determining structural dynamics. Based on a network of FRET variants, the conformational states and their exchange dynamics are determined, which are then used to identify the structural models. T4L variants with mutations altering their enzymatic activity relate the structural models to enzymatic states. Based on the gathered information, the enzymatic cycle can be modeled.

## 2. Results

### 2.1 Detecting T4L's states by MFD

In our smFRET-experiments, we monitor the distance between a donor (D) and acceptor (A) attached to specific amino acids of a T4L variant (see Methods). We designed a network of 33 distinct T4L variants to probe hinge-bending motions of T4L from different spatial directions (Figure 2A) that cover the whole protein.

In Figures 2B and C, we present MFD-histograms with the two FRET-indicators for two exemplary variants of our FRET network, highlighted in Figure 2A by solid lines (left panel). Three peaks are identified in the MFD-histogram. In both histograms, a major and a minor

FRET peak are present. The peak located at a low FRET-efficiency $E$ corresponds to molecules without, or with an inactive acceptor, fluorophore (DOnly).

For an *open* (PDB ID: 172L, blue) and a *closed* (PDB ID: 148L, magenta) conformation, FRET efficiencies $E$ are predicted by experimentally calibrated dye models (see Methods, section 5.5) [33,34]. These $E$ values are shown as horizontal lines in the marginal distributions of Figure 2B,C. A comparison with the major peak (Figure 2B: $0.3 < E < 0.7$, Figure 2C: $E > 0.7$) demonstrates that they are similar but not identical to known structural models. Next, we will show how dynamic exchange explains the observed peaks.

In MFD-histograms, FRET-lines serve as a unique guide to visualize conformational dynamics by peak shifts and splitting like in NMR relaxation dispersion measurements (Supplementary Information 1.1-1.3). A static FRET-line relates $E$ and $\langle\tau_{D(A)}\rangle_F$ for molecules in the absence of dynamics (magenta line, Figure 2B,C). States that exchange on a time scale much slower than the observation time (quasi-static case) are separated in a MFD-histogram and follow the static FRET-line. However, a shift of a peak to the right with respect to the static FRET-line is a model-free indication for sub-ms dynamics [28], because the FRET-efficiencies in a MFD-histogram are averaged over the observation time of the molecules (~ms). Thus, very fast exchanging states result in single average peak that is shifted to the right of the static FRET-line. These peaks can be described by dynamic FRET-lines, which connect the exchanging states. For a visual representation of the possible transitions, the dynamic FRET-lines of the identified exchanging states are displayed in the MFD-diagrams (orange, cyan and green). A dynamic FRET-line connecting high FRET states with the DOnly population (grey) demonstrates the lack of significant photobleaching or blinking of acceptor dyes.

In the presented data, the major populations are shifted to the right of the static FRET-line (Figure 2B,C). This gives clear evidence for a dynamic exchange faster than ms. For molecules in very rapid (µs) exchange between an *open* and a *closed* conformation, we expected to detect a single averaged peak in MFD-histograms. Hence, taking fast exchange into account, the major peak of the smFRET data is in agreement with known X-ray structures [11,35] and kinetic data [4,10,17,18,20,21,26,36-38], most likely corresponding to the dynamic averaging of the hinge bending mechanism.

However, in 18 out of 33 MFD-histograms, we visually identify additional minor populations, which are in slow exchange with the major populations. Surprisingly, these minor populations ($E > 0.8$, Figure 2B) and ($0.2 < E < 0.6$, Figure 2C) can neither be assigned to the predicted average *open* and *closed* conformation (Figure S3). This is a first indicative for the existence of a third, conformationally *excited*, structurally distinct conformer.

In conclusion, MFD-histograms identify three conformers in T4L referred to as $C_1$, $C_2$ and $C_3$. The conformers $C_1$ and $C_2$ are likely in fast exchange, while $C_3$ is in slow exchange with $C_1$ or $C_2$. These conformers may represent limiting states in the exchange dynamics [39 40].

Following the workflow presented in Figure 1E, we next determine the kinetic signatures via fFCS and resolve remaining ambiguities by simulations of MFD-experiments (Section 2.2).

### 2.2 Connectivity of states in a kinetic network

The variant S44pAcF/I150C is used as pseudo-wildtype ("wt**") to construct a reaction scheme of T4L's enzymatic cycle. At first, we performed control experiments by comparing for this variant (DA)-labeled and reversely (AD)-labeled T4L variants. In this way, we could exclude potential dye artifacts (Figure S2A-C in Supplementary Information 2.1.2) because the kinetic behavior was independent of the labeling scheme. The MFD-histogram of S44pAcF/I150C (Figure 3A) reveals a typical pattern: a major population $C_1/C_2$ (0.2 < $E$ < 0.6) and a minor $C_3$ population ($E$ > 0.8) similar to the variants presented in Figure 2B,C.

To unravel the kinetic behavior of an enzyme, one has to be aware that an enzymatic cycle with multiple states can be described by a transition rate matrix, which contains all exchange rate constants of the states. To recover T4L's transition rate matrix, we determine a set of relaxation times by fFCS (Section 2.2.1) and the species fractions of the states by analysis of the fluorescence decays (for details see section 2.3). This analysis results in ambiguous solutions, which are resolved by simulating MFD experiments making use of the information contained in smFRET experiments (Section 2.2.2).

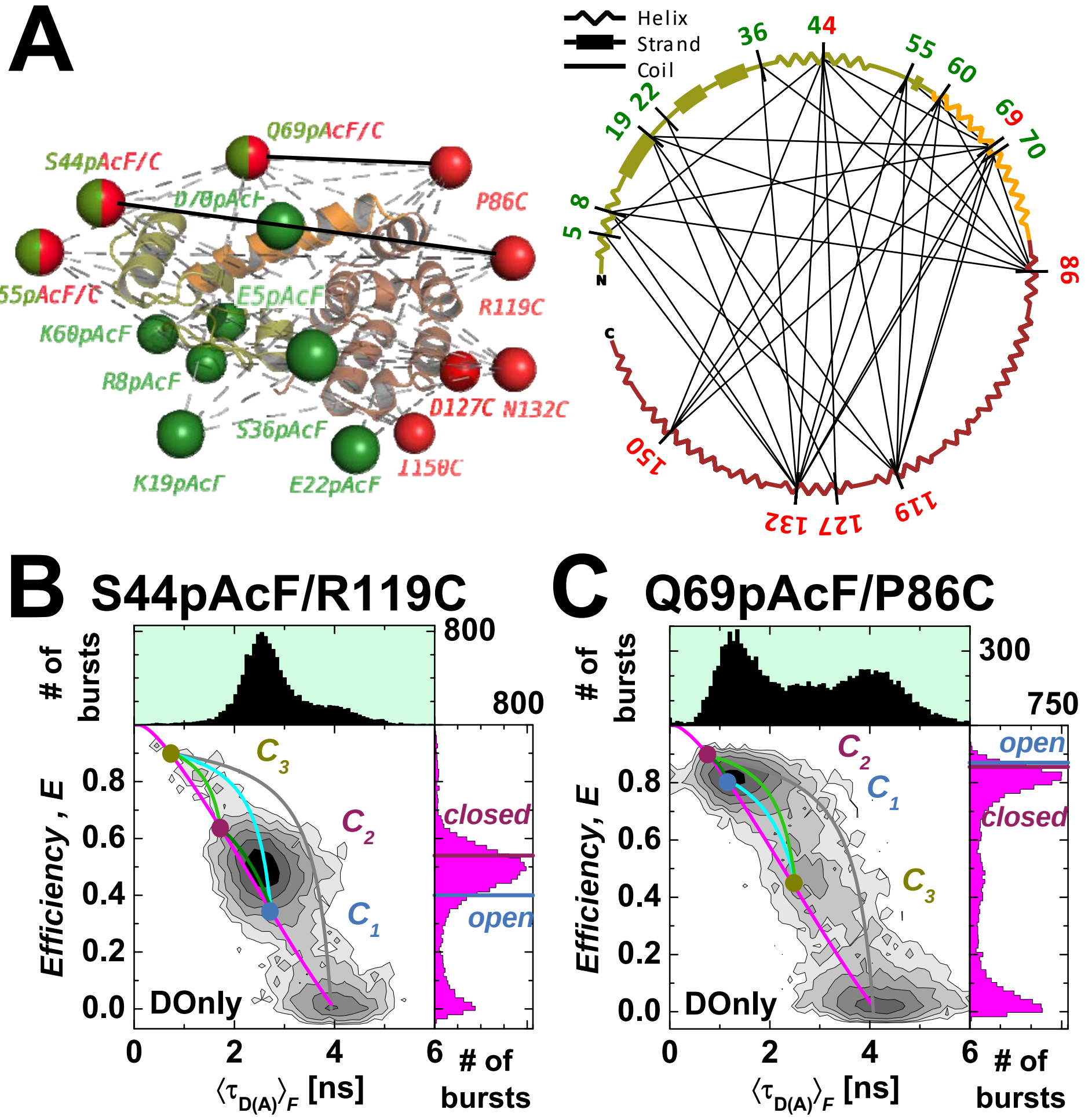

**Figure 2 FRET studies probe T4L structure and dynamics. (A)** Network representing 33 measured distinct FRET-variants of T4L. The donor (D) Alexa488-hydroxylamine and acceptor (A) Alexa647-maleimide are coupled to *para*-acetylphenylalanine (pAcF) and cysteine (C), respectively. The spheres in the left panel represent the average donor (green) and acceptor (red) position for a structure of T4L (PDB ID: 172L) that is determined by the FRET Positioning System (FPS) [41]. Positions 44, 55, and 69 are used for donor and acceptor labeling. Right panel shows the secondary structure elements (helix, strand or coil) of T4L. The labeling positions are indicated and network pairs are connected. Color scheme is the same as for the left panel. **(B, C)** The FRET-efficiency, $E$, and fluorescence lifetime of D in the presence of A, $\langle\tau_{D(A)}\rangle_F$, of the DA-labeled T4L variants (**B**) S44pAcF/R119C and (**C**) Q69pAcF/P86C are shown in two-dimensional MFD-histograms (center) with one-dimensional projections of $E$ (right) and $\langle\tau_{D(A)}\rangle_F$ (top). The magenta lines (static FRET-line) describe those molecules with a single conformational state. The limiting states (circles) of the dynamic FRET-lines ($C_1$, blue; $C_2$, purple; $C_3$, yellow) are identified by eTCSPC (Supplementary Information 1.5-1.7 and 2.3-2.4). The dynamic FRET-lines are shown in dark green ($C_1$-$C_2$), cyan ($C_1$-$C_3$), and green ($C_2$-$C_3$) (Supplementary Information 1.2). The grey lines trace molecules of high FRET-efficiency with a bleaching A. For comparison, the FRET-efficiencies of the X-ray structures for an *open* (blue, PDB ID: 172L) and *closed* (violet, 148L) state (determined by FPS) are shown as horizontal lines in the FRET-efficiency histograms; for S44pAcF/R119C $E(open)$ = 0.40 and $E(closed)$ = 0.54 and for Q69pAcF/P86C $E(open)$ = 0.87 and $E(closed)$ = 0.86.

### 2.2.1 Kinetic network of conformational states resolved by fFCS.

By fFCS, we probe transitions in T4L on all relevant timescales [31 32]. fFCS uses species-specific information encoded as a characteristic pattern within the ns-regime of the polarization-resolved fluorescence decays [31,42]. This amplifies the contrast compared to conventional FCS for resolving relaxation times with high precision. We find very good agreement between the normalized species cross-correlation functions (*sCCF*) of the (AD)- and (DA)-labeled molecules. A global analysis of the *sCCF*s and the species auto-correlation functions (*sACF*) requires at least two relaxation times ($t_{R1}$ = 4 µs and $t_{R2}$ = 230 µs, Figure 3B, Supplementary Information 2.2).

In summary, the two relaxation times obtained by *sCCF*s independently support the hypothesis of the interconversion between three states at sub-ms timescales. Moreover, in line with the MFD-histograms, we find a fast and a slow relaxation time.

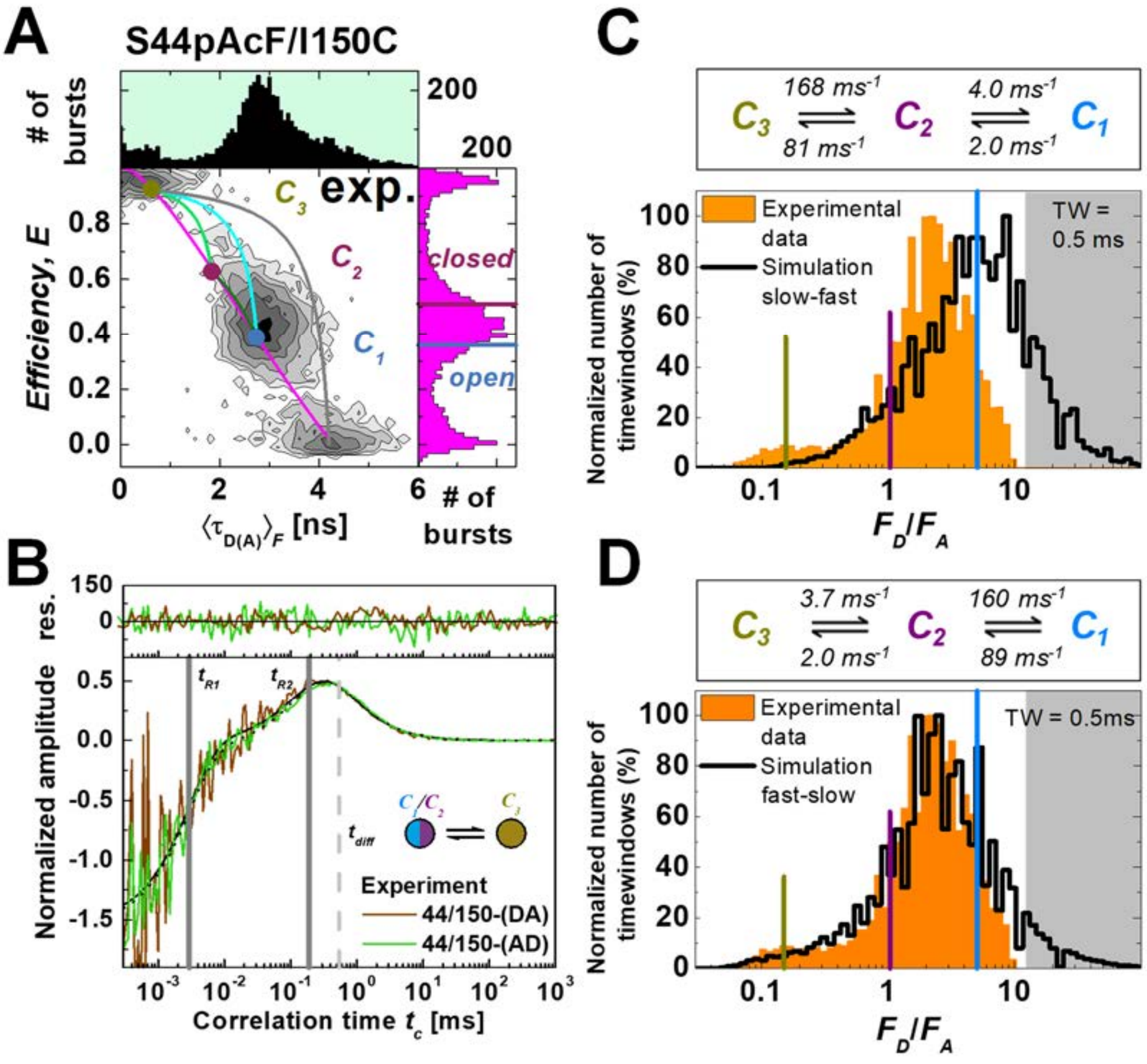

**Figure 3 Kinetic connectivity. (A)** The FRET-efficiency, $E$, and fluorescence lifetime of D in the presence of A, $\langle\tau_{D(A)}\rangle_F$, of the DA-labeled T4L variant S44pAcF/I150C (or wt**) is shown in the two-dimensional MFD-histograms (center) with the one-dimensional projections of $E$ (right) and $\langle\tau_{D(A)}\rangle_F$ (top). The magenta line (static FRET-line) represents the molecules with a single conformation. The dynamic FRET-lines represent the molecules changing their conformational state under monitoring. The limiting states (circles) of the dynamic FRET-lines ($C_1$, blue; $C_2$, purple; $C_3$, yellow) were identified by eTCSPC (see Figure 4). The dynamic FRET-lines are shown in dark green ($C_1$-$C_2$), cyan ($C_1$-$C_3$), and green ($C_2$-$C_3$) (Supplementary Information 1.2). The grey line represents the trace molecules of high FRET-efficiency with a bleaching A. The FRET-efficiencies of X-ray structures for an *open* (blue, PDB ID: 172L) and *closed* (violet, 148L) state (determined by FPS) are shown for comparison, and represented as horizontal lines in the FRET-efficiency histograms. **(B)** Overlay of the normalized *sCCF*s (eq. S17) of S44pAcF/I150C-(DA) and S44pAcF/I150C-(AD). The global fit with other variants shows two relaxation times ($t_{R1}$ = 4.0 ± 2.3 µs, $t_{R2}$ = 230 ± 28 µs) and a diffusion time $t_{diff}$ = 0.54 ms. **(C, D)** Overlay of time-window analysis, slow-fast in (C) and fast-slow in (D),

respectively. The shaded area in gray indicates the region of donor only. See Supplementary Information 2.6 for data on the model and the simulation.

### 2.2.2 Simulation of the kinetic network of T4 Lysozyme.

The three identified conformers $C_1$, $C_2$, and $C_3$ are assigned by their characteristic species fractions (see Section 2.3) to the corresponding structural states *open*, *closed*, and *excited*, respectively.

Three distinct kinetic linear reaction schemes are possible: $C_1 \leftrightharpoons C_2 \leftrightharpoons C_3$, $C_2 \leftrightharpoons C_1 \leftrightharpoons C_3$, $C_1 \leftrightharpoons C_3 \leftrightharpoons C_2$. The FRET-lines in the MFD-diagram (Figure 3A) serve as kinetic boundaries; nonetheless, the sequential closing, from the most open (lowest FRET-efficiency in the variant S44pAcF/150C) state to the most compact (highest FRET-efficiency in the variant S44pAcF/150C) is depicted by $C_1 \leftrightharpoons C_2 \leftrightharpoons C_3$, so that we can discard the models $C_1 \leftrightharpoons C_3 \leftrightharpoons C_2$ and $C_2 \leftrightharpoons C_1 \leftrightharpoons C_3$. With the relaxation times determined by fFCS and the species fractions obtained by analysis of the fluorescence decays (Section 2.3), we calculate the exchange rate constants and find two competing solutions (Supplementary Information 2.6, eq. S36). The exchange between $C_1$-$C_2$ can be either slow (Figure 3C) or fast (Figure 3D).

To solve this ambiguity, we simulate sm-experiments of the two possible solutions [39] (Supplementary Information 2.6). The obtained simulations are compared with experimental histograms and fFCS (Figure 3C,D, Figure S8). The corresponding $p$-values ($C_1$-$C_2$ fast vs. $C_1$-$C_2$ slow) of the respective 1D ($p = 1$ for $\langle \tau_{D(A)} \rangle_F$, $p = 0.734$ for $E$) and 2D MFD-histograms ($p = 1$) clearly demonstrate a better agreement of the experimental data with a fast exchange between $C_1$ and $C_2$ (Supplementary Information 2.6, Table S4A,B).

To conclude, we experimentally determine all reaction rate constants that define the reaction network, and the resulting species fractions. This description covers μs – ms and captures the relevant global motions of T4L.

**2.3 Characterization of the third conformer by eTCSPC**.

As demonstrated by fFCS analysis, T4L is highly dynamic. Hence, the FRET-efficiencies in smFRET-histograms only represent dynamic averages of states [43]. Therefore, for resolving the limiting states of the system, we record high-precision fluorescence decays by eTCSPC and analyze the distribution of the photon arrival times, *t*, with respect to the excitation pulse in fluorescence decays. This analysis benefits from polarization-free effects resulting from measuring at magic-angle detection, low background fluorescence, and the absence of photobleaching. Moreover, it can reveal DA-distance distributions and species populations [44]. To dissect the donor quenching by FRET (i.e. FRET-induced donor decay), we jointly analyze the DA- and DOnly-dataset, where the fluorescence lifetime distribution is shared with the DA-dataset. For physically meaningful analysis results, we explicitly consider the DA-distribution broadening due to the linkers by normal distributions [44].

The analysis results of all 33 FRET-datasets are discussed using the variant S36pAcF/P86C shown in Figure 4A (for other variants see Supplementary Information 2.3, Figure S5A). We display the experimental data by fluorescence decays of the DA- and the corresponding DOnly-sample (Figure 4A). In agreement with the MFD-histograms and the fFCS data, 1-component models result in broad DA-distributions and/or are insufficient to describe the data (Figure 4A, weighted residuals, violet). For S36pAcF/P86C, we obtain both, an unphysical distribution width and significant deviations in the weighted residuals, a strong indication that more than one conformer is found.

The analysis of the fluorescence decays by a 2-component model yields an inconsistent assignment by the species fractions (Figure S5B-C). This is characterized by significant differences that are evident among the species fractions (Table S2B). Moreover, the DA-distances disagree with known structural models (compare Table S2D,E).

In our effort to seek a consistent description of all measured fluorescence decays, we develop a joint/global model function. For such description, we treated all fluorescence decays as a single dataset sharing common species fractions for the states. This reduces the number of free parameters and dramatically stabilizes the optimization algorithm. Because the global 2-component model (Figure 4, cyan, Table S2C) shows no agreement with the data, we consequently used a 3-component model (Figure 4A orange, Table S2D-F) to describe the data.

To analyze the precision of this fit, the uncertainties $\Delta R_{\mathrm{DA}}$ of the obtained distances, $\langle R_{DA} \rangle$, from the 3-component model need to be determined. $\Delta R_{\mathrm{DA}}$ depends on statistical uncertainties and systematic errors. We used the known shot noise of the fluorescence decays to estimate the statistical uncertainties, $\Delta R_{\mathrm{DA}}(k_{\mathrm{FRET}})$, of the FRET-rate constant $k_{FRET}$ (Figure 4B, Supplementary Information Tables S2D-F). Moreover, we recorded polarization-resolved fluorescence decays of the donor and acceptor by eTCSPC to analyze the time-resolved anisotropy (Supplementary Information Tables S5A,B) for estimating systematic errors, $\Delta R_{\mathrm{DA}}(\kappa^2)$, due to the orientation factor $\kappa^2$. In conclusion, we can demonstrate that $\Delta R_{\mathrm{DA}}(\kappa^2)$ dominates the overall uncertainty of $\Delta R_{\mathrm{DA}}$ (eq. 6, Supplementary Information Tables S2D-F). Moreover, we sampled the model parameters of a 3-component model for individual datasets by a Markov chain Monte Carlo (MCMC) method. This demonstrates that, for given state populations, the mean distances $\langle R_{DA,1} \rangle$, $\langle R_{DA,2} \rangle$, and $\langle R_{DA,3} \rangle$ are very well defined (compare red to black in Figure 4B). This also shows that a global model, which interrelates the state populations among datasets, improves the capability to resolve interdye distances.

A global 3-component model has too many degrees of freedom (Supplementary Information 1.7) to be exhaustive when sampling by MCMC. Hence, we vary the state population of the minor state, $x(C_3)$, while optimizing all other model parameters (support plane analysis). This way, we determine the dependency of $x(C_3)$ on the quality parameter $\chi^2_r$ of all measurements (Figure 4C). This analysis (*1*) shows that the minor state population is in the range of 0.1 to 0.27 and best agrees with the data for 0.21 (Figure 4C, $p$-value = 0.68), (*2*) provides an estimate for $\Delta R_{DA}(k_{FRET})$ (Table S2D-F), and (*3*) demonstrates that $\Delta R_{DA}(\kappa^2)$ dominates $\Delta R_{DA}$ (eq. 6).

In summary, only a 3-component analysis describes all FRET samples and reference samples in a global model. This analysis recovers a set of physically meaningful average DA-distances that are grouped automatically and unbiased by their state populations. Additionally, the 3-component model is consistent with the fFCS data and with the dynamic FRET-lines displaying dynamically averaged sm-subpopulations in MFD (Figure 2, Figure S3).

The integrated results are consistent with a view that T4L adopts three states ($C_1$, $C_2$, and $C_3$), as opposed to the expected two conformational states based on structural pre-knowledge.

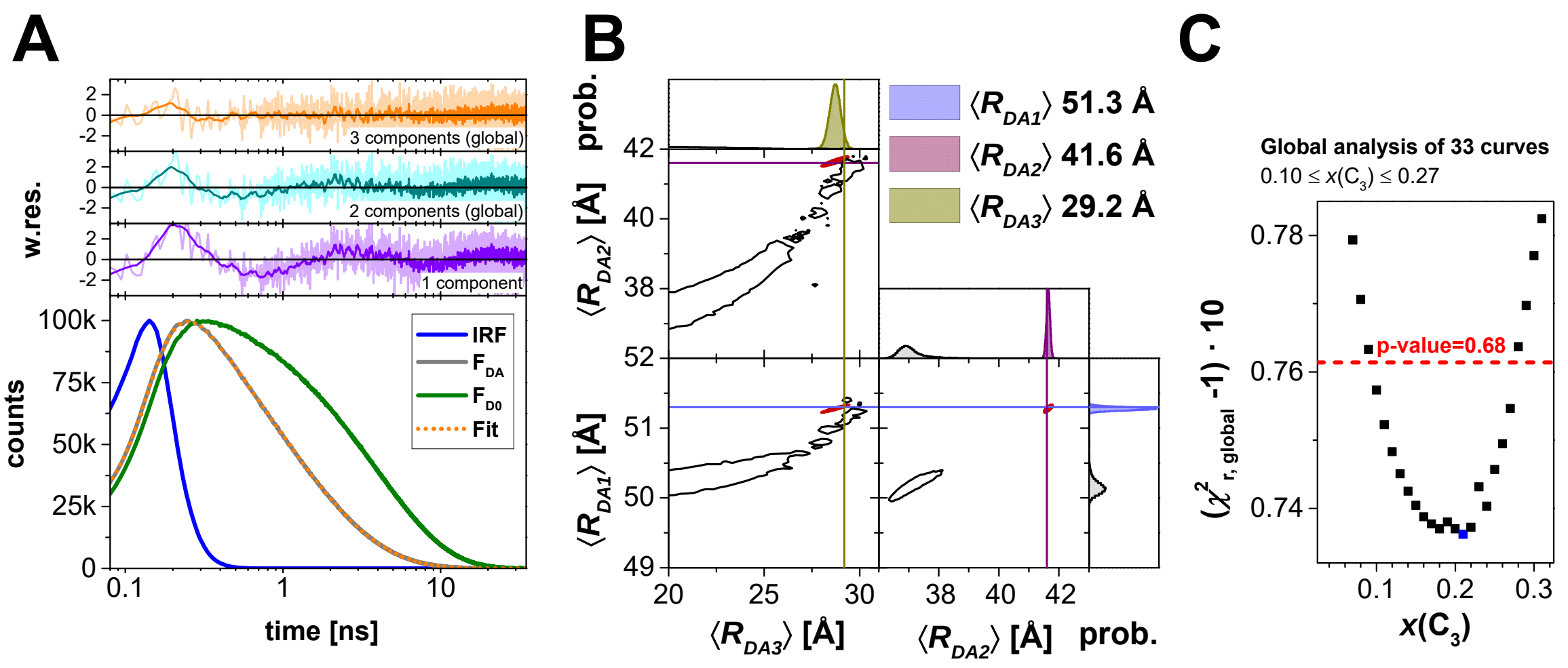

**Figure 4 Uncovering the third conformational state by eTCSPC.** The eTCSPC measurements of all FRET variants are analyzed by superposition of normal distributed DA-distances (due to the coupling of dyes with long and flexible linkers, details see Supplementary Information 1.7). **(A)** Experimental fluorescence decay of the variant S36pAcF/P86C ($F_{DA}$, gray), the corresponding DOnly-reference sample ($F_{D0}$, green), and instrumental response function (IRF, blue). At the top, the weighted residuals (w.res.) of different analyses by models with DA-distance distributions composed of one (violet), two (cyan), and three (orange) normal distributions are shown in corresponding colors. In the 2- and 3-component analysis, all FRET-measurements are jointly analyzed (global), and the species fractions of the states are shared among all 33 datasets. For the 1-component model of the variant S36pAcF/P86C, we find a mean DA-distance of 45.7 Å with a width of 17.6 Å. The analysis results of the 3-component model are summarized in Table S2D-F. **(B)** Uncertainties, $\Delta R_{DA}(k_{FRET})$, of the 3-component analysis for the variant S36pAcF/P86C. Other contributions to the total uncertainty, $\Delta R_{DA}$, i.e. $\kappa^2$, are not shown for clarity. To the sides, the marginalized (projected) histograms of the sampled model parameters are shown. The lines highlight the most likely combination of distances. **(C)** Uncertainty estimation of the species fraction of the third (minor) state, $x(C_3)$, for the three-component analysis. The fraction $x(C_3)$ was varied from 0-0.32 followed by a subsequent minimization of all other model parameters. This yields the global, reduced $\chi^2_{r,global}$ of all 33 FRET eTCPSC measurements in dependence of $x(C_3)$. This curve has a minimum at $x(C_3) = 0.21$ ($\chi^2_{r,min}$ =1.074). Points above the red line ($\chi^2_r$ =1.0761, *p*-value = 0.68) are significantly worse than the best analysis result as judged by an *F*-test (225 parameters, 100,000 channels).

## 2.4 Structural features of conformational states

To compare the experimental distances $\langle R_{DA,exp}\rangle$ obtained from the fluorescence decays – under consideration of their respective uncertainties $\Delta R_{DA}$ – (Section 2.3) to the structural models deposited in the PDB, we clustered all available 578 structures of T4L and aligned them. We observed that the structural models of T4L group into *open*, *ajar*, and *closed* clusters (based on the proximity of the CTsD and NTsD, Table S3 in Supplementary Information 2.4) with an intra-cluster root mean-squared displacement of less than 1.8 Å. The representative structures of these clusters are given by PDB IDs 172L, 1JQU, and 148L for the *open*, *ajar*, and *closed* conformations, respectively (Figure 5A).

Next, we applied the FRET Positioning System (FPS) [41] to compute an error function ($\chi^2_{r,FPS}$) that compares the three sets of 33 distances $\langle R_{DA,exp}\rangle$ to the modeled distances $\langle R_{DA,model}\rangle$ by FPS. In $\chi^2_{r,FPS}$, we consider explicitly the uncertainties, $\Delta R_{DA}$, of the distance $\langle R_{DA,exp}\rangle$ [33]. The overall agreement (minimum $\chi^2_{r,FPS}$) for the distance sets for $C_1$ and $C_2$ is best for 172L and 148L, respectively (Figure 5B). In Figure 5C, $\langle R_{DA,model}\rangle$ for 172L and 148L are compared to $\langle R_{DA,exp,}\rangle$ of $C_1$ and $C_2$, respectively. A linear regression (red line) with a slope close to one demonstrates the absence of significant systematic deviations.

Structurally, the *ajar* state is more closed than the *open* state and more open than the *closed* state, most likely representing an intermediate conformation or it could arise from structural instabilities introduced by specific mutations such as W158L [45]. The deviation from the *open* and *closed* state is clearly reflected in the (slightly) elevated $\chi^2_{r,FPS}$. Consequently, we can safely assign $C_1$ as *open* and $C_2$ as *closed* state. However, none of the cluster representatives can be assigned to the $C_3$ state as judged by the disagreement with the data (Figure 5B, $\chi^2_{r,FPS}$). Thus, we conclude that $C_3$ is an *excited* conformational state of currently unknown structure.

A

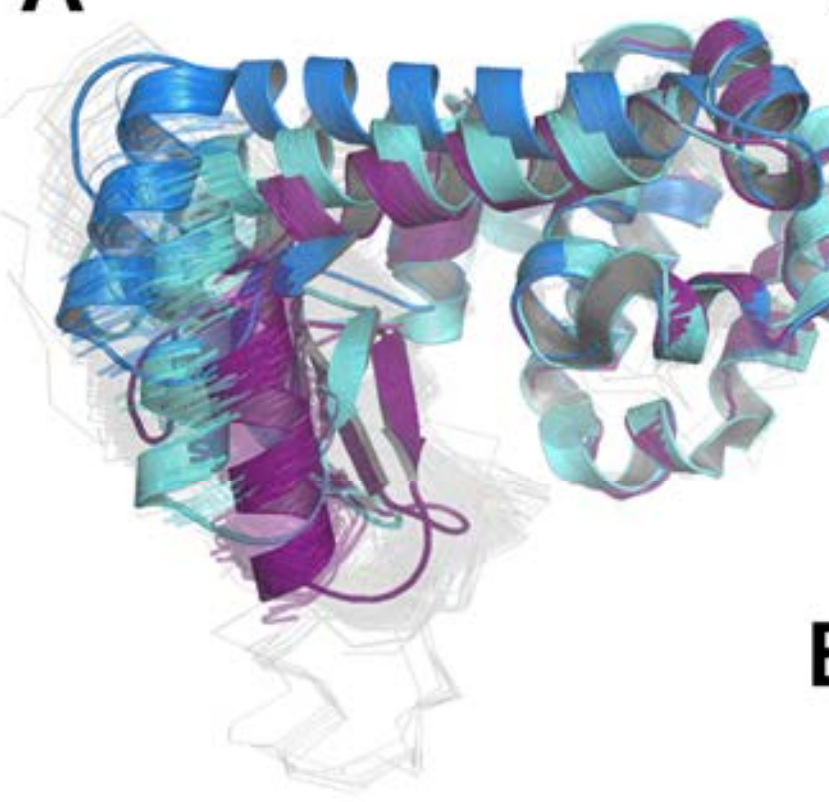

C

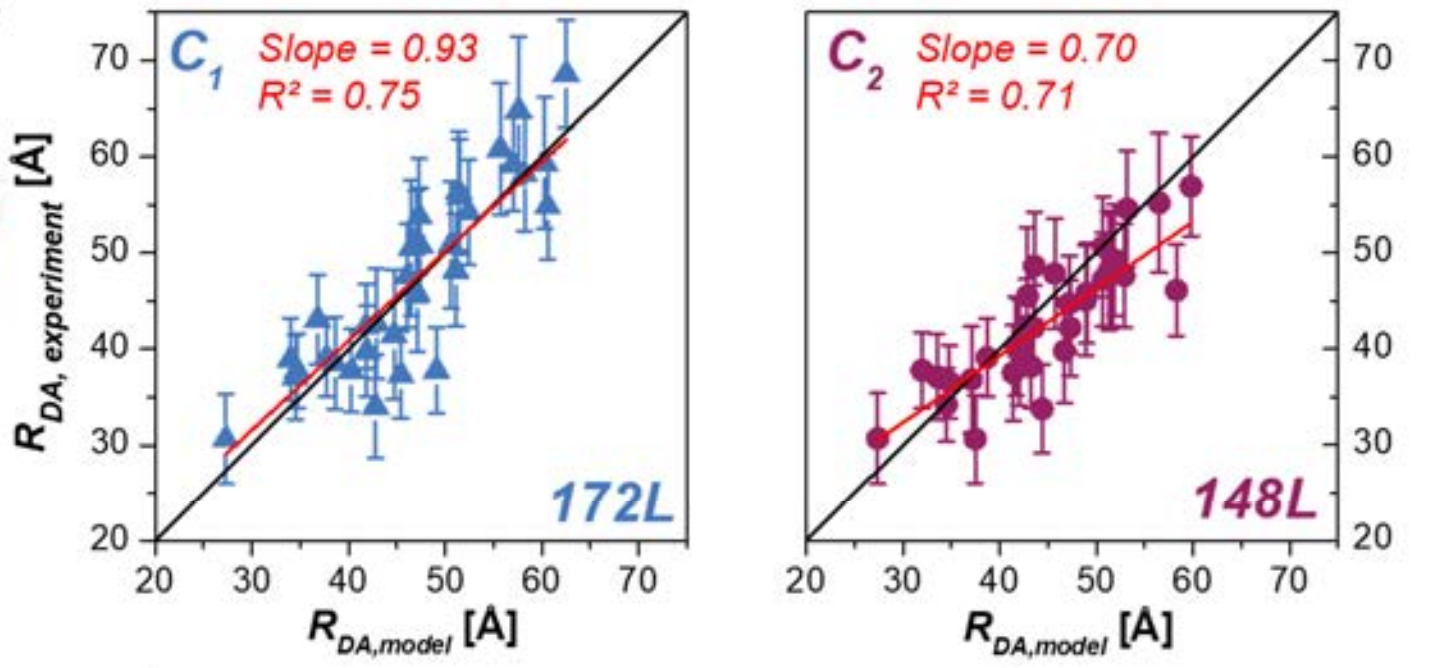

B

| data set \ PDB ID | reduced $\chi_r^2$ 172L Open | 1JQU Ajar | 148L closed |
|---|---|---|---|
| C1 | 0.79 | 1.03 | 1.54 |
| C2 | 1.65 | 0.99 | 0.84 |
| C3 | 11.00 | 9.39 | 8.59 |

**Figure 5 PDB screening. (A)** Overlay of the PDB structures used for screening. Blue, light blue and violet cartoons show the cluster representative of the *open* (172L), *ajar* (1JQU), and *closed* (148L) conformation of T4L. **(B)** Reduced $\chi^2_r$ for each distance set compared to the expected distances from the selected cluster representative. **(C)** The experimental distances $R_{DA,experimental}$ of the $C_1$ and $C_2$ dataset are plotted against the model distances $R_{DA,Model}$ from the best PDB structure representative and fitted linearly (red lines). The black lines show a 1:1-relationship.

## 2.5 Functionally relevant states within T4L's enzymatic cleavage cycle

### 2.5.1 Detection of $C_3$ by EPR

We use Double Electron-Electron Resonance (DEER) to provide additional support for the $C_3$ state. The variant S44pAcF/I150C (wt**) is labeled with the appropriate spin label HO-225 to produce the variant 44R1/150R1, which is chosen because it contains spin labels in each domain. The distribution of interspin distances is broad for this variant both in the wt* background and for the variant with the covalently linked substrate adduct T26E(+), which resembles the enzyme-product-complex *EP* within the catalytic cleavage cycle of T4L (Figure 6A, solid lines, Figure S5D). Some of the most frequently observed distances, which fall within the range of 40 and 60 Å, may correspond to the various substates of the *open* ($C_1$) and *closed* state ($C_2$), as expected from the enzyme hinge-bending motion. In addition to these distances, a second population of distances of ~ 35 Å is observed. To ensure that this small population is not an artifact of the Tikhonov regularization algorithm [46,47] or due to the rotamer populations of the spin label-carrying side chain, we lower the pH to influence the conformational equilibria of the states [48]. The FRET-experiment with the variant S44pAcF/I150C shows an increase in the population of $C_3$ at pH 2 (Figure 6B), and the analogous DEER experiment at pH 3 shows a remarkably similar redistribution of interspin distances. Compared to physiological pH conditions (Figure 6A, dashed trace), these distances exhibit a shortening that is consistent with the $C_3$ state, thus validating our conclusion that EPR and FRET do show the *excited* state $C_3$.

### 2.5.2 Trapped reaction states of T4L

To mimic functional enzymatic states, we mutated the residues E11 and T26 at the active site using the backbone of the S44pAcF/I150C variant, also named wt** [11,35,49]. We use wt** because of the advantage in clearly resolving all three conformations of the free enzyme (*E*) by FRET. These mutations help identifying the role of $C_3$ during enzyme catalysis: E11A, which inactivates T4L, causes the enzyme to bind its substrate *S* (peptidoglycan from *Micrococcus luteus*) while obviating the expected hydrolysis reaction [49]. Thus, in the presence of excess substrate, this mutation mimics the enzyme-substrate complex (*ES*). We monitor the effect of the substrate binding for the E11A mutation by FCS and compare the characteristic translational diffusion times, $t_{diff}$, in both the absence and presence of substrate. While $t_{diff}$ is small (0.54 ms, Figure 6C, green curve) without the substrate, it increases by several orders of magnitude when the large substrate is introduced (Figure 6C, yellow curve). Moreover, the shift towards the larger donor anisotropy values upon incubation with substrate also provides additional evidence for substrate binding without cleavage (Figure S2E).

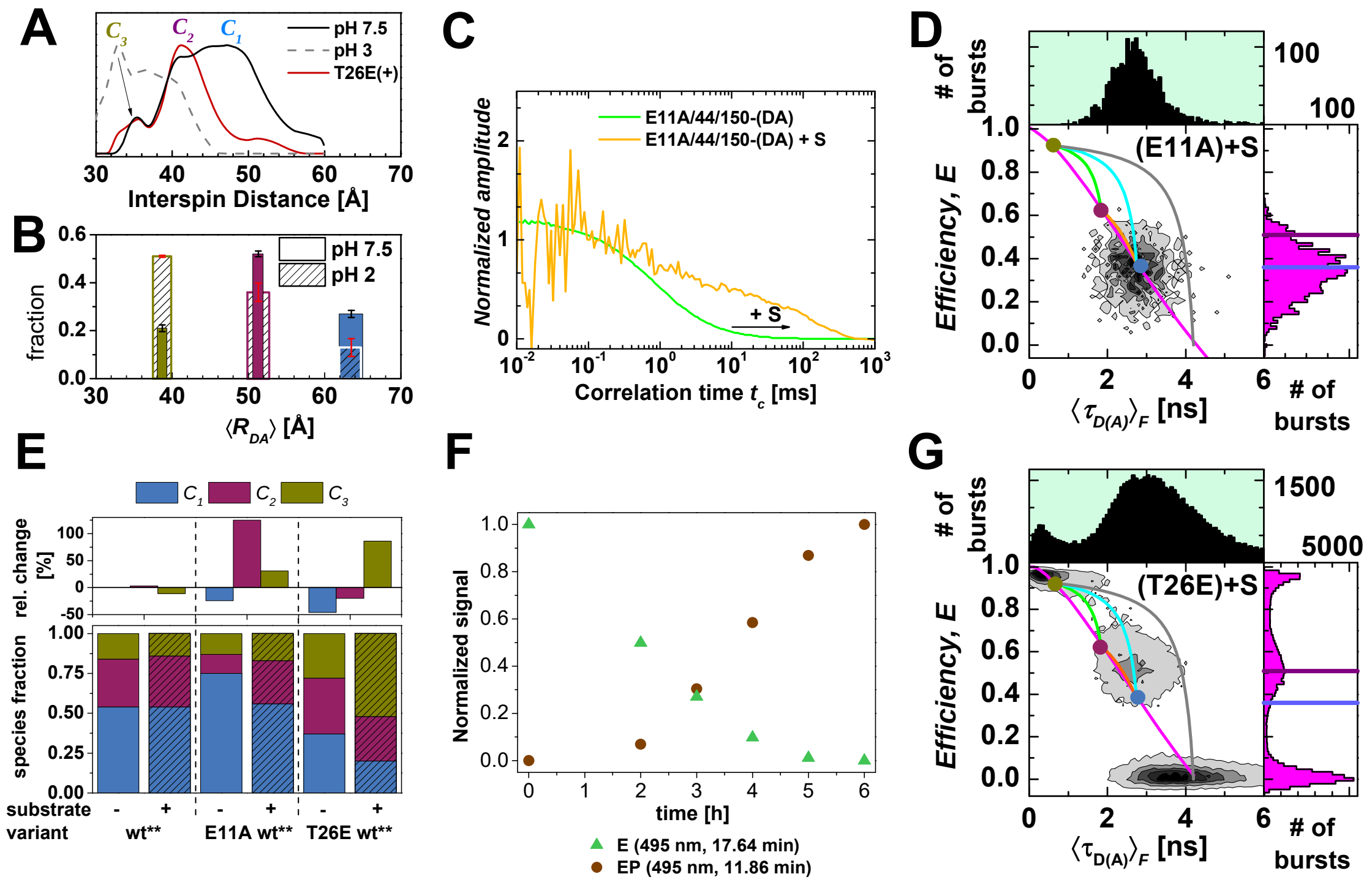

**Figure 6 Functional states of T4L. (A, B)** Normalized interspin distance distributions for S44(*R1*)/I150(*R1*) and T26E/S44(*R1*)/I150(*R1*) from DEER experiments, and DA-distance distributions for S44pAcF/I150C

from eTCSPC. The linker effects prevent a direct comparison of the interlabel distances. The shoulder of the distribution (as indicated by the arrow) at a short distance agrees with the short interdye distance of state $C_3$. For eTCSPC data, the width of the bar represents the uncertainty in distance (filled bars: pH 7.5, hatched bars: pH 2). The error bars represent the statistical uncertainty in the amplitudes (black: pH 7.5, red: pH 2). $C_3$ increases in population at low pH. **(C, D)** The effects of the substrate on E11A/S44C/I150C. **(C)** Overlay of normalized *sCCF* of E11A/S44C/I150C with and without substrate. Consistent with the larger rotational correlation, we observe a shift of $t_{diff}$ towards longer times for the variant E11A/S44C/I150C when incubated with the substrate. (**D**) MFD-histograms for the variant E11A/S44C/I150C with substrate. Upon addition of the substrate, we observe a shift towards lower *E* values. **(E)** Species fractions of the variants S44pAcF/I150C (wt**), E11A/S44C/I150C (E11A wt**), and T26E/S44pAcF/I150C (T26E wt**) used to mimic free enzyme (*E*), enzyme-substrate complex (*ES*), and enzyme-product bound state (*EP*) without (-) and with (+) peptidoglycan. At the top, the relative change in fractions upon addition of peptidoglycan is shown. **(F, G)** Effects of the substrate on the variant T26E/S44pAcF/I150C. (**F**) Reverse-phase HPLC is used to monitor the adduct formation of the labeled T4L with peptidoglycan via plotting of the normalized signal at 17.64 min (*E*) and 11.86 min (*EP*) after background subtraction. (**G**) MFD-histograms for the variant T26E/S44pAcF/I150C with substrate. An accumulation of the high FRET state is observed.

Sub-ensemble TCSPC analysis of the DA-subpopulation of the *ES* state (E11A/wt** in the presence of substrate, Figure 6D, Figure S7A-D in Supplementary Information 2.5) reveals an increase of 125 % in the population corresponding to $C_2$ compared to the free enzyme state *E*, with a concomitant reduction of $C_1$. In contrast, no effect of substrate binding for wt**-(DA) is observed because *ES* is not trapped (Figure 6E).

Although the variant T26E cleaves the substrate, the formation of a covalent adduct (PDB ID 148L) prevents a release of the formed product [11]. Therefore, we use this intermediate adduct to mimic the product-bound enzyme state (*EP*). To confirm the adduct formation under our measurement conditions, we monitor the adduct formation of labeled T4L (T26E/wt** variant) by HPLC (Figure 6F). T4L without substrate (*E*) elutes at ~18 min. After incubation with the substrate, the peak of *E* drops, and a new elution peak at ~12 min is detected with increasing incubation time (Figure 6F, S6), indicative of the adduct form of T4L (*EP*). Both ensemble (Figure 6E) and sm MFD-measurements (Figure 6G, Figure S7E-H in Supplementary Information 2.5) show a significant increase of the relative fraction of the $C_3$ state, an effect also observed in the EPR measurements (Figure 6A). In the T26E variant, the accumulation of the $C_3$ state is connected to the inability of this variant to release a part of the product [11]. We conclude that the new *excited* conformational state must be involved in this step.

## 3. Discussion

### 3.1 The $C_3$ state and its structural properties

To corroborate the existence of $C_3$, we detail our experiments in *i*) the kinetic behavior in sm-experiments, *ii*) the error statistics of data analysis, *iii*) the structural validation of the obtained FRET parameters, and *iv*) the effect of specific mutations.

*Kinetic behavior.* Considering 18 out of 33 variants with FRET-pairs, the sm-experiments directly show the presence of an additional DA-subpopulation in the MFD-histograms, which differs significantly from $C_1$ and $C_2$ (Figures 2, 3, S3). This DA-subpopulation is either populated or depopulated with specific mutations that alter the overall catalytic activity of T4L. Moreover, our global fluctuation analysis recovers at least two relaxation times that are shared throughout all studied variants (Figures 3, S4). Applying kinetic theory, two relaxation times indicate at least three states in equilibrium, which are reproduced by Brownian dynamics simulations (Figure S8).

*Error statistics*. Key to the analysis and determination of $C_3$ by ensemble fluorescence decays is the use of global analysis of all 33 variants, which reduces the number of free parameters, increases fitting quality (Supplementary Information 1.7), and gives a consistent description with sm-experiments. Moreover, assuming that CTsD and NTsD are rigid, there are six independent degrees of freedom in the system, which we significantly oversample by measuring 33 variants.

*Structural validation*. In contrast to our 3-component model, the global analysis of eTCSPC data using a 2-component model yields two distance sets, which cannot describe the expected interdye distances of the known conformers ($C_1$ (172L) and $C_2$ (148L)). Furthermore, for the 2-component model, we do not observe the expected linear correlation between the modeled and experimental interdye distances, as shown in Figure S5B-C.

*Specific mutations*. The variant Q69pAcF/P86C is especially informative, as the donor is placed in the middle of helix c (Orange Figure 1A), which connects both domains, while the acceptor is located in the middle of helix d, which is part of the CTsD (Brown Figure 1A). According to FPS, the interdye distances for $C_1$ (172L) and $C_2$ (148L) states are hardly distinguishable by FRET, $\langle R_{DA} \rangle$ of 34 and 35 Å, respectively. Assuming that both domains preserve their secondary structure, the compaction of T4L in $C_3$ can only proceed by kinking the helix c. Given the location of the dyes and the extension of the dye linkers, expected dye orientations will lead to an increase of the interdye distance for $C_3$, i.e., a greater interdye distance is expected for $C_3$ compared to the experimental distances for $C_1$ (39 Å) and $C_2$ (37 Å). The additional observed distance of 52.4 Å agrees with this hypothesis (Table S2F, Figure 2C).

An additional inactive variant (R137E) [50,51], which disturbs the salt bridge between residues 22 and 137, reduces the population of $C_3$ by ~ 50 % (Figure S2D, Table S2G), a phenotype also observed for the inactive variant E11A/S44C/I150C.

In conclusion, the existence of $C_3$ demonstrates a greater level of complexity of the domain motions of T4L than a single hinge-bending motion, which is in agreement with recent indirect observations [17,21,26]. The complex exchange dynamics between the conformations with relaxation times of 4 and 230 µs and the small population of $C_3$ may explain the difficulties of other experimental biophysical methods and MD simulations in identifying this exchange.

### 3.2 Relating conformational states and enzyme function

A three-step process characterizes the T4L hydrolysis of peptidoglycans. First, the glycosidic bond of the substrate (*S*) is protonated by E11 followed by the simultaneous nucleophilic attack of water molecules, which are hydrogen-bonded to residues D20 and T26, on the C-1 carbon of *S*. As a result, the covalent adduct (*ES*) is observed in PDB ID 148L [11]. Second, the proton is presumably returned from D20 to E11 via solvent transfer. The third and final step is the product release from the active site to regenerate the enzyme to the original state.

In view of the structural dynamics and to link T4L's functional cycle to our three observed conformations, we use an extended MMm (eMMm) as suggested by Kou *et al.* [27] (eq. 1).

$$\boldsymbol{E}+\boldsymbol{S} \underset{}{\overset{I}{\rightleftharpoons}} \boldsymbol{ES}\,[\overset{II}{\rightleftharpoons} \boldsymbol{EP}\,] \xrightarrow{III} \boldsymbol{E^*}+\boldsymbol{P}\,; \quad \boldsymbol{E^*} \xrightarrow{\delta} \boldsymbol{E} \qquad (1)$$

Here, the substrate *S* binds reversibly to the enzyme *E* to form an enzyme-substrate complex *ES* and is converted to the product *P*, resulting in an *EP* complex with the product still bound to the enzyme. A transition of *E* to an excited state $E^*$ then releases *P* from the complex, followed by a relaxation of the free enzyme $E^*$ to *E*.

Next, we consider our results in the light of the eMMm framework. First, by using the S44/I150C backbone and two key functional T4L variants (E11A, T26E), we create the relationship between the conformational ($C_1$, $C_2$, $C_3$) and the above described reaction states (*E*, *ES*, *EP*) for purposes of elucidating the functional role of $E^*$ (Figure 7). We observed a significant difference of the populations of the conformers (Figure 6) between the three reaction states.

To connect conformational equilibria with catalysis, we analyzed the relative changes in species fractions observed across the functional variants in both the absence and presence of substrate (upper panel of Figure 6E) by generating a 3 x 3 state matrix (Figure 7A). As indicated in the matrix, specific conformational states are favored in each enzyme reaction state (Figure 7A, B). For this representation, we use the relative population changes as compared to the wt** to monitor the conformational populations of the different enzyme states.

In the free enzyme state *E*, the open conformation $C_1$ is mostly populated to enable substrate binding, which initiates the catalytic cycle through the formation of *ES*. Through this cycle, the closed conformation $C_2$ now becomes the most abundant conformation [11]. In this conformation, the glycosidic bond can be cleaved such that $C_2$ connects both *ES* and *EP*. In our studies, we determined that the product release may occur in the compact conformation $C_3$, a population that is greatly increased in *EP.* Thus, $C_3$ links *EP* and *E* so that the original enzyme *E* is regenerated from *EP*, which closes the enzymatic cycle. As a consequence, the compact state $C_3$ now corresponds to the *excited* conformational state $E^*$ (eq. 1) (Figure 7A, C), the function of which is to disperse the product [15,26] (Figure 7A, C). Most strikingly, even under saturating conditions, which favor the *ES* and *EP* states, the enzyme was observed to remain in dynamic equilibrium between the conformational states, rather than transforming entirely into a single conformational state.

To visualize the relative energetic changes of the enzyme during the various steps of the catalytic cycle, we use the species fractions and reaction rate constants to compute the relative energy landscape based on the Arrhenius equation (Figure 7C, Figure S9 in Supplementary Information 2.7). We observe a sequential increase of the $C_3$ population consistent with a ratchet model for purposes of providing directionality on the reaction [52] [53] [54] beyond the directionality introduced by the excess of *S*.

All our evidence suggests that the conformational state $C_3$, which appears to be more compact than any other structure known of T4L, is compulsory *after rather than before* *catalysis*. Thus, the compact nature of this structure suggests a functional role that is related to product release via an excited state $E^*$ (eq. 1). This mechanism can be an evolutionary advantage when processivity is required for function. On the contrary, considering a system with only two conformational states and without an active cleaning mechanism, the stochastic dissociation of the product can become rate-limiting given a high affinity of the product to the enzyme in the *EP* state. Indeed, a large surfeit of substrate is always characteristic of the *in vivo* conditions for T4L. Thus, the use of a three-state system to decouple the substrate access and product release can mitigate the occurrence of substrate inhibition in a two-state system when the route to the active site is clogged by excess substrate concentrations [55].

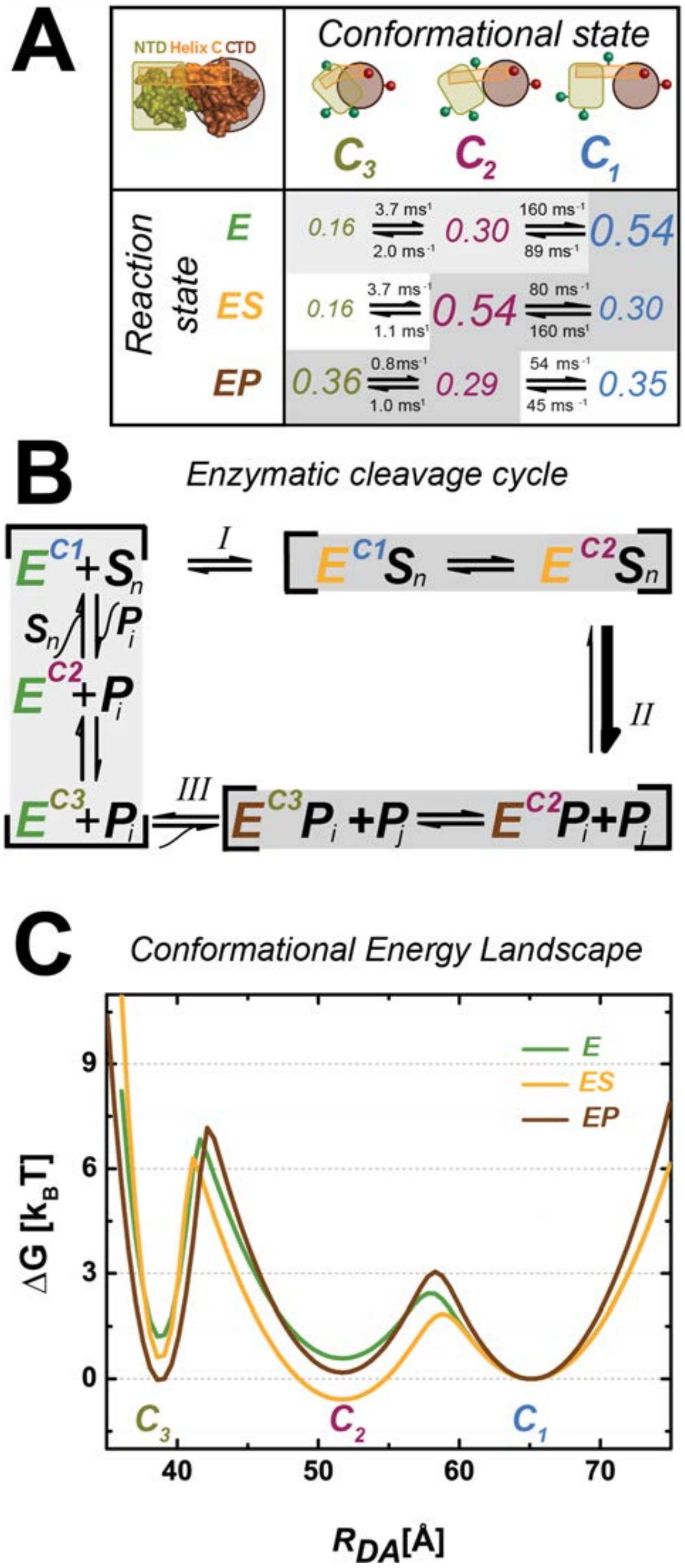

**Figure 7 Energy landscape of T4L.** **(A)** T4L interconverts between three major $C_1$, $C_2$, and $C_3$ conformational states. The population fractions of $C_1$, $C_2$, and $C_3$ are normalized to the variant S44AcF/I150C in the absence of substrate using the relative changes in Figure 6E to correct for the influence of specific mutations in that absence. The different font sizes represent the species fractions $x_i$ for each conformer according to Table S2G and satisfy $x_1$+$x_2$+$x_3$=1. The three enzyme states were monitored via the following three enzyme variants: (*i*) the free enzyme state *E* via S44AcF/I150C; (*ii*) the enzyme-substrate state *ES* via

the inactive E11A/S44C/I150C with bound substrate; and (*iii*) the enzyme product state *EP* via the product adduct with T26E/S44AcF/I150C after substrate cleavage. The reaction rate constants were calculated according to the process detailed in Supplementary Information 2.6, the confidence intervals of which are shown in Table S4C. (**B**) The peptidoglycan chain with $n$ subunits ($S_n$) is cleaved into two products ($P_i$ and $P_j$ with $n=i+j$) by T4L, both of which can be further processed by T4L until only the dimer of *N*-acetylglucosamine and *N*-acetylmuramic acid remains. The gray shaded steps indicate the conformational/reaction states observed. (**C**) Relative Gibbs free energy landscapes were calculated using $\Delta G^0 = -k_B T \ln\left(\frac{k_{ji}}{k_{ij}}\right)$, where $k_B$ is the Boltzmann constant and $T$ is the temperature; $k_{ij}$ are the reaction rate constants between states $C_i$ and $C_j$ for the data presented in panel A. The activation energies were calculated according to $\Delta G^{0+} = -k_B T \ln\left(\frac{k_{ji}}{k_0}\right)$ assuming $k_0 = 10^3$ ms$^{-1}$ as an arbitrary constant. The distributions consider $C_1$, $C_2$, and $C_3$ to follow a Gaussian distribution as a function of the interdye distance $R_{DA}$. The Gaussian widths ($\sigma_i$) were adjusted to satisfy the energy differences and calculated activation energies. Each energy landscape is independently normalized to $C_1$.

## 4. Conclusion

To oversample the anticipated simple hinge-bending motion of T4L, we effectively used 33 distinct FRET-pairs and identified three substrate-dependent fluorescence states that are in fast kinetic exchange. Inverting the positions of the dyes, e.g. S44pAcF/I150C-DA *vs*. S44pAcF/I150C-AD (Figure 3B), rules out specific interactions of the dyes with T4L. Functional variants change the relative populations of the fluorescence states that are determined in a substrate-dependent manner (Figure 6). A detailed enzymatic reaction scheme is developed to relate the fluorescence states to the three reaction states and provide a meaningful description of the data. Mutagenesis and stability studies indicate the stability of CTsD and a flexibility of the NTsD [4,10,36-38,56], which may be a necessary principle for the construction of enzymes undergoing conformational changes during catalysis. The combination of known structural models and fluorescence data is used to create a proposed novel structural state in the catalytic cycle of T4L involving a rearrangement of the reactive NTsD with respect to the CTsD, which is deemed consistent with the eMMm for enzyme kinetics. For a complete structural insight, we are now using the obtained FRET-restraints to present a potential model of $C_3$, which is left for a future report.

We anticipate that the presented fluorescence spectroscopic toolkit can accelerate the development of dynamic structural biology by resolving the behaviors of long- and short-lived excited states for purposes of characterizing their functional relevance. The elucidation of these conformational states is necessary for initiating a thorough in-depth understanding of the mechanisms of enzymes.

## 5. OnlineMethods

### 5.1 Sample preparation

T4L cysteine and amber (TAG) mutants were generated via site-directed mutagenesis as previously described in the pseudo-wild-type construct containing the mutations C54T and C97A (WT*), which was subsequently cloned into the pET11a vector (Life Technologies Corp.) [57] [58,59]. The plasmid containing the gene with the desired mutant was co-transformed with pEVOL [57] into BL21(DE3) *E. coli* (Life Technologies Corp.) and plated onto LB-agar plates supplemented with the respective antibiotics, ampicillin and chloramphenicol. A single colony was inoculated into 100 mL of LB medium containing the above-mentioned antibiotics and grown overnight at 37 °C in a shaking incubator. 50 mL of the overnight culture were used to inoculate 1 L of LB medium supplemented with the respective antibiotics and 0.4 g/L of pAcF (SynChem) and grown at 37°C until an $OD_{600}$ of 0.5 was reached. The protein production was induced for 6 hours by addition of 1 mM IPTG and 4 g/L of arabinose.

Cells were harvested, lysed in 50 mM HEPES, 1 mM EDTA, and 5 mM DTT at pH 7.5 and purified using a monoS 5/5 column (GE Healthcare) with an eluting gradient from 0 to 1 M

NaCl according to standard procedures. High-molecular weight impurities were removed by passing the eluted protein through a 30 kDa Amicon concentrator (Millipore), followed by subsequent concentration and buffer exchange to 50 mM PB, 150 mM NaCl pH 7.5 of the protein flow through with a 10 kDa Amicon concentrator. For the double cysteine mutant containing E11A, the temperature was reduced to 20 °C after induction and the cells were grown additional 20 hours to increase the fraction of soluble protein. This mutant was produced and purified as described above, except that only ampicillin for selection and IPTG for induction were needed.

Site-specific labeling of T4L was accomplished using orthogonal chemistry. For labeling the Keto group of the *p*-acetyl-L-phenylalanine (pAcF) amino acid at the N-terminus, hydroxylamine linker chemistry was used for Alexa488 and Alexa647 (Life Technologies Corp.). Cysteine mutants were labeled via a thiol reaction with maleimide linkers of the same fluorescent dyes. For spin labeling, the S44C/I150C double mutant was labeled with a methanthiosulfonate nitroxide (HO-225) as previously reported [59].

Binding of labeled T4L mutants to peptidoglycan from *Micrococcus luteus* (Sigma-Aldrich) was monitored by reverse phase chromatography using a C-18 column out of ODS-A material (4 X 150 mm, 300 Å) (YMC Europe GmbH, Dinslaken, Germany). The protein was eluted with a gradient from 0 to 80% acetonitrile containing 0.01% trifluoroacetic acid for 25 min at a flow rate of 0.5 ml/min. The labeled complex elution was monitored by absorbance at 495 nm.

### 5.2 Measurement and data analysis.

#### 5.2.1 Single-molecule experiments (MFD)

For single-molecule measurements we added 40 μM TROLOX to the measurement buffer to minimize the acceptor blinking and 1 μM unlabeled T4L to prevent any adsorption to the cover glass. A custom-built confocal microscope with a dead time-free detection scheme using 8 detectors (four green (τ-SPAD, PicoQuant, Germany) and four red channels (APD SPCM-AQR-14, Perkin Elmer, Germany)) was used for MFD and fFCS measurements. A time-correlated single photon counting (TCSPC) module with 8 synchronized input channels (HydraHarp 400, PicoQuant, Germany) was used to register the detected photon counts in the Time-Tagged Time-Resolved (TTTR) mode. The data was analyzed by established MFD procedures [28,30,39] and software, a more detailed description is given in Supplementary Information 1.1-1.3. Exemplary data analysis is shown in Figure S1 in Supplementary Information 2.1, MFD-histograms of all measurements are collected in Figure S2 and Figure S3.

**5.2.2 Filtered FCS (fFCS)**

Filtered FCS (fFCS) is a derivative of fluorescence correlation spectroscopy (FCS). In fFCS, the information on the fluorescent species contained in the time- and polarization- resolved fluorescence decays (exemplary shown in Figure S4A,C) was used to amplify the transitions between the species of interest [29,31,42]. For this, we constructed species-selective filters (exemplary shown in Figure S4B,D) based on the major and minor population in the smFRET experiment and calculated species-selective auto- (*sACF*) and cross-correlation functions (*sCCF*). The in total four curves (two *sACF*'s and two *sCCF*'s) were analyzed jointly using established fitting models (eq. S15-17) [29,31,42]. Details see Supplementary Information 1.4.

**5.2.3 Fluorescence decay analysis (eTCSPC)**

Fluorescence decays of all samples and single-labeled reference samples were collected on either an IBH-5000U (IBH, Scotland) or a Fluotime 200 (Picoquant, Germany) system. We collected high-precision fluorescence decays histograms with 30 million photons to precisely determine the FRET parameters of limiting states together with their corresponding structural properties. eTCSPC has the advantage of better photon statistics, polarization-free measurements due to magic-angle detection, a keenly evolved instrumental response function (IRF), low background fluorescence, and the absence of photobleaching at low excitation powers. As the fluorophores were coupled via long and flexible linkers, this resulted in a DA-distance distribution even for single protein conformations. For our data analysis, we assumed that the dyes rotate quickly ($\kappa^2 = 2/3$) and diffuse slowly compared to the fluorescence lifetime (~ns) [34]. We validated the assumption of fast rotating dyes by time-resolved anisotropy measurements (Table S5A-C). Moreover, we interpret the broadening of the DA-distance distributions, beyond what is expected from flexible linkers, as evidence for conformational heterogeneity of the host molecule. To dissect the donor-quenching by FRET (i.e. FRET-induced donor decay), we jointly analyzed the DA- and DOnly-dataset, where the donor fluorescence lifetime distribution was shared with the DA-dataset (Supplementary Information 1.7, eq. S20, S24-S27) [44]. We compared three different fit models (Table S1 in Supplementary Information 1.7). The results are summarized in Table S2 in Supplementary Information 2.3. We estimated the statistical uncertainty of the model parameters by making use of the known shot noise of the fluorescence decays. We randomly sampled the model parameters by a Markov chain Monte Carlo (MCMC) method to estimate their uncertainties for a single dataset (Supplementary Information 1.7)[60]. The applied joint, global fit significantly reduced the overall dimensionality of the analysis but still left too many degrees of freedom (Supplementary Information 1.7) for an exhaustive sampling by MCMC. Hence, we applied a support plane analysis to estimate the model parameter uncertainties, in which we systematically varied $x_3$ while minimizing all other parameters.

**5.2.4 EPR**

For double electron electron resonance (DEER) measurements of doubly spin labeled proteins, ~200 µM spin-labeled T4L containing 20 % glycerol (v/v) was placed in a quartz capillary (1.5 mm i.d. X 1.8 mm o.d.; VitroCom) and then flash-frozen in liquid nitrogen.

Sample temperature was maintained at 80 K. The four-pulse DEER experiment was conducted on a Bruker Elexys 580 spectrometer fitted with an MS-2 split ring resonator. Pulses of 8 ns (π/2) and 16 ns (π) were amplified with a TWT amplifier (Applied Engineering Systems). Pump frequency was set at the maximum of the central resonance, and the observe frequency was 70 MHz less than the pump frequency. Dipolar data were analyzed by using a custom program written in LabVIEW (National Instruments Co.). Distance distributions were acquired using Tikhonov regularization [46].

**5.3 Recovering the reaction network: Brownian dynamics simulation**

To solve the ambiguity in the connectivity of states and kinetics of T4L, i.e. between the two possible analytical solutions of the transition rate matrix (eq. S31), we used Brownian dynamics simulation of single-molecule and fFCS experiments. Simulations of single-molecule measurements were done via Brownian dynamics [61] [62] [63,64]. The spatial intensity distribution of the observation volume was assumed a 3D Gaussian. In contrast to other simulators, freely diffusing molecules in an "open" volume are used. Transition kinetics is modeled by allowing $i \rightarrow j$ transitions. The time that molecules spend in $i$ and $j$ states ($t_i$ and $t_j$, respectively) are exponentially distributed with

$$P(t_i) = k_i^{-1} \exp(-k_i t_i) \text{ and } P(t_j) = k_j^{-1} \exp(-k_j t_j). \quad (2)$$

Simulated photon counts are saved in SPC-132 data format (Becker & Hickel GmbH, Berlin, Germany) and treated as experimental data. To quantify the difference between the two possible, simulated models and the experimental data, we calculated the relative $\chi^2$ for the one-dimensional and two-dimensional MFD-histograms (Supplementary Information 2.6, Table S4A,B).

**5.4 Simulation of interdye distances & structural modelling (FPS)**

**5.4.1 Accessible contact volume (ACV) and interdye distance**

The accessible volume considers dyes as hard sphere models connected to the protein via flexible linkers (modeled as a flexible cylindrical pipe) [34]. The overall dimension (width and length) of the linker is based on their chemical structures. For Alexa488 the five carbon linker length was set to 20 Å, the width of the linker is 4.5 Å and the dye radii $R_1$= 5 Å, $R_2$= 4.5 Å and $R_3$= 1.5 Å. For Alexa647 the dimensions used were: length = 22 Å, width = 4.5 Å and three dye radii $R_1$= 11 Å, $R_2$= 3 Å and $R_3$= 3.5 Å. Here, the dye distribution was modeled by the accessible contact volume approach (ACV) [33], which is similar to the accessible volume (AV) [34], but defines an area close to the surface as contact volume.

Similar approaches have been introduced before to predict possible positions for EPR and FRET labels [65 41 66]. The dye is assumed to diffuse freely within the AV and its diffusion is hindered close to the surface. The part of AV which is closer than 3 Å from the macromolecular surface (contact volume) is defined to have higher dye density $\rho_{Dye,trapped}$. The spatial density $\rho_{Dye}$ along $R$ is approximated by a step function: $\rho_{Dye}$ = [$\rho_{Dye,free}$, $R$ < 3 Å; $\rho_{Dye,trapped}$, R ≥ 3 Å]. The $\rho_{Dye,trapped}/\rho_{Dye,free}$ ratio is calculated from the fraction of the trapped dye $x_{Dye,trapped}$ for each labeling position separately: $\rho_{Dye,free}/\rho_{Dye,trapped} = V_{Dye,trapped}\cdot(1 - x_{Dye,trapped})/(x_{Dye,trapped}\cdot V_{Dye,free})$. For this, the fraction $x_{Dye,trapped}$ was approximated by the ratio of the residual, $r_\infty$, and fundamental anisotropy, $r_0$, determined by the time-resolved anisotropy decay of the directly excited dyes (Table S5).

To account for dye linker mobility, we generated a series of ACV's for donor and acceptor dyes attached to T4L placing the dyes at multiple separation distances. For each pair of ACV's, we calculated the distance between dye mean positions ($R_{\mathrm{mp}}$)

$$R_{mp} = \left| \left\langle \vec{R}_{D(i)} \right\rangle - \left\langle \vec{R}_{A(j)} \right\rangle \right| = \left| \frac{1}{n}\sum_{i=1}^{n} \vec{R}_{D(i)} - \frac{1}{m}\sum_{j=1}^{m} \vec{R}_{A(j)} \right|, \qquad (3)$$

where $\vec{R}_{D(i)}$ and $\vec{R}_{A(i)}$ are all the possible positions that the donor fluorophore and the acceptor fluorophore can take. However, in ensemble TCSPC (eTCSPC) the mean donor-acceptor distance is observed:

$$\langle R_{DA}\rangle = \left|\left\langle \vec{R}_{D(i)} - \vec{R}_{A(j)} \right\rangle\right| = \frac{1}{nm}\sum_{i=1}^{n}\sum_{j=1}^{m}\left|\vec{R}_{D(i)} - \vec{R}_{A(j)}\right|, \quad (4)$$

which can be modeled with the accessible volume calculation.

The relationship between $R_{mp}$ and $\langle R_{DA}\rangle$ can be derived empirically following a third order polynomial from many different simulations. The $\langle R_{DA}\rangle$ is not directly related to the distance between atoms on the backbone (Cα-Cα), except through the use of a structural model.

**5.4.2 FRET positioning & screening**

FPS is done in four steps, and its flow is based on the recent work by Kalinin *et al.* [41]. In order to do our experimental design using the available PDB structures of T4L with respect to our FRET data, FPS calculates the donor and acceptor accessible volumes for each donor-acceptor labeling scheme. We then compute an error function for each conformation $j$

$$\chi^{2\,(j)}_{r,FPS} = \frac{1}{N}\sum_{i=1}^{N}\frac{\left(\langle R_{DA}\rangle^{(i)}_{\text{experiment}} - \langle R_{DA}\rangle^{(i)}_{\text{model, j}}\right)^2}{\Delta R^2_{DA}\left(\kappa^2, k_{FRET}\right)^{(i)}}, \quad (5)$$

where $N = 33$ is the total number of FRET distances ($\langle R_{DA}\rangle$) and the overall uncertainty $\Delta R^2_{DA}\left(\kappa^2, k_{FRET}\right)$ is determined by the error-propagation rule (Eq. (S33)):

$$\Delta R^2_{DA}\left(\kappa^2, k_{FRET}\right) = \Delta R^2_{DA}\left(\kappa^2\right) + \Delta R^2_{DA}\left(k_{FRET}\right) \quad (6)$$

$\Delta R^2_{DA}\left(\kappa^2\right)$ is the uncertainty from the mutual orientation of donor and acceptor ($\kappa^2$ errors) and can be estimated by analyzing anisotropy decays accessible in eTCSPC and MFD [67]. We assume a weak dependence of $R_0$ on $\kappa^2$ [68] (Table S5D). $\Delta R^2_{DA}\left(k_{FRET}\right)$ is calculated from the spread of obtained distances for the three states of the global fit using the shortest ($\langle R_{DA,min}\rangle$) and longest distance ($\langle R_{DA,max}\rangle$) below the 1σ-threshold (Table S2D-F). As the distance $\langle R_{DA}\rangle$ of the global fit with the lowest $\chi_r^2$ ($x_3$ = 0.18) is not necessarily the average of $\langle R_{DA,min}\rangle$ and $\langle R_{DA,max}\rangle$, both $\Delta R_{DA}^2(k_{FRET})$ and the resulting $\Delta R_{DA}^2(\kappa^2, k_{FRET})$ are not symmetric. The error propagation was applied to all measured FRET distances (Table S2D-F).

We clustered all PDB models using the Cα RMSD (Root Mean Squared Deviation) as the distance measure to trace the backbone course. Clustering allowed us to sort all PDB models into three distinct groups based on the similarity of their backbone shapes. We found that the structural models of T4L group into *open*, *ajar*, and *closed* clusters (based on the proximity of the CTsD and NTsD) with an intra-cluster RMSD of less than 1.8 Å. Representative structures of these clusters are given by PDB IDs 172L, 1JQU, and 148L for the *open*, *ajar*, and *closed* conformations, respectively (Figure 5A). This was done using the agglomerative hierarchical complete-linkage clustering of the "fastcluster" [69] software.

In Table S3 we provide the complete breakdown of the three clusters.

**5.5 Controls for FRET**

As the problems inherent in the use of smFRET involve complexities associated with the labels, we performed ten controls to check for any potential label artifacts. Please refer to the SI for additional data and extended controls (Supplementary Information 2.8).

1. The labeling does not alter enzyme function with the labeled T26E variant indicating an expected adduct formation (Figure 6, S6).

2. Local quenching of the donor dye is considered when calculating distances and cross correlations (Table S2).

3. The triplet state quenchers do not affect the observed relaxation times and amplitudes on the species cross-correlation (Figure S10).

4. The acceptor cis-trans isomerization does not contribute to the signal on the species correlation analysis (Figure S10).

5. The $\kappa^2$ distributions indicated the validity of our assumption of $\kappa^2 = 2/3$. Table S5 summarizes the residual anisotropies ($r_\infty$) of D - donor, A - acceptor and A(D) by the FRET-sensitized emission of acceptor used for calculating $\kappa^2$ distributions [28].

6. The existence and the population fraction of the new conformational state $C_3$ with a confidence rate of 68 % between 10 – 27 % is consistent across our library of 33 variants with $x_3$ = 21 %. The variation of the experimental uncertainty is consistent with the determination in the literature that mutations slightly affect the conformational stability of T4L, which was measured in chemical denaturation experiments [56]. We thus attribute this variability of the species fractions to mutation effects.

7. All 33 variants provide a consistent view of the T4L conformational states, in which we determined after X-ray crystallography a consistency with the two limiting structures determined by T4L without outliers (Figure 5, Table S2D-F).

8. We oversample the FRET restraints to reduce the uncertainty introduced from each point mutation (Figure 5, Table S2D-F).

9. The thermodynamic stability and proper folding of our variants were verified by chemical denaturation with urea and by measuring CD spectra for both unlabeled and labeled T4L (data not shown).

10. We fit time resolved fluorescence decays with various models to provide a consistent view of the conformational space (Supplementary Information 1.5 and 1.7).

**Data availability**

The datasets generated during and/or analyzed during the current study are available from the corresponding authors on reasonable request.

**Code availability**

Most general custom-made computer code is directly available from http://www.mpc.hhu.de/en/software. Additional computer code custom-made for this publication is available upon request from the corresponding authors.

In-house programs were used *i*) in the confocal multiparameter fluorescence detection experiments, *ii*) to elucidate the filtered fluorescence correlation spectroscopy curves, and *iii*) to analyze the fluorescence lifetime measurements. Software for analysis of single-molecule measurements and fluorescence correlation analysis and their simulation is available at http://www.mpc.uni-duesseldorf.de and software for analysis of fluorescence decays can be downloaded from http://www.fret.at/tutorial/chisurf/.

**Acknowledgments:** We wish to acknowledge the support of Marina Rodnina, Philipp Neudecker, and M. Neal Waxham for their comments and suggestions on the manuscript, and Daniel Rohrbeck for support on the mutagenesis. We also wish to acknowledge Evan Brooks for generating several of the T4L variants and in assisting with the DEER sample preparation. This research was supported by the European Research Council through the Advanced Grant 2014 hybridFRET (671208) (to CAMS), by the NIH (grant R01EY05216 to WLH), NSF CAREER MCB (grant 1749778 to HS) and the Jules Stein Professorship Endowment (to WLH). HS wishes to acknowledge the support from the Alexander von Humboldt foundation and the Clemson Start-up funds. KH wishes to acknowledge the support of the NRW Research School BioStruct and the iGRASP$_{seed}$ International Graduate School of Protein Science & Technology.

**Author contributions:** H.S. and K.H. purified and labeled the protein. H.S., T.P., K.H., and D.R. measured and analyzed the FRET experiments. T.P., M.D. and H.G. performed structural screening. M.R.F. performed and analyzed EPR experiments. W.H. performed study design and EPR analysis. S.F. developed fluorescence analysis tools. F.K. and R.K. developed fluorescence instrumentation and gave technical support. All authors discussed the results and commented on the manuscript. H.S., K.H., W.H. and C.A.M.S. wrote the paper. H.G. contributed to the writing of the paper. C.A.M.S. supervised the project.

**Author Information:** H.S. current address: Clemson University, SC, USA; M.R.F. current address: Avanir Pharmaceuticals Inc., Aliso Viejo, California, USA; correspondence and requests should be addressed to C.A.M.S. (cseidel@hhu.de) and H.S. (hsanabr@clemson.edu).

**Supplementary Information** accompanies this paper at https://doi.org/.

**Competing interests:** The authors declare no competing interests.

# Resolving dynamics and function of transient states in single enzyme molecules

**Authors:** Hugo Sanabria[1,2,#,*], Dmitro Rodnin[1,#], Katherina Hemmen[1,#], Thomas O. Peulen[1], Suren Felekyan[1], Mark R Fleissner[3,7], Mykola Dimura[1,5], Felix Koberling[4], Ralf Kühnemuth[1], Wayne Hubbell[3], Holger Gohlke[5,6], Claus A.M. Seidel[1*]

#: contributed equally, *: corresponding authors

## *Affiliations*

[1]Institut für Physikalische Chemie, Lehrstuhl für Molekulare Physikalische Chemie, Heinrich-Heine-Universität, Düsseldorf, Germany. [2]Department of Physics and Astronomy, Clemson University, Clemson, South Carolina, U.S.A. [3]Jules Stein Eye Institute and Department of Chemistry and Biochemistry, University of California, Los Angeles, U.S.A. [4]PicoQuant GmbH, Berlin, Germany. [5]Institut für Pharmazeutische und Medizinische Chemie, Heinrich-Heine-Universität Düsseldorf, Düsseldorf, Germany. [6]John von Neumann Institute for Computing (NIC), Jülich Supercomputing Centre (JSC) & Institute for Complex Systems - Structural Biochemistry (ICS 6), Forschungszentrum Jülich GmbH, 52425 Jülich, Germany. [7]Present address: Avanir Pharmaceuticals, Aliso Viejo, California, U. S. A.

*Correspondence to: cseidel@hhu.de and hsanabr@clemson.edu

## Supporting Information:

## Supporting Figures:

## Supporting Tables:

# *1. Materials and Methods*

## 1.1 Multiparameter Fluorescence Detection (MFD)

MFD for confocal single molecule Förster Resonance Energy Transfer (smFRET) measurements was done using a 485 nm diode laser (LDH-D-C 485 PicoQuant, Germany, operating at 64 MHz, power at objective 110 µW) exciting freely diffusing labeled T4L molecule that passed through a detection volume of the 60X, 1.2 NA collar (0.17) corrected Olympus objective. The emitted fluorescence signal was collected through the same objective and spatially filtered using a 100 µm pinhole, to define an effective confocal detection volume. Then, the signal was divided into parallel and perpendicular components at two different colors ("green" and "red") through band pass filters, HQ 520/35 and HQ 720/150, for green and red respectively, and split further with 50/50 beam splitters. In total eight photon-detectors are

used- four for green (τ-SPAD, PicoQuant, Germany) and four for red channels (APD SPCM-AQR-14, Perkin Elmer, Germany). A time correlated single photon counting (TCSPC) module (HydraHarp 400, PicoQuant, Germany) was used for data registration.
For smFRET measurements samples were diluted (buffer used 50 mM sodium phosphate, pH 7.5, 150 mM NaCl, 40 μM TROLOX and 1 μM unlabeled T4L) to pM concentration assuring ~ 1 burst per second. Collection time varied from several minutes up to 10 hours. To avoid drying out of the immersion water during the long measurements an oil immersion liquid with refraction index of water was used (Immersol, Carl Zeiss Inc., Germany). NUNC chambers (Lab-Tek, Thermo Scientific, Germany) were used with 500 μL sample volume. Standard controls consisted of measuring water to determine the instrument response function (IRF), buffer for background subtraction and the nM concentration green and red standard dyes (Rh110 and Rh101) in water solutions for calibration of green and red channels, respectively. To calibrate the detection efficiencies, we used a mixture solution of double labeled DNA oligonucleotides with known distance separation between donor and acceptor dyes.

### 1.2 MFD burst analysis: Multiparameter FRET histograms and FRET lines

Bursts were selected by 2σ criteria out of the mean background value with cut off times that vary from sample to sample with a minimum of 60 photons for each burst. Each burst was then processed and fitted using a maximum likelihood algorithm [1] using in house developed software (LabVIEW, National Instruments Co.). Fluorescent bursts were plotted in 2D histograms (Origin 8.6, OriginLab Co).
The relation FRET-efficiency $E$ and the species weighted average donor lifetimes $\langle\tau\rangle_x$ depends on the fluorescence quantum yields of the dyes ($\Phi_{FD(0)}$ and $\Phi_{FA}$ for donor and acceptor respectively) and implicitly on background ($\langle B_G\rangle$ and $\langle B_R\rangle$ for green and red channels), detection efficiencies ($g_G$ and $g_R$ for green and red respectively) and crosstalk ($\alpha$):

$$E = \frac{1}{1+\dfrac{F_D/\Phi_{FD(0)}}{F_A/\Phi_{FA}}} = 1-\frac{\langle\tau_{D(A)}\rangle_x}{\langle\tau_{D(0)}\rangle_x} \tag{S1}$$

The corrected fluorescence ($F_D$ and $F_A$) depends on the detection efficiencies of green ($g_G$) and red ($g_R$) channels as follows:

$$F_D = \frac{S_G-\langle B_G\rangle}{g_G}, \tag{S2}$$

$$F_A = \frac{S_R-\alpha F_G-\langle B_R\rangle}{g_R}, \tag{S3}$$

where the total signal in green and red channels are $S_G$ and $S_R$, respectively. The ratio ($F_D/F_A$) is weighted by the species fractions.
In eq. S1, brackets $\langle\ldots\rangle_x$ represent averaging over all lifetime components. For the species $\tau^{(i)}$ weighted by its population fraction$x^{(i)}$, these averages are given by:

$$\langle\tau_{D(0)}\rangle_x = \sum_i x^{(i)}\tau_{D(0)}{}^{(i)} \quad \text{and} \quad \langle\tau_{D(A)}\rangle_x = \sum_i x^{(i)}\tau_{D(A)}{}^{(i)} \tag{S4}$$

Above $\langle\tau_{D(A)}\rangle_x$ and $\langle\tau_{D(0)}\rangle_x$ are the species averaged fluorescence lifetimes of the donor in presence and absence of an acceptor, respectively.

In sm FRET experiments approximately ~100 green photons per burst are detected. Hence, only the average time since excitation is reliably determined experimentally by the maximum likelihood estimators (MLE) for individual bursts. This time is weighted by the fluorescence intensity and hence, relates to the fluorescence lifetime components by:

$$\left\langle \tau_{D(0)} \right\rangle_f = \frac{\sum_i x^{(i)} \tau_{D(0)}^{(i)\,2}}{\sum_i x^{(i)} \tau_{D(0)}^{(i)}} \quad \text{and} \quad \left\langle \tau_{D(\mathrm{A})} \right\rangle_f = \frac{\sum_i x^{(i)} \tau_{D(A)}^{(i)\,2}}{\sum_i x^{(i)} \tau_{D(A)}^{(i)}}, \tag{S5}$$

We call these lifetimes fluorescence weighted average lifetimes.

The two averaged observables $E$ (in eq. S1) and $\langle \tau_{D(A)} \rangle_f$ can be related to each other. We call a line describing a theoretical relation of the two a "FRET-line". Such FRET-lines are projections of a parametrization of a multi-dimensional lifetime distribution to a two-dimensional plane using either the transfer-efficiency $E$ or $F_D/F_A$ as one and $\langle \tau_{D(\mathrm{A})} \rangle_f$ as second axis.

Fluorophores are moving entities coupled to biomolecules at specific places via flexible linkers. Therefore, for single protein conformations a DA-distance distribution has to be considered. For simplicity, we use normal distributions to describe the DA-distance distributions. If the donor and acceptor interfluorophore average distance is $\langle R_{\mathrm{DA}} \rangle$, the corresponding DA-distance distribution is:

$$p(R_{DA}) = \frac{1}{w_{\mathrm{DA}} \sqrt{\pi/2}} \exp\left( -2 \left[ \frac{R_{\mathrm{DA}} - \left\langle R_{DA} \right\rangle}{w_{DA}} \right]^2 \right), \tag{S6}$$

Here, $w_{DA}$ is the width of the DA-distance distribution attributed to the broadening due to the linker-flexibility set to a physical meaningful value of 12 Å [2]. Using the Förster-relationship $\tau_{D(\mathrm{A})}(R_{DA}) = \tau_{D(0)} \cdot \left(1 + \left(R_0/R_{DA}\right)^6\right)^{-1}$ and the following integrals:

$$\left\langle \tau_{D(\mathrm{A})} \right\rangle_x = \int \tau_{D(\mathrm{A})}(R_{DA}) p(R_{DA}) dR_{DA}, \tag{S7}$$

$$\left\langle \tau_{D(\mathrm{A})} \right\rangle_f = \frac{\int \left(\tau_{D(\mathrm{A})}(R_{DA})\right)^2 p(R_{DA}) dR_{DA}}{\left\langle \tau_{D(\mathrm{A})} \right\rangle_{x,L}}, \tag{S8}$$

This (eq. S6) distribution can be projected to a point in the $E$-$\langle \tau_{D(\mathrm{A})} \rangle_f$ plane. If the average DA distance $\overline{\langle R_{DA} \rangle}$ is varied within a given range (i.e. [0,∞]) a line within the $E$-$\langle \tau_{D(\mathrm{A})} \rangle_f$ plane is obtained. Such a line we call a static FRET-line, as it is valid for all molecules with given (single) conformation, irrespectively of the mean DA-separation, $\overline{\langle R_{DA} \rangle}$.

To describe molecules, which are interconverting between two states with mean distances $\overline{\langle R_{DA} \rangle}^{(1)}$ and $\overline{\langle R_{DA} \rangle}^{(2)}$ and fractions $x^{(1)}$ and $x^{(2)}=1-x^{(1)}$, by a line in the $E$-$\langle \tau_{D(\mathrm{A})} \rangle_f$ plane we use the following distance distribution:

$$p(R_{DA}) = \frac{x_{DA}^{(1)}}{w_{\mathrm{DA}} \sqrt{\pi/2}} \exp\left( -2 \left[ \frac{R_{\mathrm{DA}} - \left\langle R_{DA}^{(1)} \right\rangle}{w_{DA}} \right]^2 \right) + \frac{1 - x_{DA}^{(1)}}{w_{\mathrm{DA}} \sqrt{\pi/2}} \exp\left( -2 \left[ \frac{R_{\mathrm{DA}} - \left\langle R_{DA}^{(2)} \right\rangle}{w_{DA}} \right]^2 \right), \tag{S9}$$

To obtain a “dynamic” FRET-line, which is valid for a molecule in exchange between these two states, the fraction $x^{(1)}$ is varied within the range [0, 1] and the position in the $E$-$\langle\tau_{D(A)}\rangle_f$ plane is calculated using the eq. S7, S8 and eq. S1.

**1.3 Guidelines for reading MFD histograms**

Several guidelines are needed to properly read MFD histograms. A short list is presented here.
*I*) Donor only population is shown at low $E$ with lifetime ~ 4 ns (donor-only for Alexa488).
*II*) High FRET appears at shorter lifetimes when the fluorescence of acceptor is high ($E \rightarrow 1$).
*III*) Static FRET states follow a theoretical line that accounts for dye linker mobility called "static FRET line" [3].
*IV*) A molecule that exchanges conformations at timescales faster than the diffusion time emits a burst of photons whose mixed fluorescence is characterized by the fluorescence average lifetime. Elongation of peaks in $E$-$\langle\tau_{D(A)}\rangle_f$-histograms and deviation from static FRET-lines are an indication for slow conformational dynamics processes on the hundreds of microseconds.

We inspect the signal over the duration of the measurement. Typically, we find stable signal over 1 hr and 10 hrs. Additionally, we minimize unlikely effects of multimolecule events by comparing the difference in the burst duration in donor and acceptor channels ($|T_G|-|T_R|$) or aggregates (e. g. via burst duration or the diffusion time component of the correlated molecular bursts) and impurities due to e.g. free, unattached fluorophores (e.g. by plotting the scatter-corrected anisotropy vs. $\langle\tau_{D(A)}\rangle_f$).

**1.4 Filtered Fluorescence Correlation Spectroscopy**

In fluorescence correlation spectroscopy (FCS) information on fluctuating systems is obtained by calculating the correlation function [4,5]:

$$^{A,B}G(t_c) = 1 + \frac{\left\langle \delta^{A}S(t) \cdot \delta^{B}S(t+t_c) \right\rangle}{\left\langle {}^{A}S(t) \right\rangle \cdot \left\langle {}^{B}S(t) \right\rangle}. \tag{S10}$$

where $t_c$ is the correlation time, $^{A,B}S(t)$ represents the detected intensity signal (number of detected photons per time interval) at channels $A$ or $B$, and $\delta^{A;B}S(t)$ corresponds to the deviation of the signal from the time average signal denoted as $\left\langle {}^{A;B}S(t) \right\rangle$. $^{A,B}G(t_c)$ is an auto-correlation function (*ACF*) if $A = B$ otherwise $^{A,B}G(t_c)$ is a cross-correlation (*CCF*).
The correlation function [6 7] of a mixture of $n$ molecular static species is given by an weighted average:

$$G(t_c) = 1 + \frac{1}{N} \cdot \frac{\sum_i^n x^{(i)} \cdot (Q^{(i)})^2 \cdot G_{diff}^{(i)}(t_c)}{\left( \sum_i^n x^{(i)} \cdot Q^{(i)} \right)^2}, \tag{S11}$$

where $G_{diff}^{(i)}(t_c)$ describes molecular diffusion. For a 3-dimensional Gaussian detection probability, $W(x,y,z) = \exp\left(-2\left(x^2+y^2\right)/\omega_0^2\right) \cdot \exp\left(-2\,z^2/z_0^2\right)$, $G_{diff}^{(i)}(t_c)$ is given by:

$$G_{diff}^{(i)}(t_c)=\left(1+\frac{t_c}{t_{diff}^{(i)}}\right)^{-1}\cdot\left(1+\left(\frac{\omega_0}{z_0}\right)^2\cdot\frac{t_c}{t_{diff}^{(i)}}\right)^{-\frac{1}{2}}, \quad \text{(S12)}$$

The $1/e^2$ radii in $x$ and $y$ or in $z$ direction are denoted by $\omega_0$ and $z_0$, respectively. The characteristic diffusion time $t_{diff}^{(i)}$ relates to the diffusion coefficient of each species $i$ $t_{diff}^{(i)}=\omega_0^2/4D^{(i)}$ . The amplitude of the correlation is scaled with the reciprocal of the average number of fluorescent particles $N$ in the confocal volume. Each molecular fraction $x^{(i)}=c^{(i)}/\sum_i c^{(i)}$ has a concentration $c^{(i)}$, and brightness $Q^{(i)}$.

To separate species, we use filtered FCS (fFCS) [8,9]. fFCS differs from standard FCS [4] and FRET-FCS [10] by interrogating the "species" (conformational states) fluctuations instead of photon count rates [10]. We define the species cross- correlation function (*sCCF*) as

$$G^{(i,m)}(t_c)=\frac{\left\langle F^{(i)}(t)\cdot F^{(m)}(t+t_c)\right\rangle}{\left\langle F^{(i)}(t)\right\rangle\cdot\left\langle F^{(m)}(t+t_c)\right\rangle}=\frac{\left\langle\left(\sum_{j=1}^{d\cdot L} f_j^{(i)}\cdot S_j(t)\right)\cdot\left(\sum_{j=1}^{d\cdot L} f_j^{(m)}\cdot S_j(t+t_c)\right)\right\rangle}{\left\langle\sum_{j=1}^{d\cdot L} f_j^{(i)}\cdot S_j(t)\right\rangle\cdot\left\langle\sum_{j=1}^{d\cdot L} f_j^{(m)}\cdot S_j(t+t_c)\right\rangle}, \quad \text{(S13)}$$

where ($i$) and ($m$) are two selected "species" or "pseudospecies" in a mixture, where pseudospecies correspond to the equilibrium of two or more mixed species that are in fast exchange. A set of filters $f_j^{(i)}$ that depend on the arrival time of each photon after each excitation pulse is used. The signal $S_j(t)$, obtained via pulsed excitation, is recorded at each $j$ = 1 ... $L$ TCSPC-channel. The signal and filters per detector, $d$, are stacked in a single array with dimensions $d\cdot L$ for global minimization according to [8]. Filters are defined in such a way that the relative "error" difference between the photon count per species ($w^{(i)}$) and the weighted histogram $f_j^{(i)}\cdot H_j$ is minimized as defined in Eq. (S14).

$$\left\langle\left(\sum_{j=1}^{d\cdot L} f_j^{(i)}\cdot H_j - w^{(i)}\right)^2\right\rangle\rightarrow \min, \quad \text{(S14)}$$

where brackets represent time averaging.

The requirement is that the decay histogram $H_j$ can be expressed as a linear combination of the conditional probability distributions $p_j^{(i)}$, such as $H_j=\sum_{i=1}^{n(=2)} w^{(i)}p_j^{(i)}$, with $\sum_{j=1}^{d\cdot L} p_j^{(i)}=1$. Hence, the species cross- correlation $G^{(i,m)}(t_c)$ provides maximal contrast for intercrossing dynamics [8]. One major advantage of *sCCF* is that, if photophysical properties are decoupled from species selection, the intercrossing dynamics [10] is recovered with great fidelity.

To properly fit the species cross-correlation function, we used [8]

$$G(t_c)=1+\frac{1}{N}\cdot G_{diff}(t_c)\cdot\left[1-G_K(t_c)\right], \quad \text{(S15)}$$

where $G_K(t_c)$ is

$$G_K(t_c)=\sum_{t_{Ri}}^{t_{Rn}} A_{Ki}\exp(-t_c/t_{Ri}). \quad \text{(S16)}$$

In Eq. (S16) the summation is over $n$ reaction times $t_{Rn}$.
The same 3-dimensional Gaussian shaped volume element is assumed. We assume that $G_{diff}(t_c) = G_{diff}^{(i)}(t_c) = G_{diff}^{(m)}(t_c)$ take the form of Eq. (S15). The normalized correlation function is presented as:

$$g(t_c) = N \cdot \left(G(t_c) - 1\right). \tag{S17}$$

Filtered FCS requires prior knowledge of the time-resolved fluorescence and polarization decays for each species or pseudospecies. For a mixture of more than two species, we generated two decays corresponding to two "pseudo-species". Using the scatter profile as the excitation pulse, the parallel and perpendicular decay components ($F_{\parallel}(t)$ and $F_{\perp}(t)$) for each "pseudo-species" were generated as

$$\begin{aligned} F_{\parallel}(t) &= F(t) \cdot (1 + (2 - 3l_1) \cdot r(t)) / 3 \\ F_{\perp}(t) &= F(t) \cdot (1 - (1 - 3 \cdot l_2) \cdot r(t)) / 3 \end{aligned}, \tag{S18}$$

where $F(t)$ is the time-resolved fluorescence decay at magic angle, and $l_1$ = 0.01758 and $l_2$ = 0.0526 are correction factors [11] [12]. The anisotropy decay $r(t)$ is given by

$$r(t) = r_{0,ov} \exp(-t / \rho_{overall}) + r_{0,ba} \exp(-t / \rho_{backbone}) + r_{0,li} \exp(-t / \rho_{linker}). \tag{S19}$$

Background signal consists of dark counts (uniformly distributed over TCSPC channels) and scatter contribution.

### 1.5 Ensemble Time Correlated Single Photon Counting with high precision

Ensemble Time Correlated Single Photon Counting (eTCSPC) measurements were performed using either an IBH-5000U (IBH, Scotland) or a Fluotime 200 (Picoquant, Germany) system. The excitation source of the IBH machine were a 470 nm diode laser (LDH-P-C470, Picoquant, Germany) operating at 10 MHz for donor excitation and a 635 nm (LDH-P-C635, Picoquant, Germany) for acceptor excitation. The excitation and emission slits were set to 2 nm and 16 nm, respectively. The excitation source of the Fluotime200 system was a white light laser (SuperK extreme, NKT Photonics, Denmark) operating at 20 MHz for both donor (485 nm) and acceptor (635 nm) excitation with excitation and emission slits set to 2 nm and 5 nm, respectively. Additionally, in both systems, cut-off filters were used to reduce the amount of scattered light (>500 nm for donor and >640 nm for acceptor emission).
For green detection, the monochromator was set to 520 nm and for red detection to 665 nm. All measurements were conducted under magic angle conditions (excitation polarizer 0°, emission polarizer 54.7°), except for anisotropy where the position of the emission polarizer was alternately set to 0° (VV) or 90° (VH).
In the IBH system, the TAC-histograms were recorded with a bin width of 14.1 ps within a time window of 57.8 ns, while the Fluotime200 was set to a bin width of 8 ps within a time window of 51.3 ns. Photons were collected up to a peak count of 100'000 corresponding in average to a total number of $30 \cdot 10^6$ photons. The instrument response function IRF (~230 ps FWHM for the IBH, ~ 150 ps for the Fluotime200) was collected under the same recording settings at the excitation wavelength of the sample without cutoff-filters using a scattering Ludox-dispersion, which yielded a comparable count rate as the later on measured samples.
For the IBH system, it was needed was performed before each measurement session a reference measurement with a continuous light signal to account for the differential non-linearity of the counting electronics. The recorded uncorrelated photons yield a reference histogram that is ideally constant. After recording of this measurement, the average number of photons in each time-bin is calculated. Next, the measurement was smoothed by a window function using a Hanning-filter with a window-size of 17 bins. The smoothed decay histogram was normalized to the previously calculated average number of photons. Instead of correcting the experimental histogram the model function is multiplied by the smoothed and normalized reference histogram to preserve the Poissonian statistics of the measured fluorescence intensity histograms of interest.

### 1.6 Donor and Acceptor fluorescence quantum yields

Depending on the labeling position, the donor and acceptor fluorescence quantum yields vary and have been estimated for each sample (Table S2A). We estimate $\Phi_{FD(0)}$ and $\Phi_{FA}$ of the fluorescent species by the species-averaged fluorescence lifetime $\langle\tau\rangle_x$ of donor or acceptor, respectively. As reference samples we used Alexa488-labeled DNA $\langle\tau_{D(0)}\rangle_x = 4.0$ ns, $\Phi_{FD(0)} = 0.8$ and for the acceptor, we used Cy5-labeled DNA with $\langle\tau_A\rangle_x = 1.17$ ns and $\Phi_{FA} = 0.32$ [13]. This FRET pair has a Förster distance of 52 Å.

### 1.7 Time-resolved fluorescence decay analysis

*Model*

We model the fluorescence decay of the donor in the absence of FRET $F_{D(0)}(t)$ by a multi-exponential decay to account for sample specific differences of the donor reference samples

$$F_{\mathrm{D(0)}}(t) = \sum_i x_{\mathrm{D(0)}}^{(i)} \exp(-t/\tau_{\mathrm{D(0)}}^{(i)}). \tag{S20}$$

Here, $\tau_{\mathrm{D(0)}}^{(i)}$ is the donor fluorescence lifetime and $x_{\mathrm{D(0)}}^{(i)}$ are the pre-exponential factors.

Sample specific differences were considered in the analysis of the FRET samples by joint analysis where all donor species are quenched by the same FRET rate constant $k_{RET}$. Such model is correct if quenching does not change the donor radiative lifetime and if FRET is uncorrelated with quenching of the donor by its local environment. Under these conditions the donor fluorescence intensity decay in the presence of FRET $F_{D(A)}(t)$ factorizes into the donor fluorescence decay in absence of FRET and the FRET-induced donor quenching $\varepsilon_{D(A)}(t)$

$$F_{D(A)}(t) = F_{D(0)}(t) \cdot \varepsilon_{D(A)}(t). \tag{S21}$$

We relate the FRET-induced donor quenching to the DA-distance distribution by the rate-constant of energy transfer as defined by Förster

$$k_{RET} = k_F \cdot \kappa^2 \cdot \left(\frac{R_{0J}}{R_{DA}}\right)^6 \tag{S22}$$

Here, $R_{0J}$ is a reduced Förster-radius, $k_F$ is the radiative rate constant of fluorescence and $\kappa^2$ is the orientation-factor. This reduced Förster-radius is given by

$$R_{0J} = \left[\frac{9(\ln 10)}{128\pi^5 \cdot N_A} \cdot \frac{J}{n^4} \cdot\right]^{\frac{1}{6}} = 0.2108 \cdot \text{Å} \cdot \left[\frac{1}{n^4} \cdot \left(\frac{J(\lambda)}{\mathrm{mol}^{-1} \cdot \mathrm{dm}^3 \cdot \mathrm{cm}^{-1} \cdot \mathrm{nm}^4}\right)\right]^{\frac{1}{6}}, \tag{S23}$$

where $N_A$ is Avogadro's constant, $n$ is the refractive index of the medium and $J = \int f_D(\lambda)\cdot\varepsilon_A(\lambda)\cdot\lambda^4 \cdot d\lambda$ is the overlap integral between $f_D(\lambda)$, the donor emission spectrum and $\varepsilon_A(\lambda)$, the acceptor absorption spectrum. The FRET-induced donor decay relates to the distance distribution $p(R_{DA})$ by

$$\varepsilon_{D(A)}(t) = \int p(R_{DA})\cdot\exp\left(-t\cdot\left\langle\kappa^2\right\rangle\cdot k_{\mathrm{F}}\left[1+(R_{0\mathrm{J}}/R_{\mathrm{DA}})^6\right]\right)dR_{DA}\,. \qquad \text{(S24)}$$

We use an average orientation factor of $\langle\kappa^2\rangle \approx 2/3$ (justified by the anisotropy studies compiled in Tables S3A-S3D). We used a reduced Förster-radius of $R_{0J}$=56.4 Å which was determined for the donor with a radiative rate constant $k_F$=0.224 ns$^{-1}$. As previously described [14] we propagate potential errors of the $\langle\kappa^2\rangle \approx 2/3$ approximation to our experimental distances (Table S5D).

We use a superposition of normal distributions to describe a mixture of states:

$$p(R_{DA}) = \sum_{i=1}^{N} x_{DA}^{(i)} \frac{1}{w_{\mathrm{DA}}\sqrt{\pi/2}} \exp\left(-2\left[\frac{R_{\mathrm{DA}}-\left\langle R_{DA}^{(i)}\right\rangle}{w_{DA}}\right]^2\right). \qquad \text{(S25)}$$

Here, $N$ is the number of states (2 or 3) with $\left\langle R_{DA}^{(i)}\right\rangle$ being the mean of the state *(i)* distance distribution with species fraction $x_{DA}^{(i)}$ and a width $w_{DA}$ set to a physical meaningful value of 12 Å (flexible dye-linkers) [2].

We analyze our data by substituting eq. S25 into eq. S24. Next, eq. S24 is inserted into eq. S21. Finally, we analyze the fluorescence intensity decay of the donor in presence and absence of FRET (eq. S26 or eq. S27) in a joint fit, in which the fluorescence lifetimes and corresponding species fractions of the donor only reference sample were identical to the respective parameters in the FRET sample. By this procedure the photon counting statistics of both the reference- and fluorescence-decay in presence of FRET is preserved. Thus, the counting statistics are clearly defined (Poisson distribution). This allows for an analysis with proper error-estimates. By the global (joint) analysis of the reference sample and the FRET-sample the photophysical properties (dynamic quenching) are taken into account. To further reduce the number of free model parameters, we combined the donor only and FRET measurements of all 33 FRET samples into a joint single data set, in which the species fractions of the DA-distance distribution were shared among all 33 variants. The so achieved reduction in degrees of freedom of a joint/global fit stabilizes the fit dramatically.

*Ensemble Time Correlated Single Photon Counting*

The experimental fluorescence intensity decays were fitted using the iterative re-convolution approach, where the model-decay curves are convoluted with the experimental instrument response function (*IRF*). Additionally, we consider a constant offset $c$ of the fluorescence intensity and correct the instrumental differential non-linearity by a time-dependent function $Lin(t)$. With these corrections, the experimental time-resolved fluorescence intensities of the FRET-sample and the donor reference sample are proportional to:

$$\begin{aligned} F_{\mathrm{FRET}}(t) &= \left(N_0\cdot\left[(1-x_{\mathrm{DOnly}})F_{\mathrm{D(A)}}(t)+x_{\mathrm{DOnly}}F_{\mathrm{D(0)}}(t)\right]\otimes IRF + sc\cdot IRF + c\right)\cdot Lin(t) \\ F_{\mathrm{Ref}}(t) &= \left(N_0\cdot F_{\mathrm{D(0)}}(t)\otimes IRF + sc\cdot IRF + c\right)\cdot Lin(t) \end{aligned} \quad . \qquad \text{(S26)}$$

Here, *sc* is due to scattered light from the sample. The model fluorescence intensity histograms were scaled to the experimental measured number of photons to reduce the number of free fitting parameters (the initial amplitude $N_0$ is not fitted).

*Sub-ensemble Time Correlated Single Photon Counting*

In sm-measurements we determine the number of fluorescent photons $N_F$ and the number of background photons $N_{BG}$ using buffer reference measurements as reference. Given the known number of fluorescence and background photons the fluorescence decays were modeled by:

$$\begin{aligned} F_{\mathrm{FRET}}(t) &= N_F \cdot \left[ (1 - x_{\mathrm{DOnly}}) F_{\mathrm{D(A)}}(t) + x_{\mathrm{DOnly}} F_{\mathrm{D(0)}}(t) \right] \otimes IRF + N_{BG} \cdot IRF \\ F_{\mathrm{Ref}}(t) &= N_F \cdot F_{\mathrm{D(0)}}(t) \otimes IRF + N_{BG} \cdot IRF \end{aligned} \quad . \qquad \text{(S27)}$$

This procedure reduced the number of the number of free parameters compared to the eTCSPC measurements.

*Summary fit models*

In total, we used three different fit models to describe our data. They differ in their number of states and the number of joint (global) and free relevant parameters, which are given below. The fit of the two globally linked states was obtained within the procedure to estimate the fraction of the third state (see below).

**Table S1**. List of evaluated fit models. The fit models are differentiated by their number of states and free parameters. The average $\chi^2_r$ and the table with the listed results is given.

| ***N*-states** | **parameter** | | | **free parameters** | | | |
|---|---|---|---|---|---|---|---|
| | **local** | **global** | **Constraints** | **per sample** | **total** | **Table** | **Average $\chi^2_r$** |
| 2 | $\langle R_{DA}^{(1)} \rangle$, $x_{DA}^{(1)}$ $\langle R_{DA}^{(2)} \rangle$, $x_{DA}^{(2)}$ | | $x_{DA}^{(1)} + x_{DA}^{(2)} = 1$ | **3** | **101** =(33*3+3-1) | S3B | 1.0825 |
| 2 | $\langle R_{DA}^{(1)} \rangle$, $\langle R_{DA}^{(2)} \rangle$ | $x_{DA}^{(1)}$, $x_{DA}^{(2)}$ | $x_{DA}^{(1)} + x_{DA}^{(2)} = 1$ | **2** | **67** =(33*2+2-1) | S3C | 1.0985 |
| 3 | $\langle R_{DA}^{(1)} \rangle$, $\langle R_{DA}^{(2)} \rangle$, $\langle R_{DA}^{(3)} \rangle$ | $x_{DA}^{(1)}$, $x_{DA}^{(2)}$, $x_{DA}^{(3)}$ | $x_{DA}^{(1)} + x_{DA}^{(2)} + x_{DA}^{(3)} = 1$ | **3** | **101** =(33*3+3-1) | S3D-S3F | 1.0736 |

*Fitting of functional variants*

Functional variants were fitted globally, i. e. distances for states $C_1$ and $C_2$ were linked over all three variants used to mimic free enzyme *E*, enzyme-substrate complex *ES* and enzyme product complex *EP* while the distance for $C_3$ was only linked for *E* and *ES* to allow for the different (covalent) nature of this state in *EP*. The experimental fluorescence decays were fitted by the conventional Levenberg–Marquardt minimization algorithm using custom software written in Python.

*Uncertainty estimation*

The statistical errors of the DA-distances were determined by sampling the parameter space [15,16] and applying the F-distribution at a confidence level of 95% given the minimum determined $\chi^2$. The maximum allowed $\chi^2_{r,\max}$ for a given confidence-level (*P;* e.g. for 2σ *P* = 0.95) was calculated by

$$\chi^2_{r,\max}(P) = \chi^2_{r,\min} \cdot \left[1 + n/\upsilon \cdot \mathrm{cdf}^{-1}(F(n,\upsilon,P))\right], \qquad \text{(S28)}$$

where $\mathrm{cdf}^{-1}(F(\mathrm{n},\nu,P))$ is the inverse of the cumulative distribution function of the *F*-distribution for *n* number of free parameters, and with ν degrees of freedom. $\chi^2_{r,\min}$ is the minimum determined $\chi_r^2$ [17].

To estimate the species fraction $x_3$ of the third state, we performed a support plane analysis for the global fit [18].

## 2. Supporting Results

### 2.1 Single-molecule and fluorescence correlation spectroscopy

### 2.1.1 Example data analysis of MFD experiments

The smFRET data analysis is done in accordance to previously published methods e.g. Kalinin *et al.*, Sisamakis *et al.* or Kudryavtsev *et al.*[19,20].

Briefly, the photons emitted from single molecules traversing through the confocal detection volume are selected from background photons using the inter-photon time and a threshold for the maximum inter photon time [21]. Additionally, only burst containing a minimum number of 60 photons were selected for further analysis.

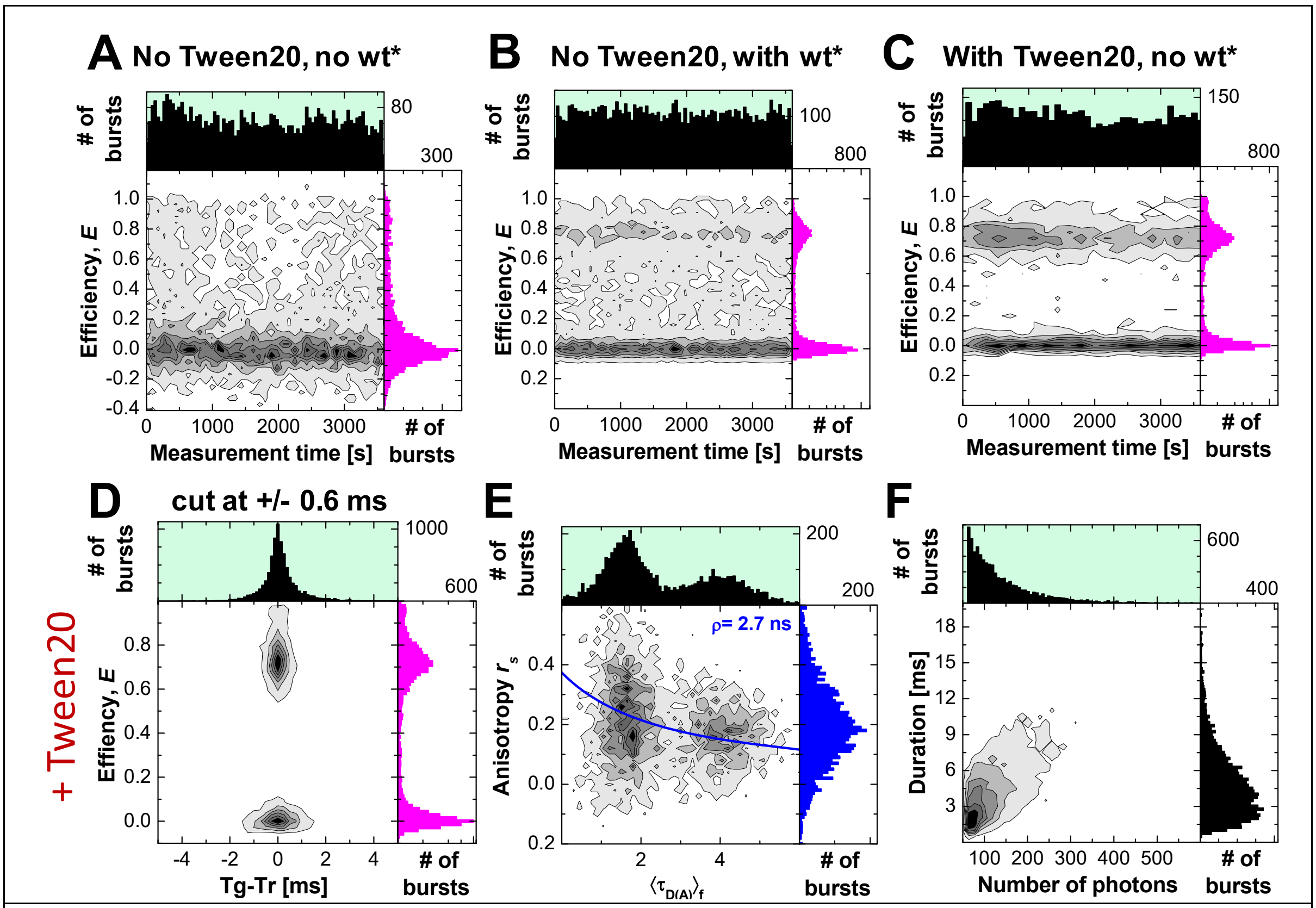

**Figure S1 Exemplary data analysis for variant K60pAcF I150C-(DA). (A-C)** FRET efficiency E vs. the measurement time for 60-150 in the absence of coating (A), coating of the measurement chamber with 0.1 % (v/v) Tween20 (B), or 1 µM unlabeled protein (C). **(D)** In the case of acceptor photobleaching, the green signal trace will be longer (Tg) than the red signal trace (Tr). The difference |Tg-Tr| is in ideal case randomly and sharply distributed around 0. Molecules, in which |Tg-Tr| exceeds 0.6 are removed from further analysis. **(E)** Free dye molecules can be recognized by a low anisotropy ~ 0, labeled single molecules follow the Perrin equation. **(D)** Aggregates will show a high brightness and long burst duration, which do not scale linearly.

After burst selection, our MFD histograms are checked for e. g. signal stability (Figure S1A-C), photobleaching (Figure S1D), contamination with free dye (Figure S1E), multimolecule events (Figure S1D, F) or aggregates (Figure S1F). Without coating the measurement chamber either with the tensile Tween20 (Figure S1B) or unlabeled protein (Figure S1C), the signal of our FRET-labeled molecules is lost within the 10 min needed to start the experiment (Figure S1A). Photobleaching of the acceptor will appear as tilting in the 1D projection of |Tg-Tr| (Figure S1D); the photon trace of the acceptor will be shorter than for the donor and thus, shift the – in ideal case sharp and randomly around 0 distributed – plot. Fluorophores coupled to a biomolecule have high anisotropy a follow the Perrin equation (Figure S1E), free dye has a very low anisotropy ~ 0. Thus, it be visible below the biomolecule population. The presence of multimolecule events or aggregates can (*i*) be detected during burst selection, (*ii*) large |Tg-Tr| and/or (*iii*) by a large number of photons within the burst and a long burst duration, respectively (Figure S1F).

### 2.1.2 Additional SMD and fFCS

To test for possible influences of the dyes on the protein, two distinct labeling configurations (DA) and (AD) were prepared as previously described. In the sub-μs to ms range the dynamics of T4L is independent of the labeling-configuration. However, we can see some small differences in the two samples. For example, the species fractions in eTCSPC for S44pAcF/I150C-(DA) and –(AD) are not identical; although, one can clearly identify the same conformers corresponding to the states $C_1$, $C_2$, and $C_3$.

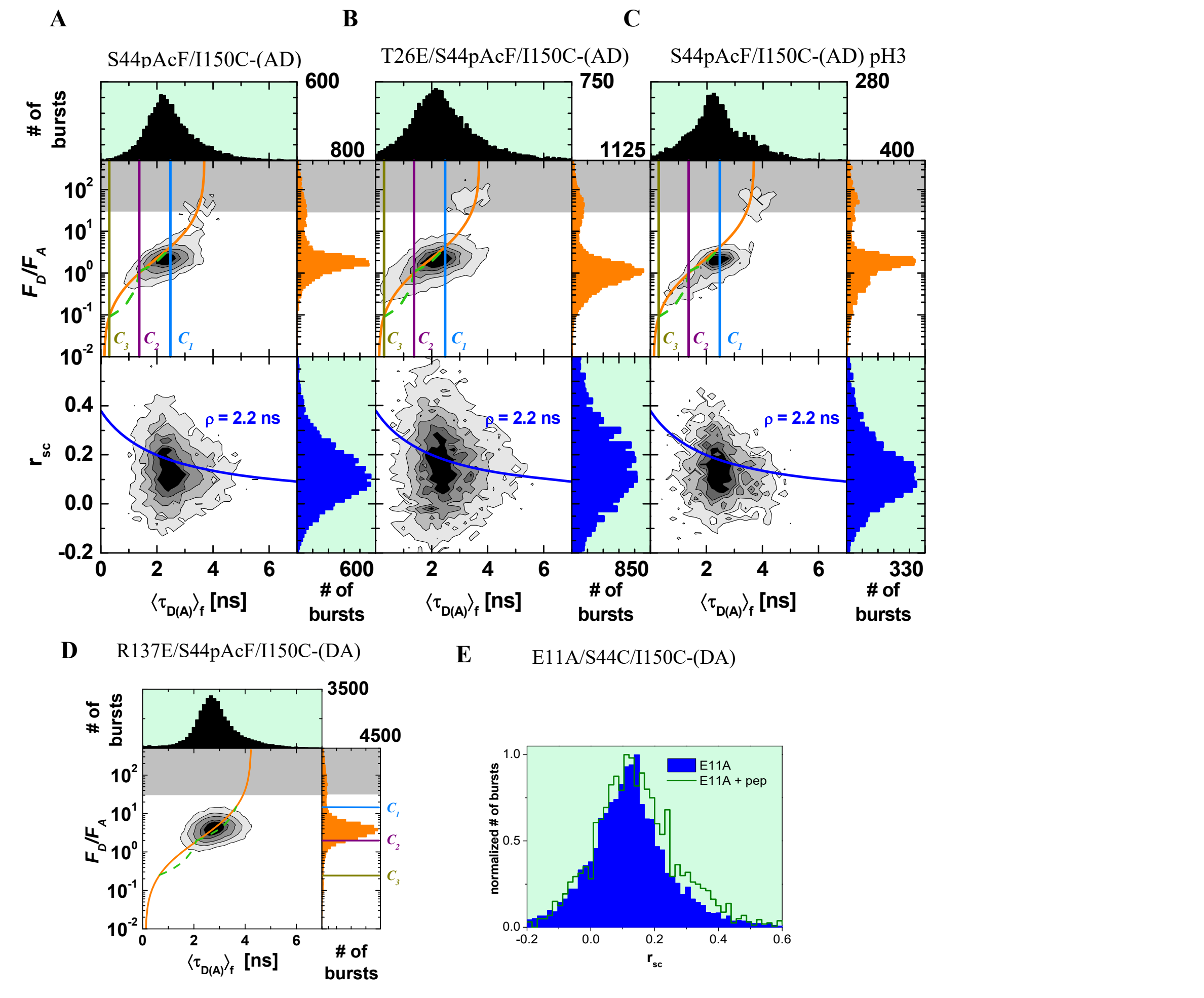

**Figure S2 MFD analysis of further samples. (A)** MFD histogram of S44pAcF/I150C-(AD) labeled T4L. Two dimensional histogram $F_D/F_A$ vs. lifetime of donor in the presence of acceptor $\langle\tau_{D(A)}\rangle_f$, and anisotropy vs. $\langle\tau_{D(A)}\rangle_f$ for S44pAcF/I150C-(AD). One dimensional projections for $F_D/F_A$, $\langle\tau_{D(A)}\rangle_f$ and anisotropy are also shown. Static FRET line is shown in orange. Pure donor and acceptor fluorescence ($F_D$ and $F_A$) are corrected for background ($\langle B_G\rangle$ = 1.8 kHz, $\langle B_R\rangle$ = 0.7 kHz), spectral cross-talk ($\alpha$ = 1.3%) and detection efficiency ratio ($g_G/g_R$ = 0.77). Shaded area in gray is the region of donor only. On the anisotropy vs. $\langle\tau_{D(A)}\rangle_f$ histograms the Perrin's equation with rotational correlation $\rho$ = 2.2 ns is shown as blue line. Vertical guidelines for states $C_1$, $C_2$, and $C_3$

according to the eTCSPC results of the same sample are added as references. Ignoring the donor only population a single unimodal distribution is observed in all $F_D/F_A$ vs. $\langle\tau_{D(A)}\rangle_f$; similarly to what was observed in the –(DA) sample. Two slight differences can be observed: the tilt towards the state $C_3$ is more evident and the accumulation of the $C_{3t}$ is not visible. **(B)** MFD histograms for the variant T26E/S44pAcF/I150C-(AD) without substrate. We observe a more pronounced tilt towards the $C_3$. **(C)** At pH 3.0, the MFD histograms for the S44pAcF/I150C-(AD) show very similar characteristics as the variant T26E. **(D) Functional mutant of T4L.** MFD histograms for R137E/S44pAcPh/I150C-(DA). **(E) Effect of substrate on E11A/S44C/I150C.** Upon addition of substrate we observe a higher anisotropy (green line). All samples were corrected for background, cross talk, and detection efficiencies according to experimentally determined parameters.

Slight differences were observed when comparing experiments for -(DA) and –(AD) at the single-molecule level. When comparing the mutant S44pAcF/I150C-(DA), shown in Figure S8A, to the –(AD) labeling scheme shown in Figure S2, we observe the following: *i*) There is more "donor-only" fraction in the -(AD) labeling scheme than in the -(DA), this is part of the variability in labeling. *ii*) There is no accumulation of a high FRET state in the –(AD) scheme.

However, in this situation the elongation toward higher FRET or state $C_3$ is slightly more pronounced. This elongation is also present in the T26E/S44pAcF/I150C-(AD) mutant (Figure S2B). This resembles the accumulation found in the sample T26E/S44pAcF/I150C–(DA) (Figure 6G, main text). Regardless of these differences, the 2D histograms and eTCSPC show similar states. This is clear evidence that the three conformational states are present independent of the fluorophores.

In summary, the kinetic scheme might change slightly, but not significantly given the conserved effect on the *sCCF* curves (Figure 3B, main text). The *sCCF* shows unequivocally that the transition times are present in both labeling schemes. Therefore, the specific dye-protein interactions are not responsible for the transition times between sub-µs and ms.

The major difference between the –(DA) and –(AD) is the state $C_{3t}$. This state seems to accumulate for the –(DA) configuration. However, at low pH the –(AD) shows a similar elongation towards the $C_3$ state similar as T26E/S44pAcF/I150C–(AD), also consistent with the data presented for the S44pAcF/I150C– (DA) at low pH.

Additional MFD histograms for further functional mutants are shown in Figure S7. A summary of the ensemble or sub-ensemble fits for these mutants is shown in Table S2G. Figure S3 shows MFD histograms for all 33 variants used within the T4L network.

| E5pAcF S44C | E5pAcF N132C | R8pAcF Q69C |
|---|---|---|

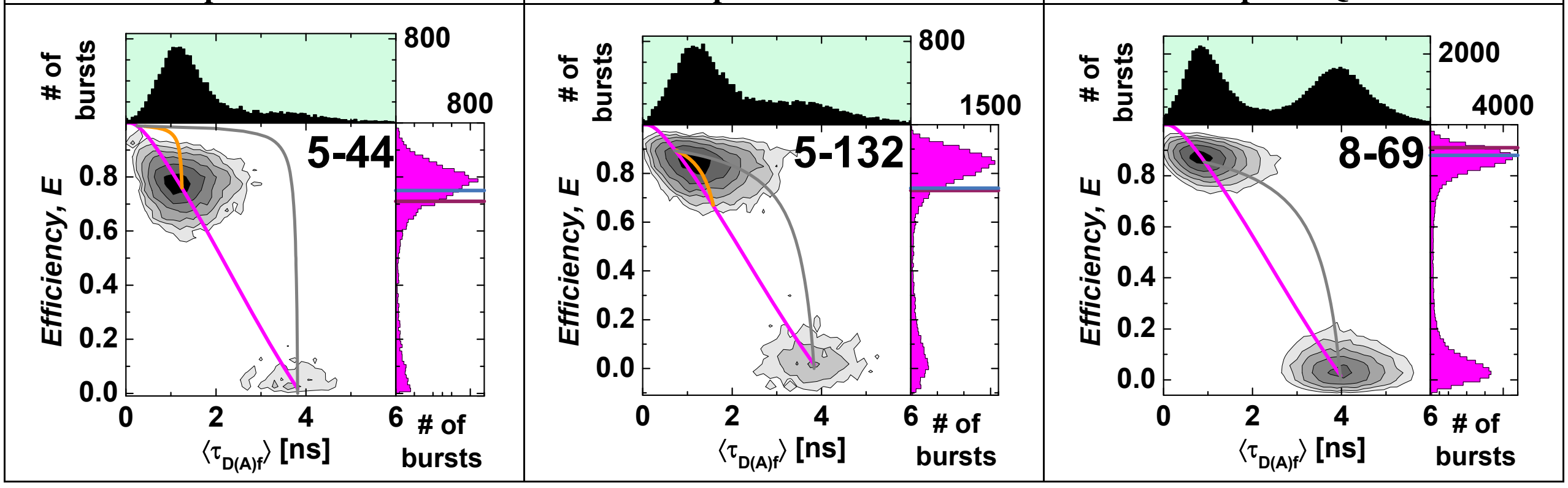

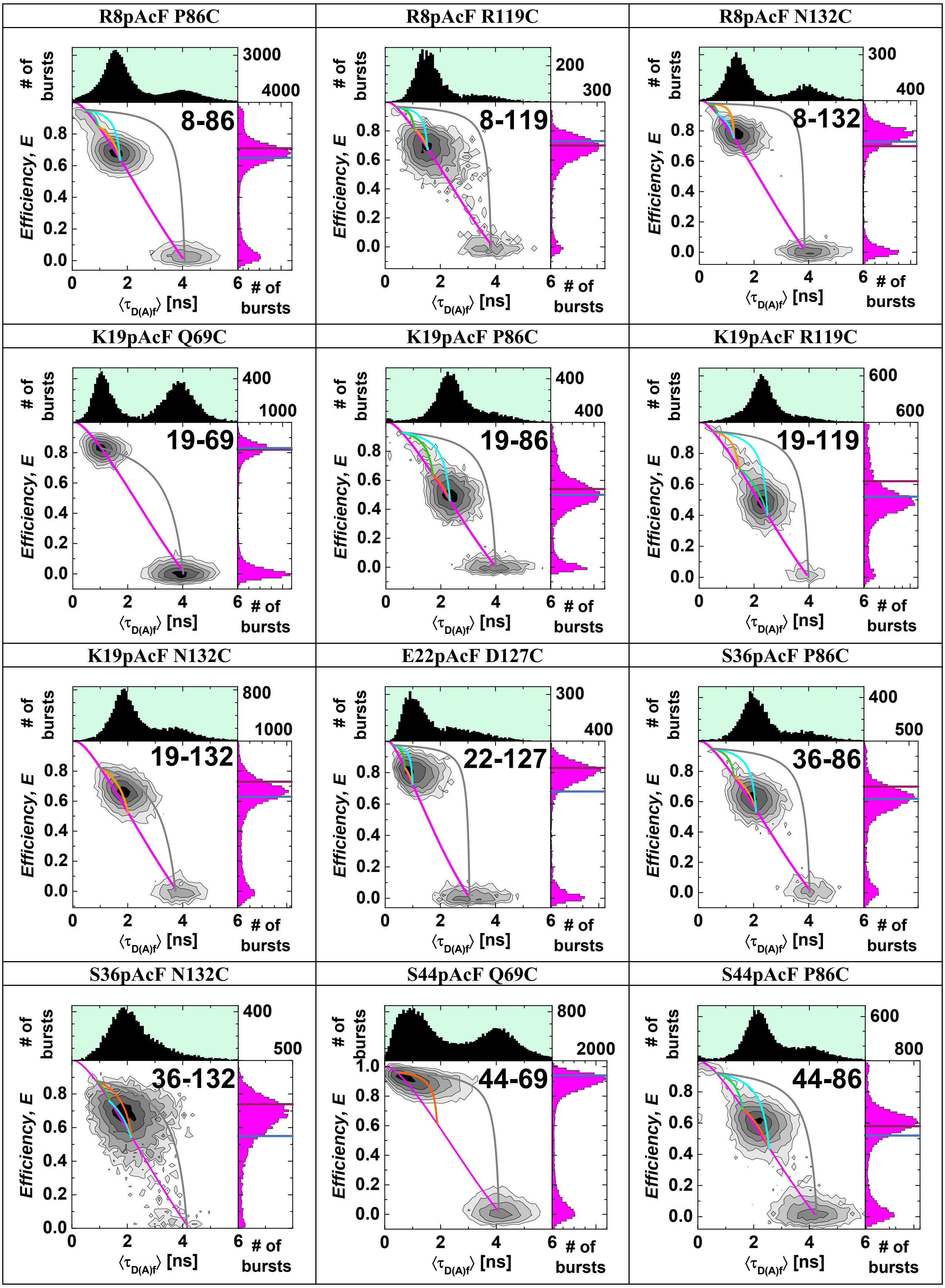

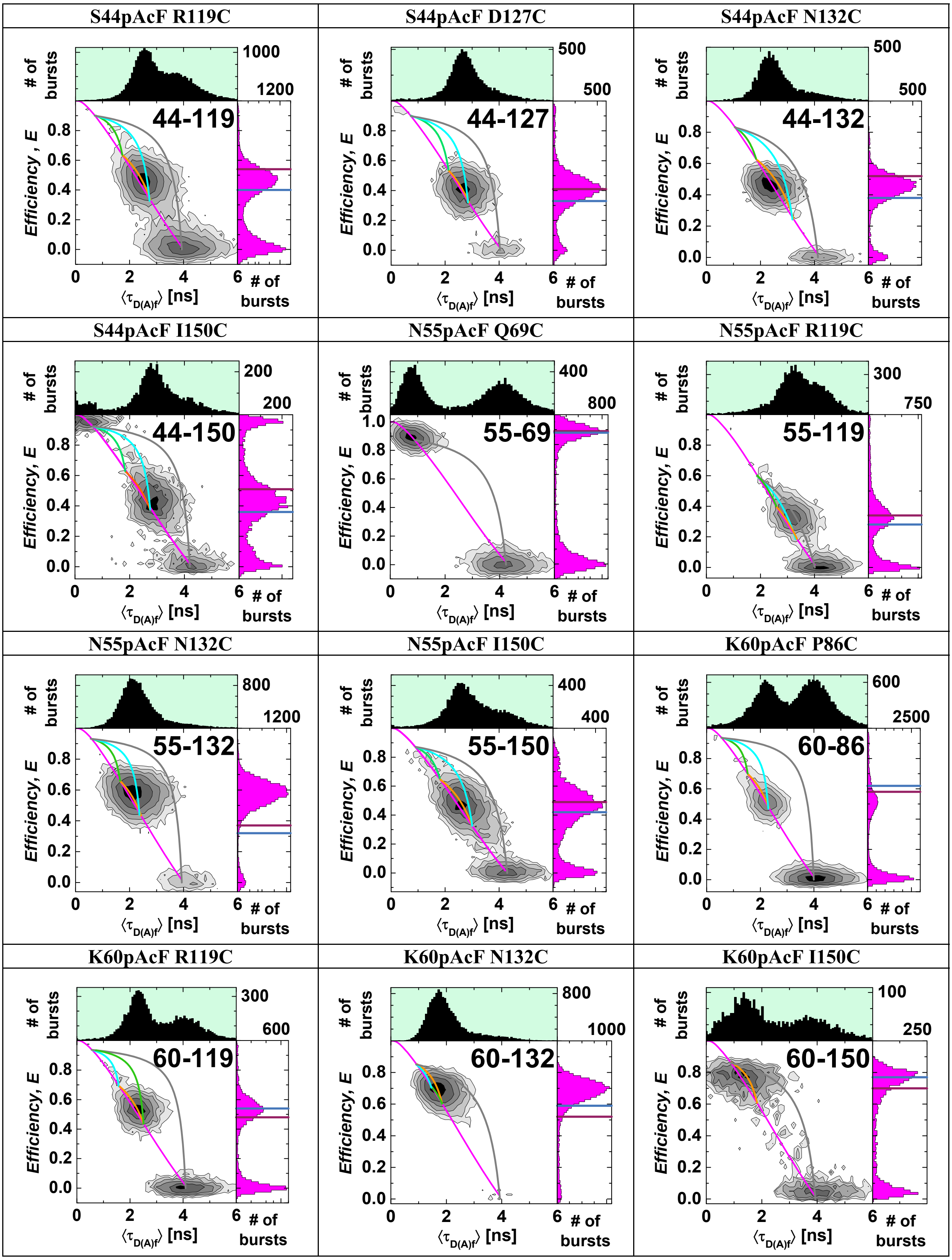

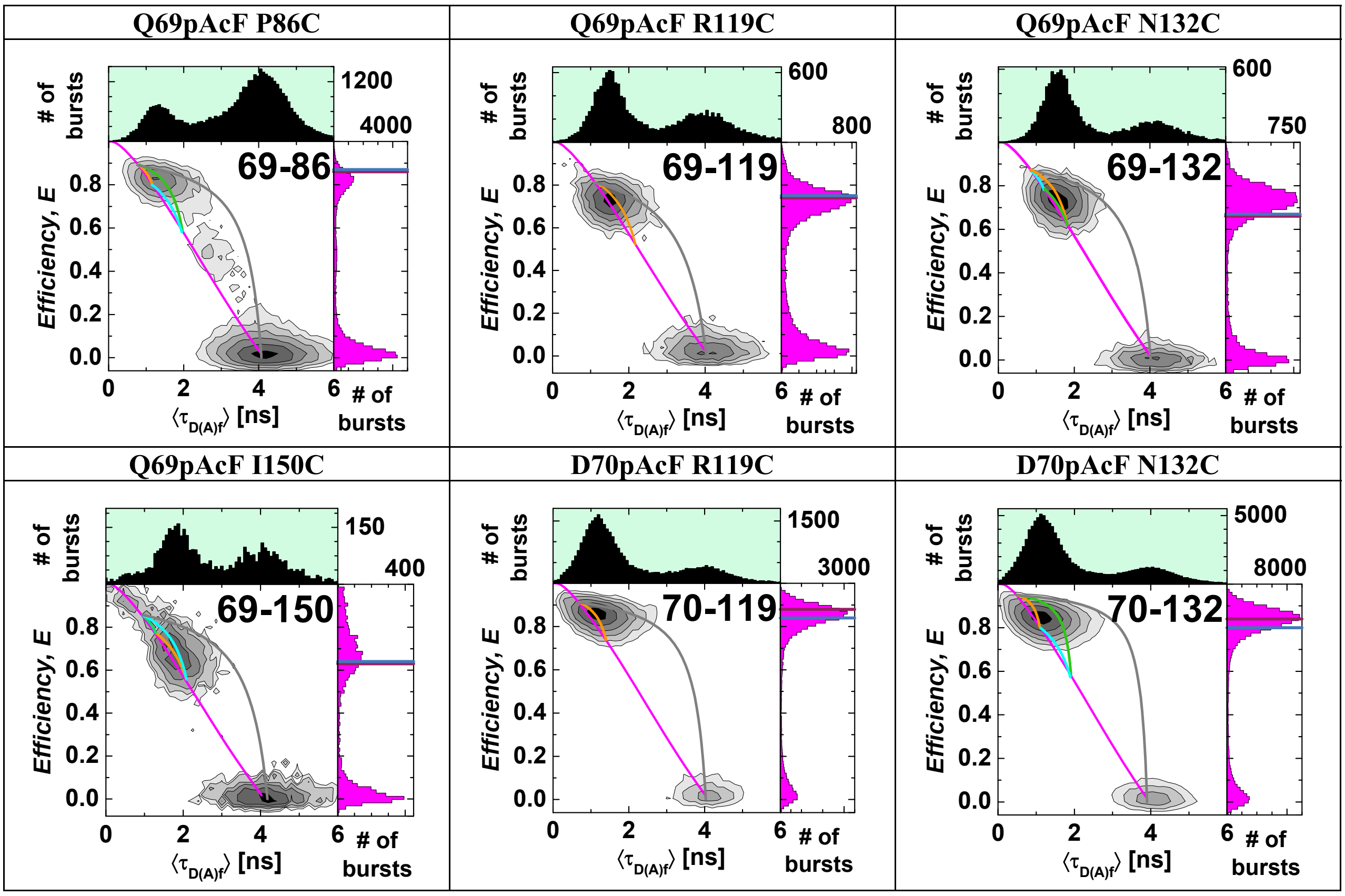

**Figure S3 MFD analysis of all 33 variants of the T4L network.** Two dimensional histogram of FRET efficiency $E$ vs. lifetime of donor in the presence of acceptor $\langle\tau_{D(A)}\rangle_f$. One dimensional projections for $E$ and $\langle\tau_{D(A)}\rangle_f$ are also shown. Static (magenta) and dynamic FRET lines connecting states $C_1$-$C_2$ (orange), $C_1$-$C_3$ (cyan) and $C_2$-$C_3$ (bright green) are also shown (S1.2, eq. S7-S9). Solid horizontal lines show the FRET efficiency expected from known X-ray structures for the *open* (blue, PDB ID 172L) and *closed* (violet, 148L) state from T4L.

## 2.2 Species Cross Correlation Function -(DA) and –(AD) labeled samples.

Theoretically, the species cross-correlation function (*sCCF*), as defined in Eq. (S13), can be extended to more than two species in solution. Practically, this suffers of technical limitations. The more species one has in solution, the more photons are required to differentiate between them. Therefore, we selected two pseudo-species that represent mixtures of the states found in solution. In addition, we added a third pseudo species that takes into consideration the contribution of scatter photons [8]. In this approach, the meaning of specific amplitudes and their relationships is lost; however, *sCCF* can extract the relaxation times as kinetic signatures of conformational transitions between all possible states.

For all data presented, we generated two pseudo-species, plus the addition of the scatter-filter. Decays were generated accordingly to Eq. (S18)-(S19), based on sub-ensemble burst analysis and eTCSPC data. In some cases, lifetimes of the pseudo-species were adjusted by 100's of ps to properly cross the y-axis of the correlation at a predetermined time for visual comparison. This procedure does not affect the recovered relaxation times.

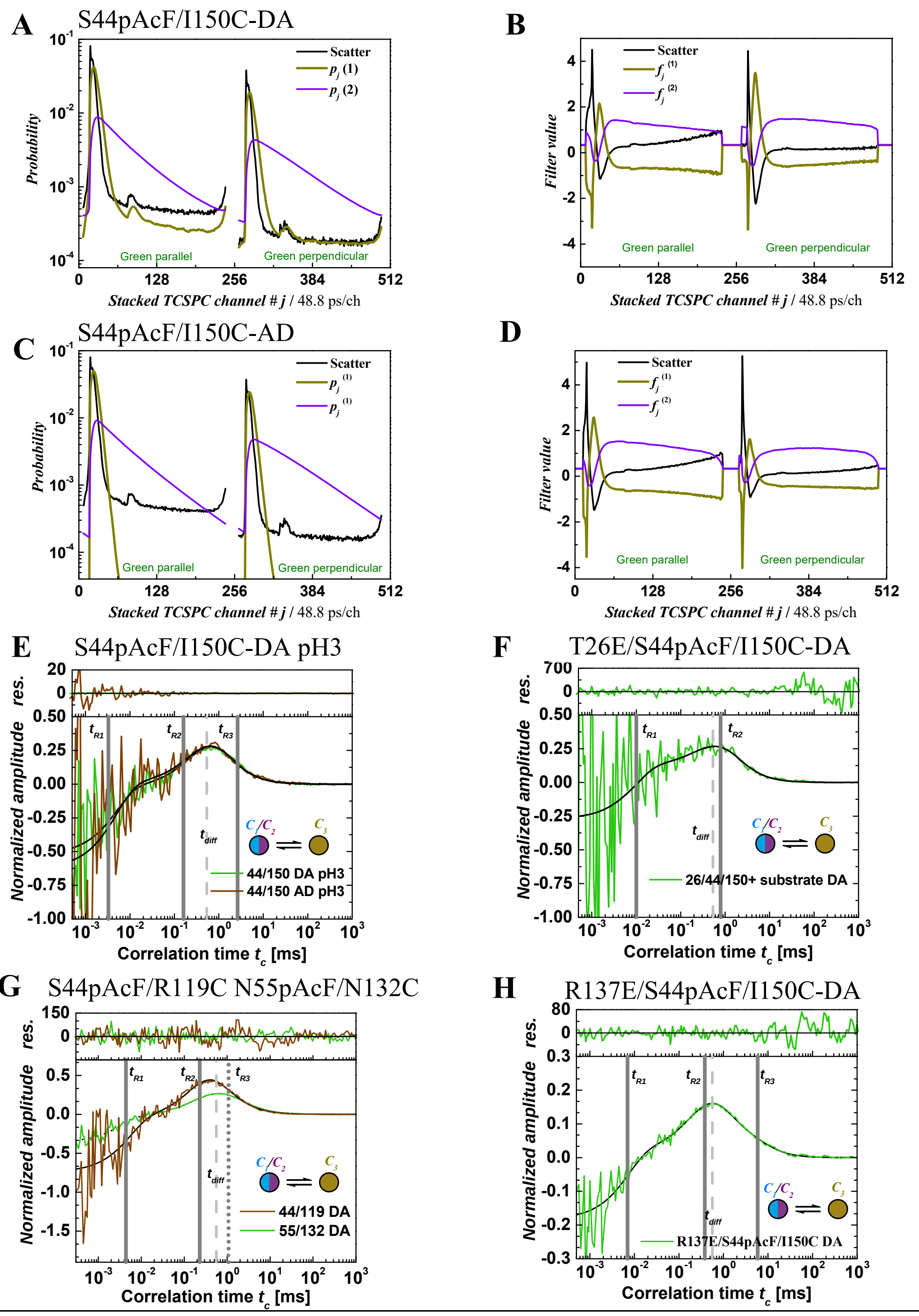

**Figure S4 fFCS results. (A)** Generated decays for two pseudo-species of S44pAcF/I150C-DA in addition to the scatter profile. The parameters of the decay generation for the first pseudo-species were $\tau_1$ = 0.25 ns, and rotational correlation time of $\rho_1$ = 3.3 ns. The second pseudo-species had a lifetime of $\tau_2$ = 2.97 ns and the same rotational correlation time. **(B)** Filters $f_j^{(i)}$ were calculated according to Eq. (S14) using the decays from graph (A). **(C, D)** Decay patterns and the corresponding filters for the S44pAcF/I150C-AD with first pseudo species lifetime $\tau_1$ = 0.25 ns, and rotational correlation time $\rho_1$ = 3.3 ns. The second pseudo species was generated with $\tau_2$ = 3.25 ns and same rotational correlation time. **(E)** *sCCF* between the mix $C_1/C_2$ and $C_3$ at pH 3.0 for the two configurations of labeling –(DA) and –(AD). The fit with eqs. (S22-S23) required three relaxation times. The diffusion time was fixed to $t_{diff}$ = 0.54 ms. **(F)** *sCCF* between the mix $C_1/C_2$ and $C_3$ for T26E/S44pAcF/I150C-(DA) incubated with substrate. Two relaxation times are found ($t_{R1}$= 10 µs, and $t_{R2}$ = 0.790 ms). **(G)** Overlay of the normalized *sCCF* of S44pAcF/R119C-DA and N55pAcF/N132C-(DA). Global fit shows two common relaxation times ($t_{R1}$ = 4 ± 2.4 µs, $t_{R2}$ = 230 ± 28 µs). The variant N55pAcF/N132C-(DA) requires an additional rate $t_{R3}$ ~ 1.1 ms. **(H)** *sCCF* for variant R137E/S44pAcF/I150C-(DA). Three relaxation times were needed to fit the curve ($t_{R1}$ = 7 µs, $t_{R2}$ = 0.38 ms and $t_{R3}$ = 5.84 ms). The diffusion time was fixed to $t_{diff}$ = 0.54 ms.

Considering the case of the double labeled mutants S44pAcF/I150C-(DA) and –(AD), the patterns $p_j^{(i)}$ that correspond to the normalized probability distributions for the –(DA) and –(AD) samples are shown on Figure S4A, C. The parameters used on the decay generation are shown in the caption. From these patterns, the filters $f_j^{(i)}$ (Eq. (S14)) were calculated. These are shown in panels B and D of Figure S4. These filters are then used to compute the *sCCF* by multiplying each photon and weighting its contribution to each state as in Eq. (S13). The patterns that are shown in Figure S4 correspond to only half of the detectors. The other half shows similar patterns. The need of another set of detectors with similar patterns and decays is to increase the amount of pair correlations and to exclude detector after-pulsing related artifacts from calculations. Finally, a full correlation containing all relaxation times and the characteristic diffusion time can be extracted. The reproducibility of the methodology is observed by the overlap of the two species cross-correlations (Figure 3B, main text), even with the fact that different parameters were used on the generation of the filters. Similar overlap is shown for the mutant S44pAcF/I150C-(DA) and –(AD) at pH 3.0 (Figure S4E). For the functional mutants (E11A, T26E, R137E), we show the *sCCF* (Figure S4F-H).

### 2.3 Fluorescence decay analysis of single and double labeled T4 Lysozyme

Selected mutants were labeled in two configurations (DA) and (AD), D for donor (Alexa488) and A for acceptor fluorophore (Alexa647). The order of the letters represents the position of the fluorophore. The first letter represents the label of the keto handle in the N-terminal subdomain and the second position corresponds to the thiol reaction for labeling in the C-terminus, except for the double cysteine mutant.

Each sample was measured in eTCSPC as described in the materials and methods section and analyzed with three different models. As the fluorophores are connected to T4L by long and flexible linkers (S1.5-S1.7) the assumption of a static, fixed interdye distance does not reflect the actual sample property. In fact, the flexible linkers assure a free rotational motion of the fluorophore, which allows to assume $\langle\kappa^2\rangle = 2/3$ (verified by corresponding anisotropy measurements of each dye, see Table S5A-D). Yet, this conformational flexibility leads to a distribution of interdye distances on the timescale of FRET. A proper model for describing our sample properties has to consider this distribution. Here, we modeled this interdye distance distribution $\langle R_{DA}\rangle$ with a Normal distribution (Eq. S24) (Table S2B-F). The best consistent model based on our experimental data and statistical analysis is that three continuous distance distributions are needed to describe all T4L variants.

**Table S2A.** Multi-exponential fit of Donor-only labeled samples, fluorescence quantum yields of the donor (Alexa488) and acceptor (Alexa647) fitted with Eq. (S20). Fluorescence quantum yields are calculated from the species averaged lifetimes $\langle\tau\rangle_x = \sum_{i=0}^{n} x_i \tau^{(i)}$ , where $x_i$'s are the species fractions. Empty cells represent parameters that are not available.

| Samples | $\tau_1$ [ns] | $x_1$ | $\tau_2$ [ns] | $x_2$ | $\tau_3$ [ns] | $x_3$ | $\langle\tau\rangle_{x,D(0)}$ [ns] | $\Phi_{FD(0)}$ | $\langle\tau\rangle_{x,A}$ [ns] | $\Phi_{FA}$ |
|---|---|---|---|---|---|---|---|---|---|---|
| E5pAcF/S44C | 4.10 | 0.77 | 2.44 | 0.17 | 0.43 | 0.06 | 3.59 | 0.78 | 1.24 | 0.36 |
| E5pAcF/N132C | 4.11 | 0.78 | 2.33 | 0.16 | 0.34 | 0.06 | 3.59 | 0.78 | 1.25 | 0.37 |
| R8pAcF/Q69C | 4.10 | 0.89 | 1.60 | 0.11 | | | 3.83 | 0.78 | 1.23 | 0.34 |
| R8pAcF/P86C | 4.16 | 0.92 | 1.89 | 0.08 | | | 3.98 | 0.79 | 1.35 | 0.40 |
| R8pAcF/R119C | 4.03 | 0.77 | 2.43 | 0.17 | 0.38 | 0.05 | 3.54 | 0.75 | 1.26 | 0.34 |
| R8pAcF/N132C | 4.08 | 0.79 | 2.27 | 0.14 | 0.34 | 0.07 | 3.55 | 0.76 | 1.25 | 0.34 |
| K19pAcF/Q69C | 4.25 | 0.78 | 2.22 | 0.13 | 0.65 | 0.09 | 3.60 | 0.80 | 1.22 | 0.33 |
| K19pAcF/P86C | 3.90 | 0.77 | 2.56 | 0.15 | 0.32 | 0.08 | 3.39 | 0.73 | 1.30 | 0.38 |
| K19pAcF/R119C | 3.91 | 0.74 | 2.51 | 0.15 | 0.52 | 0.11 | 3.27 | 0.73 | 1.32 | 0.39 |
| K19pAcF/N132C | 3.91 | 0.81 | 2.66 | 0.14 | 0.32 | 0.05 | 3.54 | 0.74 | 1.20 | 0.33 |
| E22pAcF/D127C | 3.83 | 0.36 | 1.83 | 0.47 | 0.67 | 0.17 | 2.25 | 0.60 | 1.24 | 0.34 |
| S36pAcF/P86C | 4.40 | 0.81 | 2.28 | 0.14 | 0.49 | 0.05 | 3.89 | 0.83 | 1.30 | 0.39 |
| S36pAcF/N132C | 4.39 | 0.83 | 2.23 | 0.13 | 0.46 | 0.05 | 3.91 | 0.83 | 1.25 | 0.34 |
| S44pAcF/Q69C | 4.32 | 0.94 | 1.75 | 0.06 | | | 4.17 | 0.83 | 1.25 | 0.34 |
| S44pAcF/P86C | 4.32 | 0.94 | 1.75 | 0.06 | | | 4.17 | 0.83 | 1.26 | 0.35 |
| S44pAcF/R119C | 4.32 | 0.96 | 1.44 | 0.04 | | | 4.20 | 0.84 | 1.27 | 0.35 |
| S44pAcF/D127C | 4.28 | 0.85 | 2.25 | 0.10 | 0.39 | 0.05 | 3.85 | 0.81 | 1.28 | 0.35 |
| S44pAcF/N132C | 4.32 | 0.96 | 1.44 | 0.04 | | | 4.20 | 0.84 | 1.37 | 0.37 |
| S44pAcF/I150C | 4.32 | 0.96 | 1.44 | 0.04 | | | 4.20 | 0.84 | 1.34 | 0.40 |
| N55pAcF/Q69C | 4.14 | 0.92 | 1.48 | 0.08 | | | 3.93 | 0.79 | 1.32 | 0.39 |
| N55pAcF/R119C | 4.26 | 0.78 | 2.31 | 0.15 | 0.24 | 0.07 | 3.67 | 0.80 | 1.35 | 0.41 |
| N55pAcF/N132C | 4.28 | 0.94 | 1.49 | 0.06 | | | 4.11 | 0.82 | 1.33 | 0.39 |
| N55pAcF/I150C | 4.32 | 0.69 | 3.08 | 0.25 | 0.72 | 0.06 | 3.75 | 0.79 | 1.48 | 0.40 |
| K60pAcF/P86C | 4.12 | 0.94 | 2.07 | 0.06 | | | 4.00 | 0.79 | 1.40 | 0.41 |
| K60pAcF/R119C | 4.26 | 0.91 | 1.81 | 0.09 | | | 4.04 | 0.81 | 1.34 | 0.37 |
| K60pAcF/N132C | 4.15 | 0.89 | 1.78 | 0.11 | | | 3.89 | 0.79 | 1.30 | 0.36 |
| K60pAcF/I150C | 4.09 | 0.88 | 1.76 | 0.12 | | | 3.81 | 0.77 | 1.35 | 0.37 |
| Q69pAcF/P86C | 4.20 | 0.94 | 1.52 | 0.06 | | | 4.04 | 0.81 | 1.36 | 0.37 |
| Q69pAcF/R119C | 4.20 | 0.88 | 1.64 | 0.12 | | | 3.89 | 0.79 | 1.37 | 0.38 |
| Q69pAcF/N132C | 4.20 | 0.89 | 1.47 | 0.11 | | | 3.90 | 0.80 | 1.40 | 0.38 |
| Q690pAcF/I150C | 4.20 | 0.89 | 1.88 | 0.11 | | | 3.94 | 0.80 | 0.94 | 0.35 |
| D70pAcF/R119C | 4.14 | 0.68 | 2.61 | 0.23 | 0.82 | 0.09 | 3.42 | 0.76 | 1.18 | 0.32 |
| D70pAcF/N132C | 4.08 | 0.88 | 1.12 | 0.12 | | | 3.72 | 0.78 | 1.33 | 0.36 |

**Table S2B.** Table of determined state-specific mean distances for the distribution fit (*N*=2, eq. S24) with individually fitted species fractions. Average $\chi^2_r$ for this fitting model is 1.0825.

| Samples | $\langle R_{DAmajor} \rangle$ [Å] | $x_{major}$† | $\langle R_{DAminor} \rangle$ [Å] | $x_{minor}$† | $x_{D(0)}$* | $\chi^2_r$ |
|---|---|---|---|---|---|---|
| E5pAcF/S44C | 42.2 | 0.94 | 27.9 | 0.06 | 0.074 | 1.12 |
| E5pAcF/N132C | 35.0 | 0.63 | 45.8 | 0.37 | 0.041 | 1.01 |
| R8pAcF/Q69C | 38.0 | 0.56 | 37.6 | 0.44 | 0.206 | 1.07 |
| R8pAcF/P86C | 39.5 | 0.68 | 50.1 | 0.32 | 0.178 | 1.06 |
| R8pAcF/R119C | 44.1 | 0.65 | 28.5 | 0.36 | 0.082 | 1.03 |
| R8pAcF/N132C | 17.9 | 0.84 | 40.4 | 0.16 | 0.053 | 1.02 |
| K19pAcF/Q69C | 38.9 | 0.62 | 38.9 | 0.38 | 0.513 | 1.07 |
| K19pAcF/P86C | 48.2 | 0.62 | 56.8 | 0.39 | 0.068 | 1.02 |
| K19pAcF/R119C | 49.1 | 0.55 | 49.1 | 0.45 | 0.425 | 1.05 |
| K19pAcF/N132C | 41.7 | 0.71 | 53.3 | 0.29 | 0.336 | 1.10 |
| E22pAcF/D127C | 28.1 | 0.65 | 45.2 | 0.35 | 0.581 | 0.99 |
| S36pAcF/P86C | 49.8 | 0.62 | 36.1 | 0.38 | 0.029 | 1.03 |
| S36pAcF/N132C | 38.6 | 0.54 | 50.7 | 0.46 | 0.031 | 0.96 |
| S44pAcF/Q69C | 20.9 | 0.78 | 36.9 | 0.22 | 0.068 | 1.05 |
| S44pAcF/P86C | 54.1 | 0.62 | 39.9 | 0.39 | 0.082 | 1.05 |
| S44pAcF/R119C | 56.9 | 0.68 | 41.9 | 0.32 | 0.2146 | 1.12 |
| S44pAcF/D127C | 57.0 | 0.83 | 39.9 | 0.17 | 0.101 | 1.03 |
| S44pAcF/N132C | 63.8 | 0.55 | 43.9 | 0.45 | 0.357 | 1.14 |
| S44pAcF/I150C | 55.8 | 0.73 | 41.9 | 0.27 | 0.027 | 1.09 |
| N55pAcF/Q69C | 37.3 | 0.79 | 39.3 | 0.21 | 0.231 | 1.16 |
| N55pAcF/R119C | 64.8 | 0.63 | 51.3 | 0.37 | 0.196 | 1.09 |
| N55pAcF/N132C | 53.3 | 0.68 | 40.2 | 0.32 | 0.042 | 1.04 |
| N55pAcF/I150C | 49.6 | 0.97 | 32.9 | 0.04 | 0.723 | 1.13 |
| K60pAcF/P86C | 53.2 | 0.59 | 41.7 | 0.41 | 0.338 | 1.11 |
| K60pAcF/R119C | 54.5 | 0.61 | 42.2 | 0.39 | 0.263 | 1.12 |
| K60pAcF/N132C | 48.7 | 0.52 | 38.2 | 0.48 | 0.164 | 1.04 |
| K60pAcF/I150C | 39.8 | 0.74 | 51.7 | 0.26 | 0.196 | 1.05 |
| Q69pAcF/P86C | 38.6 | 0.87 | 58.7 | 0.14 | 0.396 | 1.09 |
| Q69pAcF/R119C | 40.5 | 0.87 | 52.8 | 0.13 | 0.306 | 1.10 |
| Q69pAcF/N132C | 37.4 | 0.55 | 47.2 | 0.45 | 0.236 | 1.10 |
| Q690pAcF/I150C | 41.5 | 0.64 | 53.8 | 0.37 | 0.4 | 1.09 |
| D70pAcF/R119C | 34.1 | 0.56 | 43.1 | 0.44 | 0.115 | 1.04 |
| D70pAcF/N132C | 34.8 | 0.68 | 45.3 | 0.32 | 0.142 | 1.07 |

† Values for the FRET populations are normalized such that $x_{minor} + x_{major} = 1$. *Donor decay was fixed and $x_{D(0)}$ represents the fraction of donor only from the total.

**Table S2C.** Table of determined state-specific average distances for the distribution fit ($N$=2, eq. S24) with globally shared species fractions, $x_{minor}$ = 0.466, $x_{major}$=0.534. Average $\chi^2_r$ for this fitting model is 1.0985.

| Samples | $\langle R_{DAminor}\rangle$ [Å] | $\langle R_{DAmajor}\rangle$ [Å] | $x_{D(0)}$* | $\chi^2_r$ |
|---|---|---|---|---|
| E5pAcF/S44C | 40.3 | 43.4 | 0.08 | 1.16 |
| E5pAcF/N132C | 44.4 | 32.8 | 0.04 | 1.05 |
| R8pAcF/Q69C | 37.8 | 37.8 | 0.21 | 1.08 |
| R8pAcF/P86C | 35.7 | 47.0 | 0.17 | 1.09 |
| R8pAcF/R119C | 37.3 | 45.8 | 0.10 | 1.07 |
| R8pAcF/N132C | 28.2 | 41.4 | 0.17 | 0.97 |
| K19pAcF/Q69C | 39.1 | 39.1 | 0.05 | 1.05 |
| K19pAcF/P86C | 46.2 | 55.5 | 0.06 | 1.00 |
| K19pAcF/R119C | 49.1 | 49.1 | 0.04 | 1.07 |
| K19pAcF/N132C | 38.2 | 49.2 | 0.03 | 1.11 |
| E22pAcF/D127C | 29.7 | 43.4 | 0.07 | 1.03 |
| S36pAcF/P86C | 39.3 | 51.1 | 0.03 | 1.10 |
| S36pAcF/N132C | 36.8 | 49.6 | 0.03 | 1.00 |
| S44pAcF/Q69C | 29.0 | 38.6 | 0.18 | 1.38 |
| S44pAcF/P86C | 42.8 | 55.6 | 0.08 | 1.06 |
| S44pAcF/R119C | 45.9 | 59.8 | 0.20 | 1.15 |
| S44pAcF/D127C | 49.4 | 61.7 | 0.07 | 1.11 |
| S44pAcF/N132C | 43.4 | 61.7 | 0.42 | 1.14 |
| S44pAcF/I150C | 46.7 | 58.6 | 0.01 | 1.13 |
| N55pAcF/Q69C | 37.2 | 37.2 | 0.23 | 1.17 |
| N55pAcF/R119C | 52.7 | 67.7 | 0.17 | 1.13 |
| N55pAcF/N132C | 43.7 | 55.2 | 0.04 | 1.10 |
| N55pAcF/I150C | 49.5 | 49.5 | 0.73 | 1.13 |
| K60pAcF/P86C | 41.2 | 53.9 | 0.32 | 1.17 |
| K60pAcF/R119C | 55.6 | 44.6 | 0.05 | 1.14 |
| K60pAcF/N132C | 49.2 | 39.1 | 0.17 | 1.06 |
| K60pAcF/I150C | 47.2 | 36.7 | 0.20 | 1.04 |
| Q69pAcF/P86C | 34.4 | 42.8 | 0.42 | 1.11 |
| Q69pAcF/R119C | 45.8 | 37.0 | 0.30 | 1.12 |
| Q69pAcF/N132C | 47.0 | 37.4 | 0.23 | 1.12 |
| Q690pAcF/I150C | 51.8 | 39.8 | 0.39 | 1.09 |
| D70pAcF/R119C | 31.7 | 42.1 | 0.11 | 1.04 |
| D70pAcF/N132C | 31.1 | 42.6 | 0.14 | 1.14 |

† Values for the FRET populations are normalized such that $x_{major}+x_{minor}$ =1. *Donor decay was fixed and $x_{D(0)}$ represents the fraction of donor only from the total.

**Table S2D.** Table of determined state-specific mean distances for state $C_1$ from the distribution fit ($N$=3, eq. S24) with globally shared species fractions including statistical uncertainties as described in sections S1.5-1.7 Distances were used for FPS. We present the weighted residuals for the lowest $\chi_r^2$ for the conformation $C_1$ and the Cα- Cα distances from the model structure PDBID 172L. Minimal $\chi^2_{r,global}$ for this fitting model is 1.0736. With a 1σ-confidence interval, the fraction $x_3$ of $\langle R_{DA3}\rangle$ is 0.1 – 0.27, shown are here the distances for $x_3 = x_{middle}$ = 0.18 with $x_1$ = 0.44 and $x_2$ = 0.38. $\langle R_{DA1,min}\rangle$ and $\langle R_{DA1,max}\rangle$ are the shortest and longest distance below the 1σ-threshold. err- and err+ are the $\Delta R_{DA}^2(\kappa^2, k_{FRET})$ calculated according to eq. 6.

| Sample | $\langle R_{DA1}\rangle$ [Å] | $\langle R_{DA1,min}\rangle$ [Å] | $\langle R_{DA1,max}\rangle$ [Å] | err- [Å] | err+ [Å] | $C_\alpha$-$C_\alpha$ X-ray * [Å] | $\langle R_{DA}\rangle$ X-ray [Å] | *w.res.* *** [Å/Å] | $x_{D(0)}$ * | $\chi^2_r$ | Class |
|---|---|---|---|---|---|---|---|---|---|---|---|
| E5pAcF/S44C | 42.3 | 42.3 | 42.5 | 4.4 | 4.4 | 27.3 | 41.8 | -0.13 | 0.07 | 1.09 | $R_1 = R_2$, $R_3$ var. |
| E5pAcF/N132C | 34.7 | 33.4 | 34.7 | 5.4 | 5.4 | 25.7 | 42.8 | 1.62 | 0.04 | 1.01 | $R_1 = R_3$, $R_2$ var. |
| R8pAcF/Q69C | 37.8 | 37.8 | 37.8 | 3.9 | 3.9 | 15.1 | 34.7 | -0.80 | 0.21 | 1.08 | $R_1 = R_2 = R_3$ |
| R8pAcF/P86C | 47.6 | 46.9 | 48.2 | 5.4 | 5.3 | 26.8 | 46.1 | -0.28 | 0.16 | 1.08 | $R_2 = R_3$, $R_1$ var. |
| R8pAcF/R119C | 45.9 | 45.6 | 45.9 | 6.0 | 6.0 | 28.2 | 47.1 | 0.23 | 0.09 | 1.04 | $R_1 > R_2 > R_3$ |
| R8pAcF/N132C | 42.4 | 43.0 | 42.7 | 5.7 | 5.7 | 26 | 42.9 | 0.16 | 0.04 | 0.97 | $R_1 > R_3 > R_2$ |
| K19pAcF/Q69C | 39.1 | 39.1 | 39.1 | 4.0 | 4.0 | 20.2 | 37.7 | -0.35 | 0.53 | 1.05 | $R_1 = R_2 = R_3$ |
| K19pAcF/P86C | 54.2 | 54.2 | 54.2 | 5.5 | 5.5 | 35.9 | 52.4 | -0.33 | 0.04 | 1.00 | $R_1 > R_2 > R_3$ |
| K19pAcF/R119C | 56.4 | 56.4 | 56.4 | 6.2 | 6.2 | 37.5 | 51.4 | -0.81 | 0.25 | 1.07 | $R_1 = R_2 = R_3$ |
| K19pAcF/N132C | 50.4 | 49.8 | 51.3 | 7.0 | 7.0 | 35.3 | 46.5 | -0.58 | 0.33 | 1.11 | $R_2 = R_3$, $R_1$ var. |
| E22pAcF/D127C | 41.5 | 37.1 | 45.9 | 6.7 | 6.7 | 37.3 | 44.7 | 0.48 | 0.53 | 1.01 | $R_3 > R_1 > R_2$ |
| S36pAcF/P86C | 51.3 | 51.2 | 51.4 | 5.2 | 5.2 | 37.4 | 47 | -0.83 | 0.03 | 1.06 | $R_1 > R_2 > R_3$ |
| S36pAcF/N132C | 50.9 | 50.3 | 51.4 | 6.6 | 6.6 | 41.0 | 50.6 | -0.04 | 0.03 | 0.96 | $R_2 = R_3$, $R_1$ var. |
| S44pAcF/Q69C | 29.8 | 29.0 | 32.4 | 4.7 | 4.7 | 18.4 | 27.3 | -0.72 | 0.13 | 1.14 | $R_1 = R_2$, $R_3$ var. |
| S44pAcF/P86C | 55.8 | 55.3 | 56.2 | 6.0 | 6.0 | 39.3 | 51.5 | -0.72 | 0.07 | 1.05 | $R_1 = R_3$, $R_2$ var. |
| S44pAcF/R119C | 59.7 | 58.6 | 59.8 | 4.9 | 4.9 | 44.2 | 57 | -0.45 | 0.21 | 1.13 | $R_1 > R_2 > R_3$ |
| S44pAcF/D127C | 58.4 | 58.0 | 60.6 | 6.8 | 6.8 | 47.6 | 60.3 | 0.14 | 0.10 | 1.01 | $R_1 > R_2 > R_3$ |
| S44pAcF/N132C | 64.8 | 63.3 | 66.3 | 7.7 | 7.7 | 45.2 | 57.6 | -0.94 | 0.40 | 1.14 | $R_1 > R_2 > R_3$ |
| S44pAcF/I150C | 58.2 | 57.2 | 59.1 | 5.9 | 5.8 | 38.9 | 58.2 | 0.00 | 0.03 | 1.10 | $R_1 > R_2 > R_3$ |
| N55pAcF/Q69C | 37.1 | 37.1 | 37.1 | 4.4 | 4.4 | 20.9 | 34.5 | -0.59 | 0.23 | 1.17 | $R_1 = R_2 = R_3$ |
| N55pAcF/R119C | 68.4 | 68.0 | 69.2 | 5.6 | 5.6 | 46.5 | 62.5 | -1.09 | 0.18 | 1.13 | $R_1 > R_2 > R_3$ |
| N55pAcF/N132C | 55.2 | 54.5 | 55.2 | 5.5 | 5.5 | 46.2 | 60.6 | 1.05 | 0.04 | 1.08 | $R_1 > R_2 > R_3$ |
| N55pAcF/I150C | 60.8 | 60.3 | 61.3 | 6.8 | 6.8 | 38.7 | 55.7 | -0.75 | 0.48 | 1.13 | $R_1 = R_2$, $R_3$ var. |
| K60pAcF/P86C | 54.0 | 53.6 | 54.0 | 6.0 | 6.0 | 32.7 | 47.3 | -1.09 | 0.30 | 1.15 | $R_1 > R_2 > R_3$ |
| K60pAcF/R119C | 47.4 | 45.6 | 50.8 | 5.9 | 5.9 | 36.4 | 51.1 | 0.50 | 0.06 | 1.09 | $R_2 > R_1 > R_3$ |
| K60pAcF/N132C | 37.7 | 37.3 | 38.3 | 4.4 | 4.4 | 35.9 | 49.1 | 2.54 | 0.17 | 1.06 | $R_2 = R_3$, $R_1$ var. |
| K60pAcF/I150C | 37.8 | 37.3 | 38.2 | 4.3 | 4.3 | 27.1 | 40.3 | 0.59 | 0.20 | 1.03 | $R_1 = R_3$, $R_2$ var. |
| Q69pAcF/P86C | 38.6 | 37.2 | 40.7 | 4.3 | 4.3 | 22.0 | 34 | -1.15 | 0.42 | 1.11 | $R_1 = R_2$, $R_3$ var. |
| Q69pAcF/R119C | 39.9 | 39.1 | 40.5 | 4.7 | 4.7 | 27.9 | 41.9 | 0.45 | 0.30 | 1.10 | $R_1 = R_2$, $R_3$ var. |
| Q69pAcF/N132C | 37.3 | 36.8 | 37.8 | 4.5 | 4.5 | 31.0 | 45.4 | 1.81 | 0.24 | 1.12 | $R_1 = R_3$, $R_2$ var. |
| Q690pAcF/I150C | 50.8 | 50.2 | 51.5 | 5.9 | 5.9 | 24.7 | 47.4 | -0.58 | 0.36 | 1.09 | $R_1 = R_3$, $R_2$ var. |
| D70pAcF/R119C | 43.1 | 42.6 | 43.6 | 4.6 | 4.6 | 25.8 | 36.8 | -1.36 | 0.12 | 1.04 | $R_2 = R_3$, $R_1$ var. |
| D70pAcF/N132C | 38.8 | 36.9 | 40.3 | 4.8 | 4.8 | 28.2 | 38.6 | 0.00 | 0.13 | 1.07 | $R_3 > R_1 > R_2$ |

**Table S2E.** Table of determined state-specific mean distances for state $C_2$ from the distribution fit ($N$=3, eq. S24) with globally shared species fractions including statistical uncertainties as described in sections 1.5-1.7. Distances were used for FPS. We present the weighted residuals for the lowest $\chi_r^2$ for the conformation $C_2$ and the Cα- Cα distances from the model structure PDBID 148L. Minimal $\chi^2_{r,global}$ for this fitting model is 1.0736. With a 1σ-confidence interval, the fraction $x_3$ of $\langle R_{DA3}\rangle$ is 0.1 – 0.27, shown are here the distances $x_3 = x_{middle} = 0.18$ with $x_1 = 0.44$ and $x_2 = 0.38$. $\langle R_{DA1,min}\rangle$ and $\langle R_{DA1,max}\rangle$ are the shortest and longest distance below the 1σ-threshold. err- and err+ are the $\Delta R_{DA}^2\left(\kappa^2, k_{FRET}\right)$ calculated according to eq. 6.

| **Sample** | **$\langle R_{DA2}\rangle$ [Å]** | **$\langle R_{DA2,min}\rangle$ [Å]** | **$\langle R_{DA2,max}\rangle$ [Å]** | **err- [Å]** | **err+ [Å]** | **$C_\alpha$-$C_\alpha$ X-ray * [Å]** | **$\langle R_{DA}\rangle$ X-ray [Å]** | ***w.res.* *** [Å/Å]** | **$x_{D(0)}$*** | **$\chi^2_r$** | **Class** |
|---|---|---|---|---|---|---|---|---|---|---|---|
| E5pAcF/S44C | 42.3 | 42.1 | 42.3 | 4.4 | 4.4 | 28.9 | 43.5 | 0.30 | 0.07 | 1.09 | $R_1 = R_2$, $R_3$ var. |
| E5pAcF/N132C | 45.6 | 45.0 | 46.0 | 7.2 | 7.2 | 25.7 | 42.9 | -0.37 | 0.04 | 1.01 | $R_1 = R_3$, $R_2$ var. |
| R8pAcF/Q69C | 37.8 | 37.8 | 37.8 | 3.9 | 3.9 | 14.3 | 32 | -1.49 | 0.21 | 1.08 | $R_1 = R_2 = R_3$ |
| R8pAcF/P86C | 38.2 | 37.3 | 39.1 | 4.4 | 4.4 | 26.6 | 43.3 | 1.16 | 0.16 | 1.08 | $R_2 = R_3$, $R_1$ var. |
| R8pAcF/R119C | 39.5 | 38.3 | 41.3 | 5.4 | 5.4 | 28.3 | 46.7 | 1.27 | 0.09 | 1.04 | $R_1 > R_2 > R_3$ |
| R8pAcF/N132C | 31.1 | 35.4 | 33.8 | 4.6 | 4.6 | 25.6 | 44.4 | 2.31 | 0.17 | 0.97 | $R_1 > R_3 > R_2$ |
| K19pAcF/Q69C | 39.1 | 39.1 | 39.1 | 4.0 | 4.0 | 20.9 | 38.7 | -0.10 | 0.53 | 1.05 | $R_1 = R_2 = R_3$ |
| K19pAcF/P86C | 47.2 | 47.2 | 47.2 | 4.9 | 4.9 | 34.4 | 50.7 | -0.33 | 0.04 | 1.00 | $R_1 > R_2 > R_3$ |
| K19pAcF/R119C | 44.7 | 44.7 | 44.7 | 5.0 | 5.0 | 33.5 | 47.2 | 0.50 | 0.25 | 1.07 | $R_1 = R_2 = R_3$ |
| K19pAcF/N132C | 39.7 | 39.0 | 40.4 | 5.5 | 5.5 | 28.4 | 42 | 0.42 | 0.33 | 1.11 | $R_2 = R_3$, $R_1$ var. |
| E22pAcF/D127C | 36.8 | 33.6 | 40.0 | 5.5 | 5.5 | 28.7 | 37.1 | 0.05 | 0.53 | 1.01 | $R_3 > R_1 > R_2$ |
| S36pAcF/P86C | 41.6 | 40.2 | 44.7 | 4.9 | 4.9 | 34.3 | 42.8 | 0.08 | 0.03 | 1.06 | $R_1 > R_2 > R_3$ |
| S36pAcF/N132C | 37.6 | 36.6 | 38.4 | 5.0 | 5.0 | 30.5 | 41.4 | 0.78 | 0.03 | 0.96 | $R_2 = R_3$, $R_1$ var. |
| S44pAcF/Q69C | 29.8 | 29.0 | 32.4 | 4.7 | 4.7 | 18.9 | 27.4 | -0.70 | 0.13 | 1.14 | $R_1 = R_2$, $R_3$ var. |
| S44pAcF/P86C | 45.8 | 44.1 | 47.5 | 5.2 | 5.2 | 37.5 | 49 | 0.62 | 0.07 | 1.05 | $R_1 = R_3$, $R_2$ var. |
| S44pAcF/R119C | 50.1 | 47.5 | 53.8 | 5.2 | 5.2 | 40.0 | 50.7 | 0.01 | 0.21 | 1.13 | $R_1 > R_2 > R_3$ |
| S44pAcF/D127C | 56.1 | 51.5 | 58.9 | 7.2 | 7.2 | 43.8 | 56.5 | 0.18 | 0.10 | 1.01 | $R_1 > R_2 > R_3$ |
| S44pAcF/N132C | 47.8 | 45.5 | 50.8 | 6.2 | 6.2 | 38.3 | 51.3 | 0.52 | 0.40 | 1.14 | $R_1 > R_2 > R_3$ |

| | | | | | | | | | | |
|---|---|---|---|---|---|---|---|---|---|---|
| S44pAcF/I150C | 48.1 | 44.9 | 51.2 | 5.8 | 5.8 | 35.0 | 51.7 | 0.62 | 0.03 | 1.10 | $R_1 > R_2 > R_3$ |
| N55pAcF/Q69C | 37.1 | 37.1 | 37.1 | 4.4 | 4.4 | 21.5 | 33.5 | -0.82 | 0.23 | 1.17 | $R_1 = R_2 = R_3$ |
| N55pAcF/R119C | 56.6 | 54.6 | 59.2 | 5.2 | 5.2 | 44.8 | 59.8 | 0.56 | 0.18 | 1.13 | $R_1 > R_2 > R_3$ |
| N55pAcF/N132C | 46.8 | 45.0 | 47.2 | 4.7 | 4.7 | 42.1 | 58.3 | 2.58 | 0.04 | 1.08 | $R_1 > R_2 > R_3$ |
| N55pAcF/I150C | 47.6 | 47.2 | 48.1 | 5.4 | 5.4 | 37.1 | 52.9 | -0.75 | 0.48 | 1.13 | $R_1 = R_2$, $R_3$ var. |
| K60pAcF/P86C | 43.9 | 42.4 | 47.8 | 5.7 | 5.7 | 35.6 | 48.9 | 0.67 | 0.30 | 1.15 | $R_1 > R_2 > R_3$ |
| K60pAcF/R119C | 55.0 | 53.6 | 55.5 | 6.0 | 6.0 | 39.3 | 53.2 | -0.22 | 0.06 | 1.09 | $R_2 > R_1 > R_3$ |
| K60pAcF/N132C | 49.2 | 48.8 | 49.7 | 5.8 | 5.8 | 37.4 | 51.9 | 0.46 | 0.17 | 1.06 | $R_2 = R_3$, $R_1$ var. |
| K60pAcF/I150C | 48.5 | 47.8 | 49.5 | 5.6 | 5.6 | 29.9 | 43.6 | -0.90 | 0.20 | 1.03 | $R_1 = R_3$, $R_2$ var. |
| Q69pAcF/P86C | 36.7 | 35.6 | 37.5 | 3.8 | 3.8 | 23.2 | 34.8 | -0.46 | 0.42 | 1.11 | $R_1 = R_2$, $R_3$ var. |
| Q69pAcF/R119C | 40.0 | 38.1 | 42.4 | 5.1 | 5.1 | 28.9 | 41.7 | 0.28 | 0.30 | 1.10 | $R_1 = R_2$, $R_3$ var. |
| Q69pAcF/N132C | 47.8 | 47.5 | 48.2 | 5.7 | 5.7 | 30.1 | 45.7 | -0.38 | 0.24 | 1.12 | $R_1 = R_3$, $R_2$ var. |
| Q690pAcF/I150C | 42.2 | 41.4 | 42.9 | 5.0 | 5.0 | 24.7 | 47.3 | -0.58 | 0.36 | 1.09 | $R_1 = R_3$, $R_2$ var. |
| D70pAcF/R119C | 34.1 | 33.0 | 35.3 | 3.8 | 3.8 | 26.2 | 34.5 | 0.08 | 0.12 | 1.04 | $R_2 = R_3$, $R_1$ var. |
| D70pAcF/N132C | 30.4 | 30.2 | 31.0 | 3.6 | 3.6 | 26.6 | 37.5 | 1.92 | 0.13 | 1.07 | $R_3 > R_1 > R_2$ |

**Table S2F.** Table of determined state-specific mean distances for state $C_3$ from the distribution fit ($N$=3, eq. S24) with globally shared species fractions including statistical uncertainties as described in sections 1.5-1.7. Distances were used for FPS. Minimal $\chi^2_{r,global}$ for this fitting model is 1.0736. With a 1σ-confidence interval, the fraction $x_3$ of $\langle R_{DA3}\rangle$ is 0.1 – 0.27, shown are here the distances $x_3 = x_{middle} = 0.18$ with $x_1 = 0.44$ and $x_2 = 0.38$. $\langle R_{DA1,min}\rangle$ and $\langle R_{DA1,max}\rangle$ are the shortest and longest distance below the 1σ-threshold. err- and err+ are the $\Delta R^2_{DA}\left(\kappa^2, k_{FRET}\right)$ calculated according to eq. 6.

| Sample | $\langle R_{DA3}\rangle$ [Å] | $\langle R_{DA3,min}\rangle$ [Å] | $\langle R_{DA3,max}\rangle$ [Å] | err- [Å] | err+ [Å] | $x_{D(0)}$* | $\chi^2_r$ | Class |
|---|---|---|---|---|---|---|---|---|
| E5pAcF/S44C | 25.1 | 23.1 | 28.1 | 3.6 | 3.6 | 0.07 | 1.09 | $R_1 = R_2$, $R_3$ var. |
| E5pAcF/N132C | 35.3 | 33.9 | 37.7 | 5.9 | 5.9 | 0.04 | 1.01 | $R_1 = R_3$, $R_2$ var. |
| R8pAcF/Q69C | 37.8 | 37.8 | 37.8 | 3.9 | 3.9 | 0.21 | 1.08 | $R_1 = R_2 = R_3$ |
| R8pAcF/P86C | 30.0 | 29.3 | 30.7 | 4.6 | 4.6 | 0.16 | 1.08 | $R_2 = R_3$, $R_1$ var. |
| R8pAcF/R119C | 27.9 | 27.2 | 28.6 | 3.7 | 3.7 | 0.09 | 1.04 | $R_1 > R_2 > R_3$ |
| R8pAcF/N132C | 25.2 | 27.1 | 26.2 | 4.8 | 4.8 | 0.17 | 0.97 | $R_1 > R_3 > R_2$ |
| K19pAcF/Q69C | 39.1 | 39.1 | 39.1 | 4.0 | 4.0 | 0.53 | 1.05 | $R_1 = R_2 = R_3$ |
| K19pAcF/P86C | 32.3 | 28.7 | 36.1 | 5.0 | 5.2 | 0.04 | 1.00 | $R_1 > R_2 > R_3$ |
| K19pAcF/R119C | 31.0 | 31.0 | 31.0 | 4.6 | 4.6 | 0.25 | 1.07 | $R_1 = R_2 = R_3$ |
| K19pAcF/N132C | 39.7 | 39.0 | 41.6 | 5.7 | 5.7 | 0.33 | 1.11 | $R_2 = R_3$, $R_1$ var. |
| E22pAcF/D127C | 25.4 | 22.9 | 27.8 | 4.1 | 4.1 | 0.53 | 1.01 | $R_3 > R_1 > R_2$ |
| S36pAcF/P86C | 29.2 | 27.2 | 33.2 | 4.3 | 4.3 | 0.03 | 1.06 | $R_1 > R_2 > R_3$ |
| S36pAcF/N132C | 41.6 | 38.1 | 42.8 | 5.8 | 5.8 | 0.03 | 0.96 | $R_2 = R_3$, $R_1$ var. |
| S44pAcF/Q69C | 42.8 | 40.9 | 48.7 | 6.1 | 6.1 | 0.13 | 1.14 | $R_1 = R_2$, $R_3$ var. |
| S44pAcF/P86C | 54.0 | 51.2 | 54.9 | 5.8 | 5.8 | 0.07 | 1.05 | $R_1 = R_3$, $R_2$ var. |
| S44pAcF/R119C | 38.5 | 34.7 | 40.9 | 4.4 | 4.4 | 0.21 | 1.13 | $R_1 > R_2 > R_3$ |
| S44pAcF/D127C | 41.4 | 35.0 | 45.7 | 7.1 | 7.1 | 0.10 | 1.01 | $R_1 > R_2 > R_3$ |
| S44pAcF/N132C | 38.9 | 38.1 | 40.5 | 4.7 | 4.7 | 0.40 | 1.14 | $R_1 > R_2 > R_3$ |
| S44pAcF/I150C | 33.4 | 30.7 | 36.1 | 4.4 | 4.4 | 0.03 | 1.10 | $R_1 > R_2 > R_3$ |
| N55pAcF/Q69C | 37.8 | 37.6 | 38.2 | 4.5 | 4.5 | 0.23 | 1.17 | $R_1 = R_2 = R_3$ |
| N55pAcF/R119C | 49.0 | 47.6 | 50.3 | 4.2 | 4.2 | 0.18 | 1.13 | $R_1 > R_2 > R_3$ |
| N55pAcF/N132C | 34.4 | 29.7 | 34.4 | 4.0 | 4.0 | 0.04 | 1.08 | $R_1 > R_2 > R_3$ |
| N55pAcF/I150C | 37.3 | 30.4 | 44.2 | 8.1 | 8.1 | 0.48 | 1.13 | $R_1 = R_2$, $R_3$ var. |
| K60pAcF/P86C | 29.7 | 28.9 | 34.6 | 4.5 | 4.5 | 0.30 | 1.15 | $R_1 > R_2 > R_3$ |
| K60pAcF/R119C | 34.1 | 32.0 | 37.6 | 4.7 | 4.7 | 0.06 | 1.09 | $R_2 > R_1 > R_3$ |
| K60pAcF/N132C | 45.7 | 44.7 | 47.8 | 5.6 | 5.6 | 0.17 | 1.06 | $R_2 = R_3$, $R_1$ var. |
| K60pAcF/I150C | 38.0 | 37.2 | 38.8 | 4.4 | 4.4 | 0.20 | 1.03 | $R_1 = R_3$, $R_2$ var. |
| Q69pAcF/P86C | 52.4 | 57.5 | 54.7 | 5.2 | 5.2 | 0.42 | 1.11 | $R_1 = R_2$, $R_3$ var. |
| Q69pAcF/R119C | 50.7 | 48.4 | 55.5 | 7.0 | 7.0 | 0.30 | 1.10 | $R_1 = R_2$, $R_3$ var. |
| Q69pAcF/N132C | 41.0 | 38.3 | 41.5 | 5.0 | 5.0 | 0.24 | 1.12 | $R_1 = R_3$, $R_2$ var. |
| Q690pAcF/I150C | 38.0 | 37.3 | 38.7 | 4.5 | 4.5 | 0.36 | 1.09 | $R_1 = R_3$, $R_2$ var. |
| D70pAcF/R119C | 34.1 | 33.0 | 34.8 | 3.7 | 3.7 | 0.12 | 1.04 | $R_2 = R_3$, $R_1$ var. |
| D70pAcF/N132C | 47.1 | 45.7 | 49.6 | 5.9 | 5.9 | 0.13 | 1.07 | $R_3 > R_1 > R_2$ |

**Table S2G. Results of determined state-specific mean distances for the distribution fit for functional variants of the S44/I150C FRET pair.** Globally shared parameters are highlighted in gray cells (Eq. S24-S26).

| **Samples(showing aa #'s)** | **$\langle R_{DA1} \rangle$ [Å]** | **$x_1$†** | **± err. (%)** | **$\langle R_{DA2} \rangle$ [Å]** | **$x_2$†** | **± err. (%)** | **$\langle R_{DA3} \rangle$ [Å]** | **$x_3$†** | **± err. (%)** | **$x_{D(0)}$*** | **$\chi^2_r$** |
|---|---|---|---|---|---|---|---|---|---|---|---|
| S44pAcF/I150C(-) | 65.1 | 0.25 | 8.0 | 51.7 | 0.55 | 8.0 | 38.8 | 0.20 | 3.0 | 0.01 | 1.06 |
| S44pAcF/I150C(+)*** | 65.1 | 0.25 | 12.3 | 51.7 | 0.57 | 12.3 | 38.8 | 0.18 | 3.9 | - | 1.52** |
| T26E/S44pAcF/I150C(-)*** | 65.1 | 0.37 | 4.7 | 51.7 | 0.35 | 3.0 | 34.9 | 0.28 | 9.0 | 0.62 | 1.21 |
| T26E/S44pAcF/I150C(+)*** | 65.1 | 0.20 | 1.9 | 51.7 | 0.28 | 1.0 | 34.9 | 0.52 | 13.0 | 0.74 | 1.08 |
| E11A/S44C/I150C(-)*** | 65.1 | 0.75 | 7.8 | 51.7 | 0.12 | 7.8 | 38.8 | 0.13 | 11.8 | - | 1.98** |
| E11A/S44C/I150C(+)*** | 65.1 | 0.56 | 4.9 | 51.7 | 0.27 | 4.9 | 38.8 | 0.17 | 17.4 | - | 2.00** |
| R137E/S44pAcF/I150C | 59.3 | 0.52 | 7.3 | 49.3 | 0.37 | 2.6 | 36.2 | 0.11 | 9.1 | 0.24 | 1.07 |

† Values for the FRET populations are normalized such that $x_1+x_2+x_3=1$. *Donor decay was fixed and $x_{D(0)}$ represents the fraction of donor only from the total. ** Data from single molecule experiments shows higher $\chi^2_r$ when compared to eTCSPC, due to low photon statistics. ***Sub-ensemble fit from burst analysis. For E11A/S44C/I150C, it was not possible to measure in eTCSPC due to high donor-only (double Cys variant).

To reach to the conclusion that three continuous distance distributions are needed to describe all T4L variants, first we needed to characterize the donor and acceptor fluorescence quantum yield $\Phi_{FD(0)}$ and $\Phi_{FA}$, respectively. A summary table of these is shown in Table S2A. Table S2B summarizes the result of the two continuous distance distribution model with free amplitudes. The best fit with three continuous distance distribution is summarized in Table S2D-F; decays are shown in Figure S5A. The fit results for the functional variants are summarized in Table S2G.

Using only the two-state model and comparing the modeled distances using PDBID 172L and 148L for the two states showed that our data cannot be correlated with the structural information from the two crystallographic structures (Figure S5B, C).

**A**

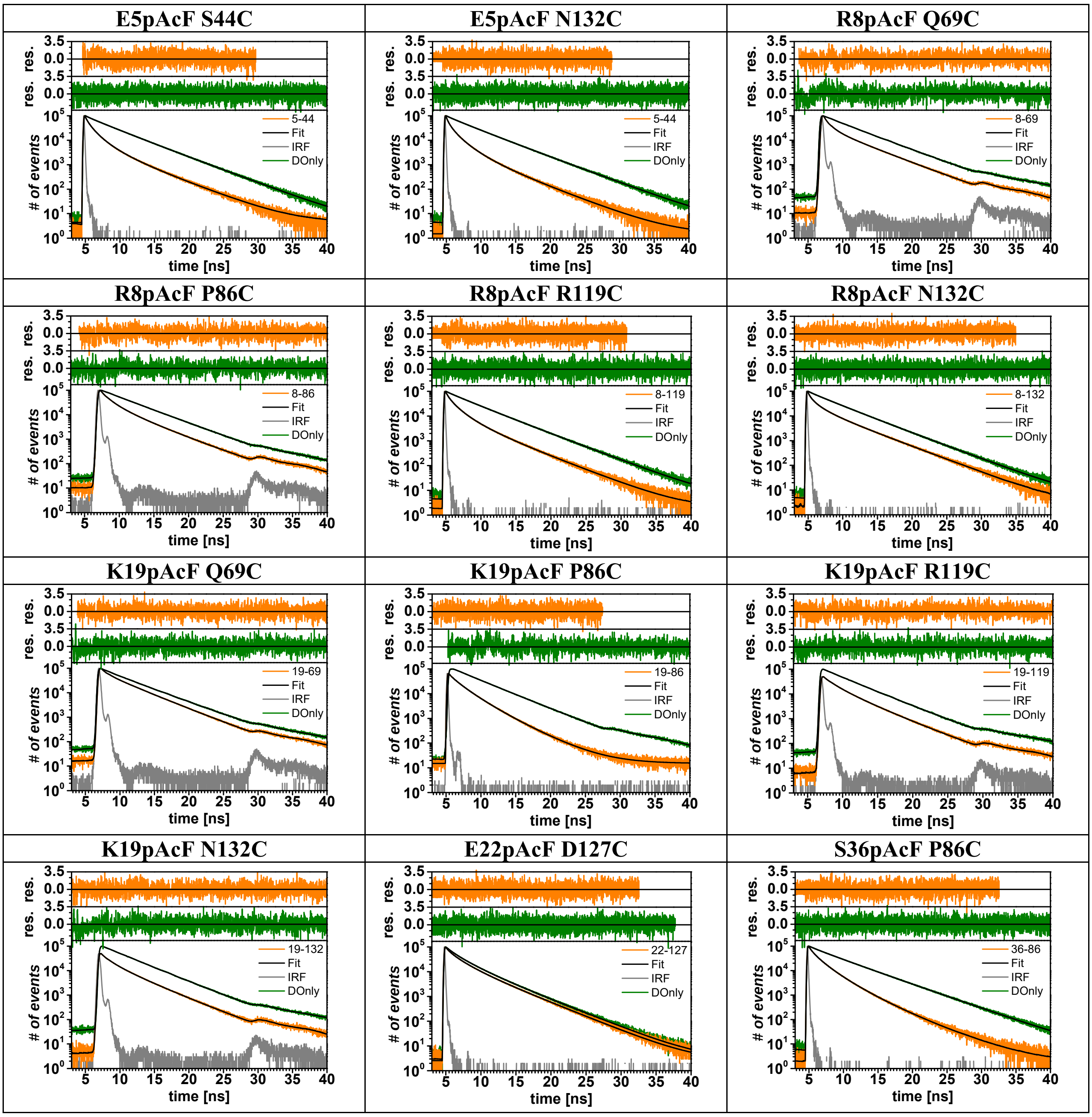

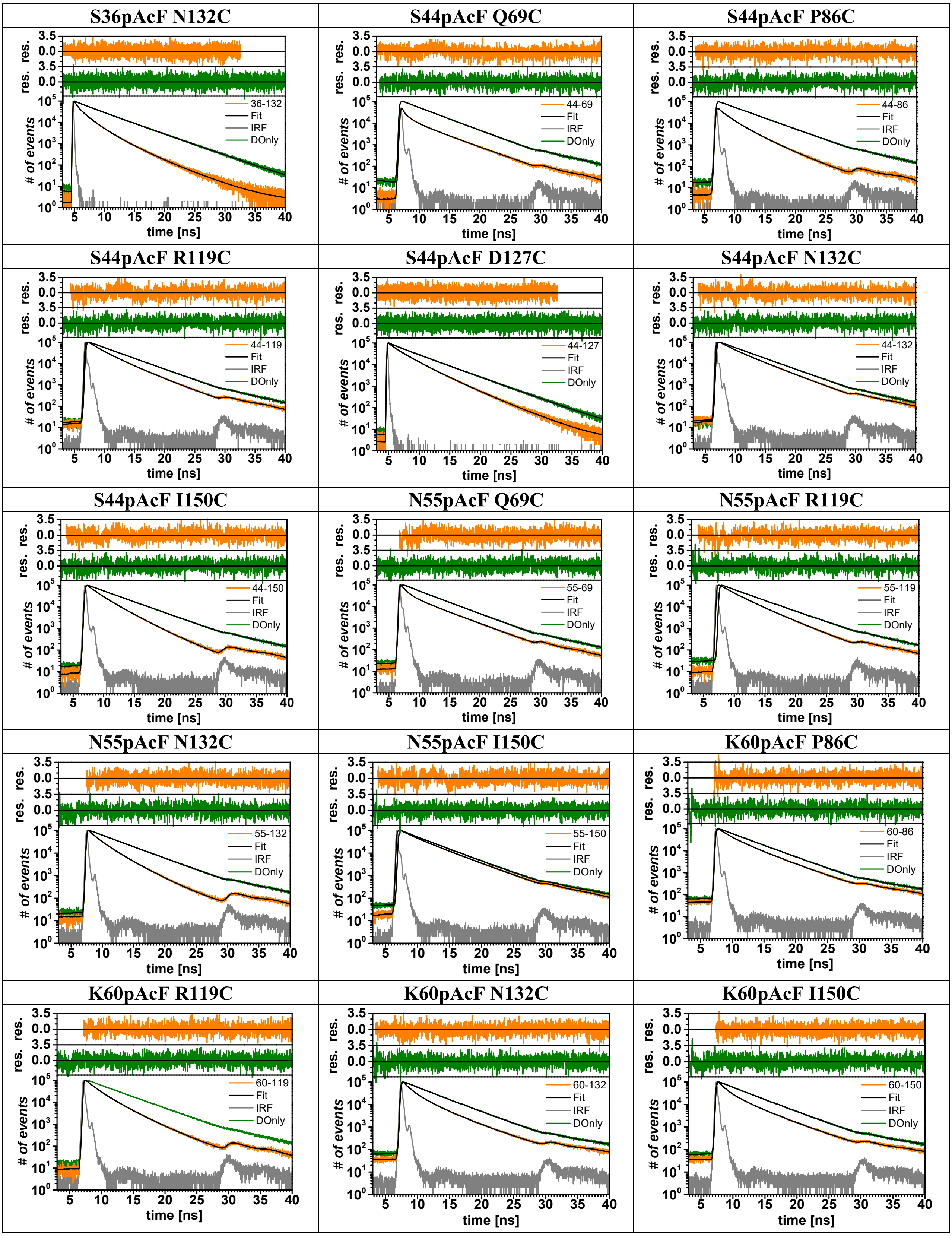

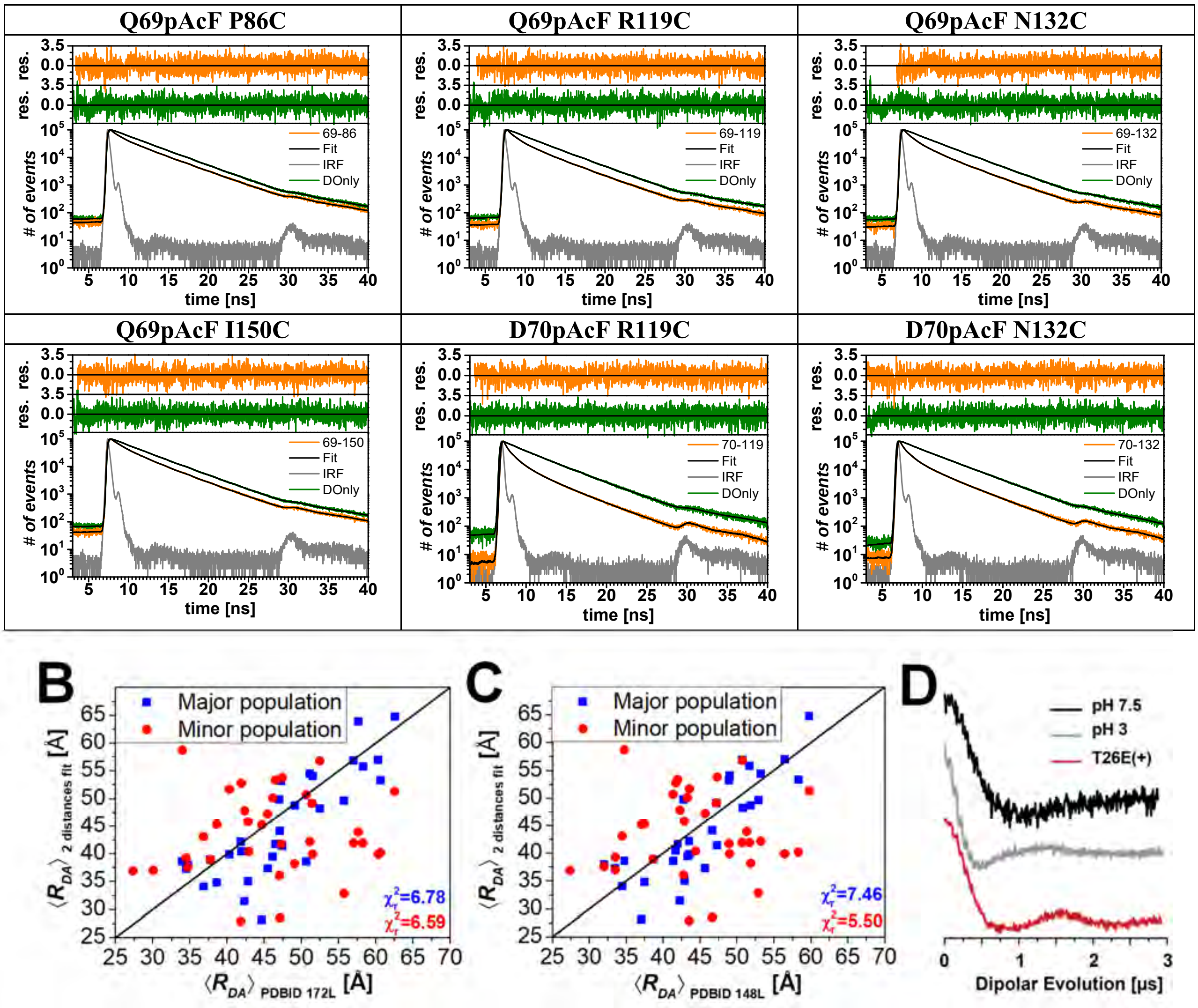

**Figure S5 eTCSPC results of T4L. (A)** Fit (black) of the experimental data of double-labeled sample (orange) and respective Donor-only labeled sample (green), weighted residuals are shown on top. Fit parameter are given in Table S2A (Donly) and S2D-F (double-labeled sample). Instrument response function (IRF) is shown in gray. **(B)** Fitted distances of two distributed states (Table S2B) fit plotted versus the distances calculated for the model X-ray structure of the open state (PDBID: 172L). "Major state" is the distance having the higher amplitude in fraction, while "minor state" is the distance with the lower fraction (**C**) Same as (B), only for the model X-ray structure of the closed state (PDBID 148L). **(D)** Experimental DEER time traces of the dipolar evolution, which were used to calculate the distance distributions shown in the main text Figure 6A.

## 2.4 Clustering results of T4l PDB structures

**Table S3**. All 578 structural models could be grouped in three clusters: *Open* (19 structures), *ajar* (26 structures and *closed* (535 structures).

| Cluster-name | PDB-ID |
|---|---|
| *Open* (19) | 172L, 151L, 168L, 169L, 173L, 174L, 178L, 1L97, 2HUK, 3EML, 3JR6, 3OE8, 3QAK, 3RZE, 3SB5, 4ARJ, 4IAP, 4K5Y, 4OO9. |
| *Ajar* (26) | 1JQU, 149L, 150L, 189L, 1P5C, 1P7S, 1QTH, 1SSY, 218L, 2HUM, 2QAR, 2QB0, 2RH1, 3PBL, 3SB9, 3SBA, 3SBB, 3SN6, 3UON, 3V2W, 3V2Y, 3VW7, 4DJH, 4EPI, 4EXM, 4GBR |
| *Closed* (535) | 148L, 102L, 103L, 104L, 107L, 108L, 109L, 110L, 111L, 112L, 113L, 114L, 115L, 118L, 119L, 120L, 122L, 123L, 125L, 126L, 127L, 128L, 129L, 130L, 131L, 137L, 138L, 139L, 140L, 141L, 142L, 143L, 144L, 145L, 146L, 147L, 152L, 155L, 156L, 157L, 158L, 159L, 160L, 161L, 162L, 163L, 164L, 165L, 166L, 167L, 170L, 171L, 175L, 176L, 177L, 180L, 181L, 182L, 183L, 184L, 185L, 186L, 187L, 188L, 190L, 191L, 192L, 195L, 196L, 197L, 198L, 199L, 1B6I, 1C60, 1C61, 1C62, 1C63, 1C64, 1C65, 1C66, 1C67, 1C68, 1C69, 1C6A, 1C6B, 1C6C, 1C6D, 1C6E, 1C6F, 1C6G, 1C6H, 1C6I, 1C6J, 1C6K, 1C6L, 1C6M, 1C6N, 1C6P, 1C6Q, 1C6T, 1CTW, 1CU0, 1CU2, 1CU3, 1CU5, 1CU6, 1CUP, 1CUQ, 1CV0, 1CV1, 1CV3, 1CV4, 1CV5, 1CV6, 1CVK, 1CX6, 1CX7, 1D2W, 1D2Y, 1D3F, 1D3J, 1D3M, 1D3N, 1D9W, 1DYA, 1DYB, 1DYC, 1DYD, 1DYE, 1DYF, 1DYG, 1EPY, 1G06, 1G07, 1G0G, 1G0J, 1G0K, 1G0L, 1G0M, 1G0P, 1G0Q, 1G1V, 1G1W, 1I6S, 1JTM, 1JTN, 1KNI, 1KS3, 1KW5, 1KW7, 1KY0, 1KY1, 1L00, 1L01, 1L02, 1L03, 1L04, 1L05, 1L06, 1L07, 1L08, 1L09, 1L0J, 1L0K, 1L10, 1L11, 1L12, 1L13, 1L14, 1L15, 1L16, 1L17, 1L18, 1L19, 1L20, 1L21, 1L22, 1L23, 1L24, 1L25, 1L26, 1L27, 1L28, 1L29, 1L30, 1L31, 1L32, 1L33, 1L34, 1L35, 1L36, 1L37, 1L38, 1L39, 1L40, 1L41, 1L42, 1L43, 1L44, 1L45, 1L46, 1L47, 1L48, 1L49, 1L50, 1L51, 1L52, 1L53, 1L54, 1L55, 1L56, 1L57, 1L58, 1L59, 1L60, 1L61, 1L62, 1L63, 1L64, 1L65, 1L66, 1L67, 1L68, 1L69, 1L70, 1L71, 1L72, 1L73, 1L74, 1L75, 1L76, 1L77, 1L79, 1L80, 1L81, 1L82, 1L83, 1L84, 1L85, 1L86, 1L87, 1L88, 1L89, 1L90, 1L91, 1L92, 1L93, 1L94, 1L95, 1L96, 1L98, 1L99, 1LGU, 1LGW, 1LGX, 1LI2, 1LI3, 1LI6, 1LLH, 1LPY, 1LW9, 1LWG, 1LWK, 1LYD, 1LYE, 1LYF, 1LYG, 1LYH, 1LYI, 1LYJ, 1NHB, 1OV5, 1OV7, 1OVH, 1OVJ, 1OVK, 1OWY, 1OWZ, 1OYU, 1P2L, 1P2R, 1P36, 1P37, 1P3N, 1P46, 1P56, 1P64, 1P6Y, 1PQD, 1PQI, 1PQJ, 1PQK, 1PQM, 1PQO, 1QS5, 1QS9, 1QSB, 1QSQ, 1QT3, 1QT4, 1QT5, 1QT6, 1QT7, 1QT8, 1QTB, 1QTC, 1QTD, 1QTV, 1QTZ, 1QUD, 1QUG, 1QUH, 1QUO, 1SSW, 1SWY, 1SWZ, 1SX2, 1SX7, 1T6H, 1T8A, 1T8F, 1T8G, 1T97, 1TLA, 1XEP, 1ZUR, 1ZWN, 1ZYT, 200L, 201L, 205L, 206L, 209L, 210L, 211L, 212L, 213L, 214L, 215L, 216L, 217L, 219L, 220L, 221L, 222L, 223L, 224L, 225L, 226L, 227L, 228L, 229L, 230L, 231L, 232L, 233L, 234L, 235L, 236L, 237L, 238L, 239L, 240L, 241L, 242L, 243L, 244L, 245L, 246L, 247L, 248L, 249L, 250L, 251L, 252L, 253L, 254L, 255L, 256L, 257L, 258L, 259L, 260L, 261L, 262L, 2A4T, 2B6T, 2B6W, 2B6X, 2B6Y, 2B6Z, 2B70, 2B72, 2B73, 2B74, 2B75, 2B7X, 2CUU, 2F2Q, 2F32, 2F47, 2HUL, 2IGC, 2L78, 2LC9, 2LCB, 2LZM, 2NTG, 2NTH, 2O4W, 2O79, 2O7A, 2OE4, 2OE7, 2OE9, 2OEA, 2OTY, 2OTZ, 2OU0, 2OU8, 2OU9, 2Q9D, 2Q9E, 2RAY, 2RAZ, 2RB0, 2RB1, 2RB2, 2RBN, 2RBO, 2RBP, 2RBQ, 2RBR, 2RBS, 3C7W, 3C7Y, 3C7Z, 3C80, 3C81, 3C82, 3C83, 3C8Q, 3C8R, 3C8S, 3CDO, 3CDQ, 3CDR, 3CDT, 3CDV, 3DKE, 3DMV, 3DMX, 3DMZ, 3DN0, 3DN1, 3DN2, 3DN3, 3DN4, 3DN6, 3DN8, 3DNA, 3F8V, 3F9L, 3FA0, 3FAD, 3FI5, 3G3V, 3G3W, 3G3X, 3GUI, 3GUJ, 3GUK, 3GUL, 3GUM, 3GUN, 3GUO, 3GUP, 3HH3, 3HH4, 3HH5, 3HH6, 3HT6, 3HT7, 3HT8, 3HT9, 3HTB, 3HTD, 3HTF, 3HTG, 3HU8, 3HU9, 3HUA, 3HUK, 3HUQ, 3HWL, 3K2R, 3L2X, 3L64, 3LZM, 3NY8, 3NY9, 3NYA, 3RUN, 3SB6, 3SB7, 3SB8, 4DAJ, 4DKL, 4E97, 4EJ4, 4EKP, 4EKQ, 4EKR, 4EKS, 4GRV, 4I7J, 4I7K, 4I7L, 4I7M, 4I7N, 4I7O, 4I7P, 4I7Q, 4I7R, 4I7S, 4I7T, 4LDE, 4LDL, 4LDO, 4LZM, 4PHU, 4TN3, 5LZM, 6LZM, 7LZM |

## 2.5Characterization of functional T4L variants

### 2.5.1 Catalytic activity of S44pAcF I150C T26E

The ability to process the selected substrate (peptidoglycan from *Micrococcus luteus*) of the mutants was monitored by reverse phase chromatography. Prior to use, the purchased peptidoglycan (Sigma-Aldrich, Switzerland) was purified as described by Maeda in 1980 [22] to remove minor fluorescent impurities. Double-labeled mutants (1 µM) were incubated with 3 mg/mL of substrate and allowed to react for several hours in 50 mM sodium phosphate buffer, 150 mM NaCl at pH 7.5. Samples at different times were monitored under a reverse phase HPLC at 495 nm. In this way we can identify the labeled lysozyme. Typical examples for the processing of substrate are shown in Figure S6. Figure S6A shows the elution profile of the peptidoglycan monitored at 215 nm. Multiple peaks from 10 to 14 min appear. In the same panel the elution of the T26E/S44pAcF/I150C-(AD) monitored at 215 nm is shown as incubated with the peptidoglycan. For better contrast of the shift in populations the absorbance was measured at the maximum for the AlexaFluor488 (495 nm). This is shown in Figure S6B. After 260 minutes this mutant is fully saturated with the substrate.

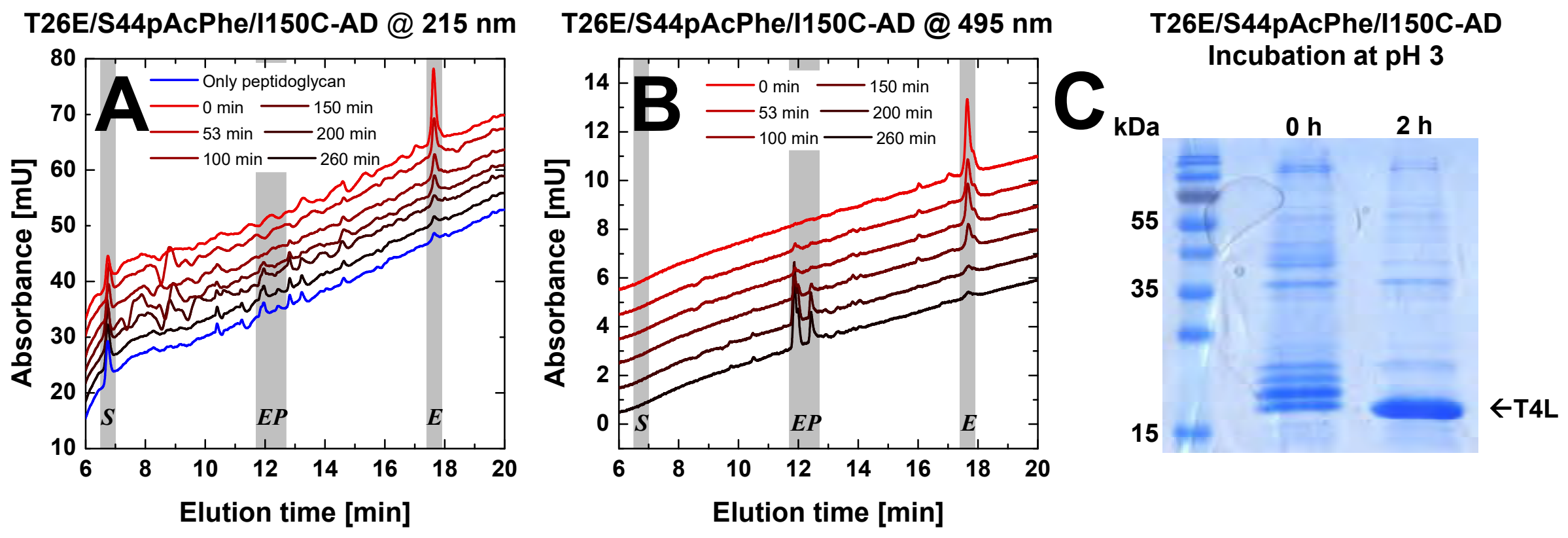

**Figure S6 T4L binding to peptidoglycan as observed by reverse phase HPLC and cleavage at low pH. (A)** The elution profile measured at 215 nm of reverse phase chromatography for T26E/S44pAcF/I150C-(AD). Samples were taken at different times during the incubation with peptidoglycan. First line shows only the elution of peptidoglycan. Note different peaks of the heterogeneity on the peptidoglycan. Offset between lines was added for clarity. Gray lines represent the free enzyme population (*E*), the product bound enzyme (*EP*) and the substrate alone (*S*). **(B)** Elution of the same sample as in (A) but monitored at 495 nm, which corresponds to the absorbance of Alexa488. Saturation of T26E/S44pAcF/I150C-(AD) with substrate is reached at ~ 4 hours of incubation. **(C)** Purification of T26E/S44pAcF/I150C from the *E. coli* cell pellet yielded a mixture of free and to cell wall pieces of different sizes bound protein. After incubation for 2 hrs at pH 3, nearly all bound peptidoglycan had been cleaved and the free enzyme could now be used for labeling and further experiments after adjusting the conditions to neutral pH again.

### 2.5.2. Single-molecule experiments in the presence of substrate

For the variant E11A/S44C/I150C we carried out two-color excitation experiments, in which we alternately excited the donor and the acceptor fluorophore (PIE) [20]. Thus, we could sort out the molecules carrying only one type of fluorophore, a disadvantage of the unspecific labeling. However, this allowed us also directly identifying the bursts stemming from the peptidoglycan. Figure S7A-B show the green to red fluorescence signal vs. the stoichiometry *S* of E11A/S44C/I150C in the absence and presence of substrate. DOnly labeled molecules are located at $S = 0$, AOnly molecules at $S = 1$ and DA labeled molecules are centered at $S = 0.5$

The peptidoglycan appears as an additional population at $S = 0.8$. Figure S7C shows that the brightness and burst duration distribution of this variant are nearly identical in the absence and presence of substrate.

For the variant T26E/S44pAcF/I150C only single-color excitation experiments were performed. The brightness and burst duration distributions of the FRET and the subset of high-FRET bursts are shown in Figure S7E-H. To avoid the contamination with bursts from aggregates, we selectively only considered bursts shorter than 5 ms in the further analysis.

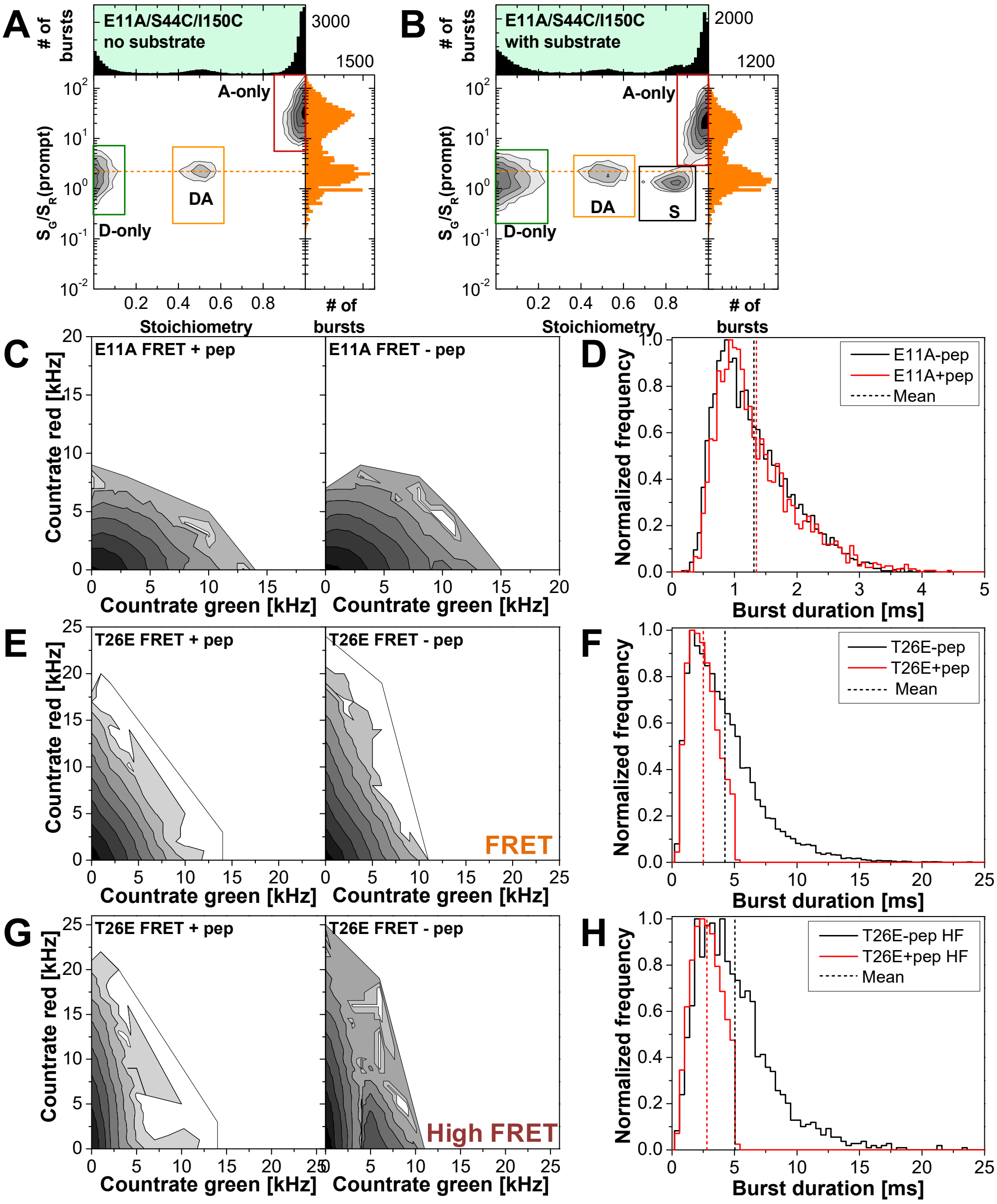

**Figure S7 Single-molecule experiments of functional variants. (A,B)** PIE experiments identify burst stemming from single and double labeled E11A/S44C/I150C and substrate alone. **(C)** Brightness distribution of bursts from double-labeled E11A/S44C/I150C. **(D)** Burst duration distribution of bursts from double-labeled E11A/S44C/I150C. **(E-H)** Same as (C,D) for the variant T26E/S44pAcF/I150C-(DA).

### 2.6 Analyzing the kinetic network of conformational states in T4L

#### 2.6.1 Detection of distinct $C_3$ species

The eTCSPC fluorescence decay of S44pAcF/I150C-(DA) was fit with a model containing three different conformational states ($C_1$, $C_2$ and $C_3$). The total population of $C_3$ corresponds to 20 % (Table S2G). From the single molecule MFD histograms it was clear to observe burst accumulation at the location where $C_3$ lies. To quantify the amount of bursts corresponding to this, we computed the area under the curve for the region of $10^{-2} < F_D/F_A < 0.3$ (Figure S8E) corresponding to 564 burst of the total 10139 burst of all single molecule events (subtracting 893 bursts from molecules missing an active acceptor). Thus, the burst accumulation of this state is 5.3% of the total number of bursts. We called this population $C_{3t}$ because it is a static accumulation of the population $C_3$ observed by eTCSPC. In order to account for the missing 15 % of $C_3$, there has to be an additional population, which exchanges with $C_1$ and $C_2$ at timescales faster than the burst duration. We called this population $C_{3d}$. Therefore; the total contribution of equals to the sum of the static plus the dynamic subpopulations of $C_3$ ($C_3$ = 20 % = $C_{3t} + C_{3d}$ = 5 % + 15 %). Because in fFCS we only observe two relaxation times from µs to ms, we ignore for the time being the existence of the 5 % of $C_{3t}$, as it is not needed to discuss the connectivity between $C_1$, $C_2$ and $C_3$ at faster timescales.

#### 2.6.2 Consolidated model of T4L

To construct the best kinetic model that describes the free enzyme in solution let us consider the experimental facts: *i*) eTCSPC resolves three different FRET states. *ii*) fFCS shows two transition times faster than 10 ms. *iii*) smFRET diagrams are better described by a unimodal distribution mixed with a small population (~ 5%) with very high FRET only for the S44pAcF/I150C-(DA) variant.

Unimodal distributions in single-molecule experiments can occur due to time-averaging. Ignoring the donor only population, the free enzyme (S44pAcF/I150C-(DA)) samples four conformational states ($C_1$, $C_2$, $C_{3d}$ and $C_{3t}$), where the $C_{3t}$ is a static population at very high FRET, and $C_1$, $C_2$, $C_{3d}$ mix at the observed times of ~4 µs and ~230 µs.

Putting aside the state $C_{3t}$, the simplest model of conformational transitions that one can build from experimental observables is

$$[C_1] \underset{k_{21}}{\overset{k_{12}}{\rightleftharpoons}} [C_2] \underset{k_{32}}{\overset{k_{23}}{\rightleftharpoons}} [C_3] \quad , \qquad \text{(S29)}$$

where $C_1$ corresponds to the most open conformer, $C_2$ is similar to the substrate-enzyme complex and $C_3$ has an interdye distance much shorter than $C_2$. With this in mind we disregarded the cyclic model

$$\begin{array}{ccc} & [C_3] & \\ \rightleftharpoons & & \rightleftharpoons \\ [C_1] & \rightleftharpoons & [C_2] \end{array} \quad . \qquad \text{(S30)}$$

due to the sequential closing of the enzyme. This limits the return process of $[C_1] \rightleftharpoons [C_3]$.

Our goal is to extract the reaction rates ($k_{12}$, $k_{21}$, $k_{23}$, $k_{32}$) from our experimental observables. To solve this, first we need to write the rate matrix $K$ for the system described in Eq. (S31).

$$K=\begin{pmatrix}-k_{12} & k_{21} & 0\\ k_{12} & -(k_{21}+k_{23}) & k_{32}\\ 0 & k_{23} & -k_{32}\end{pmatrix}. \quad \text{(S31)}$$

The two eigenvalues of $K$ correspond to the two observables measured by fFCS.

$$\frac{1}{t_{R1,2}}=\frac{1}{2}\cdot(k_{12}+k_{21}+k_{23}+k_{32})\pm\sqrt{(k_{12}+k_{21}+k_{23}+k_{32})^2-4\cdot(k_{12}k_{23}+k_{12}k_{32}+k_{21}k_{32})}. \text{(S32)}$$

The time evolution of the system on Eq. (S31) is defined by

$$\frac{d}{dt}[C_i](t)=K_{ij}[C_i](t), \quad \text{(S33)}$$

Equation (S33) has an analytical solution on the form of

$$[C_i](t)=\mathbf{C}_0\exp(K_{ij}t), \quad \text{(S34)}$$

where $C_0$ is the $i$-th eigenvector. At equilibrium, or $t\rightarrow\infty$, the equilibrium fractions for each conformer can be obtained analytically and are given by

$$[C_i]=\begin{pmatrix}\dfrac{k_{21}\cdot k_{32}}{k_{21}\cdot k_{32}+k_{12}\cdot(k_{23}+k_{32})}\\ \dfrac{k_{12}\cdot k_{32}}{k_{21}\cdot k_{32}+k_{12}\cdot(k_{23}+k_{32})}\\ \dfrac{k_{12}\cdot k_{23}}{k_{21}\cdot k_{32}+k_{12}\cdot(k_{23}+k_{32})}\end{pmatrix}. \quad \text{(S35)}$$

Note that $[C_3]=1-([C_1]+[C_2])$. These fractions are obtained by fluorescence decay analysis as done in S1.5-1.7. The reaction rates ($k_{12}$, $k_{21}$, $k_{23}$, $k_{32}$) can be expressed in terms of the equilibrium fractions ($x_1=[C_1]$, $x_2=[C_2]$, $x_3=[C_3]$) and the relaxation times ($t_{R1}$ and $t_{R2}$). The analytical solution of this system has two solutions:

$$k_{12}^{(\pm)}=\frac{[C_2]\cdot\left(\frac{1}{t_{R1}}+\frac{1}{t_{R2}}\right)\pm\left([C_2]\cdot\left(\frac{4\cdot([C_1]-1)\cdot([C_1]+[C_2])}{t_{R1}\cdot t_{R2}}+[C_2]\cdot\left(\frac{1}{t_{R1}}+\frac{1}{t_{R2}}\right)^2\right)\right)^{½}}{2\cdot([C_1]+[C_2])},$$

$$k_{21}^{(\pm)}=\frac{[C_1]\cdot\left([C_2]\cdot\left(\frac{1}{t_{R1}}+\frac{1}{t_{R2}}\right)\pm\left([C_2]\cdot\left(\frac{4\cdot([C_1]-1)\cdot([C_1]+[C_2])}{t_{R1}\cdot t_{R2}}+[C_2]\cdot\left(\frac{1}{t_{R1}}+\frac{1}{t_{R2}}\right)^2\right)\right)^{½}\right)}{2\cdot[C_2]\cdot([C_1]+[C_2])},$$

$$k_{23}^{(\pm)}=\frac{([C_1]+[C_2]-1)\cdot\left([C_2]\cdot\left(\frac{1}{t_{R1}}+\frac{1}{t_{R2}}\right)\mp\left([C_2]\cdot\left(\frac{4\cdot([C_1]-1)\cdot([C_1]+[C_2])}{t_{R1}\cdot t_{R2}}+[C_2]\cdot\left(\frac{1}{t_{R1}}+\frac{1}{t_{R2}}\right)^2\right)\right)^{1/2}\right)}{2\cdot[C_2]\cdot([C_1]-1)},$$

$$k_{32}^{(\pm)}=\frac{-[C_2]\cdot\left(\frac{1}{t_{R1}}+\frac{1}{t_{R2}}\right)\pm\left([C_2]\cdot\left(\frac{4\cdot([C_1]-1)\cdot([C_1]+[C_2])}{t_{R1}\cdot t_{R2}}+[C_2]\cdot\left(\frac{1}{t_{R1}}+\frac{1}{t_{R2}}\right)^2\right)\right)^{1/2}}{2\cdot([C_1]-1)}. \quad \text{(S36)}$$

To complete the model we need to add the static fraction of ~ 5%. We assigned this static fraction to conformer $C_{3t}$, which is identical in FRET to the state $C_{3d}$. We split the fraction of $C_3$ into these two populations. The final reaction model can be expressed as

$$[C_1] \underset{k_{21}}{\overset{k_{12}}{\rightleftharpoons}} [C_2] \underset{k_{32}}{\overset{k_{23}}{\rightleftharpoons}} [C_{3d}] \underset{k_{43}}{\overset{k_{34}}{\rightleftharpoons}} [C_{3t}] \quad . \qquad \text{(S37)}$$

Where $k_{34}$ = 0.003 $ms^{-1}$ and $k_{43}$ = 0.008 $ms^{-1}$ were empirically determined, but satisfy the condition that they have to be smaller than 0.01 $ms^{-1}$.

With all the determined rates, we did Brownian dynamics simulations as described in main text. The single-molecule MFD histograms for the simulated data shown in Figure S8 and corresponds to the experimental data shown in Figure 3.

### 2.6.3 Simulation of the FRET data in complex kinetic schemes

To describe the experimental 2D histogram a four-state scheme was used. First, we calculated FRET histograms [23] where a kinetic model with discrete conformations was assumed. The transition between the states is described by rate equations. The probability for the system to be in state $i$ at time $t$, $\boldsymbol{p}_i(t)$, satisfies a set of rate equations, which can be written in matrix notation as:

$$\frac{d\boldsymbol{p}}{dt} = \boldsymbol{K} \cdot \boldsymbol{p} \qquad \text{(S38)}$$

where $\boldsymbol{p}$ is a column vector with the components $\boldsymbol{p}_i(t)$ and $\boldsymbol{K}$ is a transition rate matrix representing the rate constants for the transitions between states $i$ and $j$. At long times, $\boldsymbol{p}(t)$ approaches its equilibrium value, $\boldsymbol{p}_{eq}$. The vector of the equilibrium populations $\boldsymbol{p}_{eq}$ is normalized to 1 and satisfies $\boldsymbol{K} \cdot \boldsymbol{p}_{eq} = 0$. For each burst, the mean averaged efficiency $\langle E \rangle$ and the average fluorescence weighted lifetime $\langle \tau_{D(A)} \rangle_f$ can be calculated by:

$$\langle E \rangle = \frac{\sum t_i \cdot E_i}{t_{burst}} \qquad \text{(S39)}$$

and

$$\langle \tau_{D(A)} \rangle_f = \frac{\sum t_i(\boldsymbol{K}) \cdot \tau_i^2}{\sum t_i(\boldsymbol{K}) \cdot \tau_i} \qquad \text{(S40)}$$

where $t_i(\boldsymbol{K})$ is time spent by a molecule in state $i$ within the duration of the burst and depends on the transition rate matrix $\boldsymbol{K}$; $E_i$ is the FRET efficiency of the $i$-th state; $t_{burst}$ is the duration of the burst and $\tau_i$ is the fluorescence lifetime of the $i$-th state. Practically, each burst has certain duration and number of photons, which were chosen arbitrary from experimentally measured $t_{burst}$ (duration time) vs. $N$ (number of photons) 2D histogram. The residence times by each molecule in different states were calculated using Gillespie algorithm for continuous-time Markov Chain. Then, the average fluorescence lifetime $\langle \tau_{D(A)} \rangle_f$ for each burst was calculated by Monte-Carlo simulation of fluorescence emission given FRET efficiencies of each state. Stationary (equilibrium) populations of states were obtained by solving interstate transition dynamics matrix and the residence times obtained on previous step. The descriptions for the vector $\boldsymbol{p}$ and the rate matrix $\boldsymbol{K}$ (resulting into the equilibrium fractions for the state $i$, $\boldsymbol{p}_{eq,i}$) and the experimental observables, $E$ and $\tau$, used in the simulations are shown in the Table S4A. For plotting, $E$ was converted to $F_D/F_A$ ratio. The simulation procedure was repeated for a high number of bursts to generate $F_D/F_A$ vs $\langle\tau_{D(A)}\rangle_f$ 2D histogram (Figure S8B-C). The resulting 1D and 2D histograms were compared to the experimental data, yielding a $\chi^2$ parameter for each simulation and histogram. To test the significance of the difference in $\chi^2$, we performed F-test as described above. The resulting values are combined in the Table S4B.

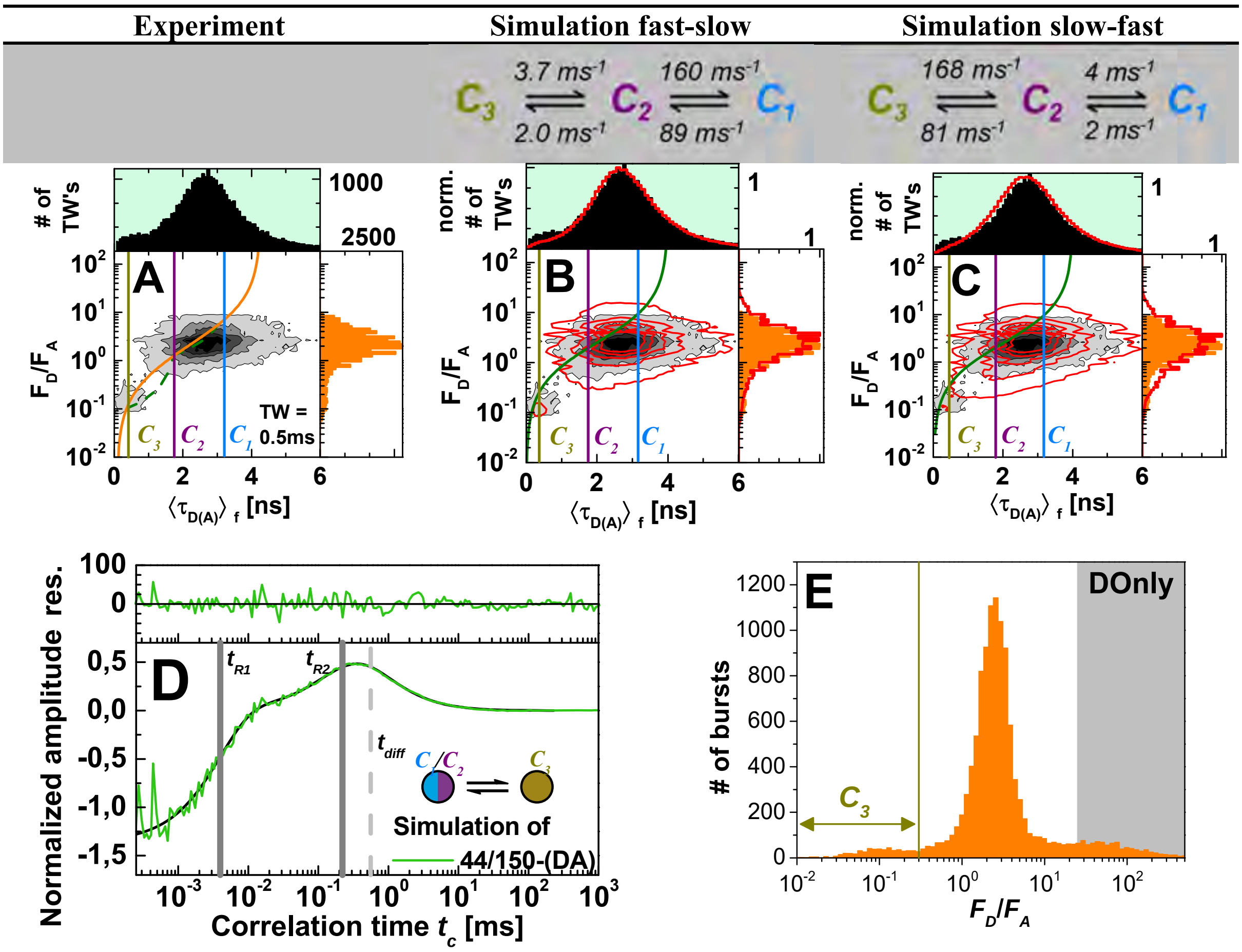

**Figure S8 Brownian Dynamic Simulations of S44pAcF/I150C-(DA).** (**A**) Two dimensional histograms from smFRET analysis ($F_D/F_A$ vs. fluorescence lifetime ($\langle\tau_{D(A)}\rangle_f$) of the raw data from the S44pAcF/I150-(DA) variant at pH 7.5 selected with 0.5 ms time-windows (TW's). FRET lines, static and dynamic are shown as orange solid and green dashed lines. $\langle B_G\rangle$ = 1.6 kHz, $\langle B_R\rangle$ = 0.8 kHz, spectral crosstalk $\alpha$ = 1.2% and ratio of green and red detection efficiencies $g_G/g_R$ = 0.77 are used for corrections. (**B, C**) Brownian dynamics simulation using the rates from Figure 3B, C was processed as the experimental data. Simulated parameters ($\langle B_G\rangle$, $\langle B_R\rangle$, $\alpha$, $g_G/g_R$) were the same as in the experiment. In addition we considered a rotational correlation of $\rho$ = 2.2 ns for conformational state. Analysis results of simulated data are presented in the same fashion as in panel (**A**). (**D**) *sCCF* between the pseudo-states consisting of the $C_1/C_2$ mix and the $C_3$ for simulated data. Fit of this *sCCF* curve returns two relaxation times of 4 µs and 220 µs, consistent with our input parameters. **(E)** One dimensional histogram of the raw data from the S44pAcF/I150-(DA) variant at pH 7.5 analyzed in burstwise mode to illustrate the region of $C_{3t}$, Donly and the dynamically mixed state.

To estimate our errors on determining the rates we considered the 2σ confidence interval in determining the population fractions (Table S2G) and the 2σ confidence interval in determining the relaxation times by fFCS (Table S4C). Taking those extremes we estimated the error and computed the reaction rate constants for Figure 7 in the main text according to Eq. (S36).

**Table S4A**. States description for the vector $\boldsymbol{p}$, equilibrium faction vector $\boldsymbol{p_{eq}}$ and the rate matrix $\boldsymbol{K}$.

| Kinetic state i | State name | Efficiency $E$ | Fluorescence lifetime $\tau$, ns [a] | Equilibrium fraction $p_{eq,i}$ |
|---|---|---|---|---|
| 1 | $C_1$ | 0.2 | 3.2 | 0.51 |
| 2 | $C_2$ | 0.5 | 2.0 | 0.29 |
| 3 | $C_3$ | 0.9 | 0.5 | 0.14 |
| 4 | $C_{3d}$ | 0.9 | 0.5 | 0.06 |

$$\boldsymbol{K}_{fastslow} = \begin{pmatrix} -158.1 & 90.8 & 0 & 0 \\ 158.1 & -94.5 & 1.8 & 0 \\ 0 & 3.7 & -1.808 & 0.003 \\ 0 & 0 & 0.008 & -0.003 \end{pmatrix}$$

$$\boldsymbol{K}_{slowfast} = \begin{pmatrix} -3.5 & 2 & 0 & 0 \\ 3.5 & -169.7 & 81.1 & 0 \\ 0 & 167.7 & -81.108 & 0.003 \\ 0 & 0 & 0.008 & -0.003 \end{pmatrix}$$

**Table S4B.** Parameters used in the evaluation of the statistical significance of different simulations.

| Type of analysis | $\chi^2$ fast-slow | $\chi^2$ slow-fast | Degrees of freedom | *F-value* | $p(C_1\text{-}C_2$ fast vs. $C_1\text{-}C_2$ slow) |
|---|---|---|---|---|---|
| 1D $\langle\tau_{D(A)}\rangle_f$ histogram | 68.6 | 187.2 | 10 | 2.72 | 1 |
| 1D $E$ histogram | 85.1 | 95.5 | 10 | 1.12 | 0.734 |
| 2D $\langle\tau_{D(A)}\rangle_f$ vs $E$ histogram | 0.4551 | 0.6069 | 10 | 1.33 | 1 |

**Table S4C.** Calculated reaction rates for several variants using Eq. (S34). Confidence intervals (2σ) are shown in squared brackets and the corresponding renormalized fractions shown below $x_1$+$x_2$+$x_{3d}$ =1*.

| **Samples** | $\boldsymbol{k_{12}}$ **[ms$^{-1}$]** | $\boldsymbol{k_{21}}$ **[ms$^{-1}$]** | $\boldsymbol{k_{23}}$ **[ms$^{-1}$]** | $\boldsymbol{k_{32}}$ **[ms$^{-1}$]** |
|---|---|---|---|---|
| 44/150-(DA) | 89 [45.5-175.6] | 160 [74.1-312.7] | 2.0 [1.5-2.3] | 3.7 [3.1-4.3] |
| 11/44/150-(DA)+pept | 217 [134.8-535.4] | 33 [19.0-86.38] | 0.8 [0.6-1.1] | 3.7 [3.1-4.3] |
| 26/44/150-(DA) + pept | 79 [49.2-193.3] | 21 [13.0-52.6] | 0.5 [0.4-0.6] | 0.9 [0.8-1.0] |
| The relaxations times used were: $t_{R1}$ = 4 ± 2.3 µs; $t_{R2}$ = 230 ± 28.4µs for 44/150-(DA) and 11/44/150-(DA)+pept. $t_{R1}$ = 10 µs; $t_{R2}$ = 790 µs (Fig. S4F) was used for 26/44/150-(DA) + pept. | | | | |

| **Chemical State** | **Samples** | $\boldsymbol{x_1}$ | $\boldsymbol{x_2}$ | $\boldsymbol{x_{3d}}$ |
|---|---|---|---|---|
| *E* | 44/150-(DA) | 0.54 | 0.30 | 0.16 |
| *ES* | 11/44/150-(DA)+pept | 0.30 | 0.54 | 0.16 |
| *EP* | 26/44/150-(DA) +pept | 0.35 | 0.29 | 0.36 |
| * Rounded to 2 digits. Renormalized fractions based on the relative changes observed in all states in the presence of substrate (Fig. 5E). Only the amino acid number of the mutagenesis is shown. | | | | |

## 2.7 Energy landscape of the catalytic cleavage cycle

The enzymatic pathway of the extended Michaelis-Menten mechanism (eq. 1 in main text) consists of three distinct reaction steps (Figure S9). *I*) The substrate $S$ binds reversibly to the enzyme $E$ to form an enzyme-substrate complex $ES$. *II*) In the $ES$ complex, $S$ is converted to the product $P$, resulting in the $EP$ complex with the product still bound to the enzyme. *III*) $P$ is released from the complex via a transition of $E$ to an excited state $E^*$. Finally, the free enzyme $E^*$ relaxes to $E$. Our observations demonstrate that a fine-tuned shift of the conformational equilibrium favors motions of active product release in T4L where the energy of product formation in step II defines the directionality of the reaction [24]. This hydrolysis reaction is irreversible and thus can be denoted as “ratchet mechanism” [25].

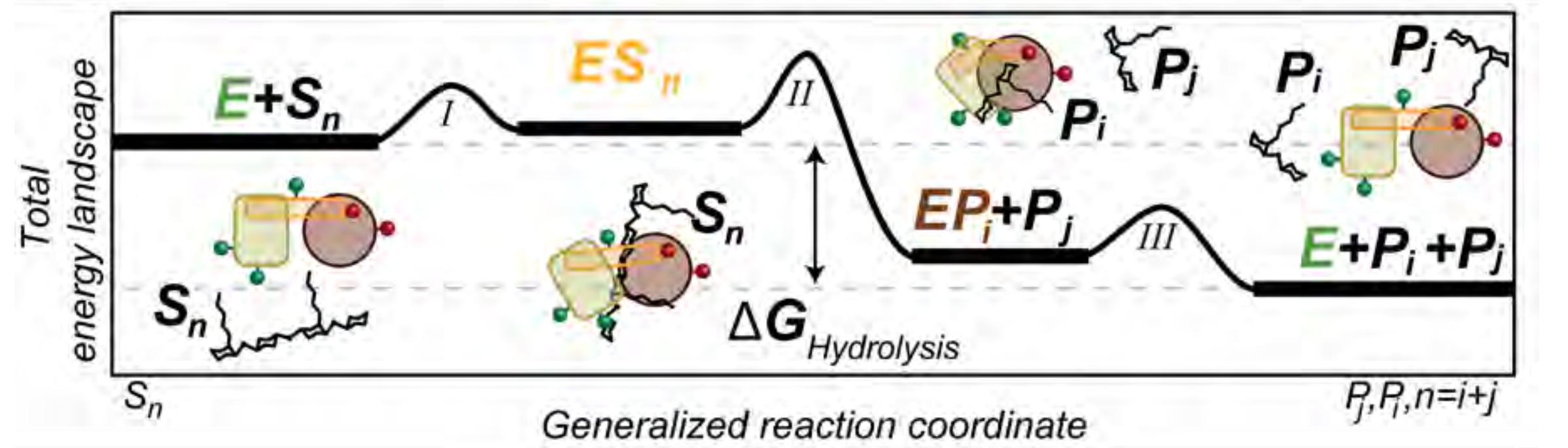

**Figure S9 Total energy landscape of the hydrolysis of T4L on a generalized reaction coordinate.** T4L cleaves the polymer chain peptidoglycan, the substrate $S$, of length $n$ ($S_n$) between the alternating residues of β-(1,4) linked N-acetylglucosamine and N-acetylmuramic acid into two shorter peptidoglycan chains, the product $P$ of chain length $i$ and $j$ ($P_i$ and $P_j$). Here, $n = i + j$.

## 2.8 Challenges of smFRET measurements and their solutions

Most of the potential problems with smFRET come from the complexities associated with the labels. We list the solution for potential label artifacts, and how our approaches and considerations allow us to draw conclusions, artifact free, of our data.

1) Labeling influence on enzymatic work:
   HPLC on the T26E/S44pAcF/150C-(DA) and –(AD) mutants show that they can process the peptidoglycan to keep the substrate bound. Non-functional mutants stayed non-functional after labeling (E11A/S44C/I150C-(DA) and R137E/S44pAcF/I150C-(DA)).
2) Local quenching of Donor:
   eTCSPC: In ensemble measurements, local quenching is observed by changes in the average lifetime of the donor. The multi-exponential fluorescence decays of the donor only labeled variants reflect the presence of quenched states. At these states, the fluorophore senses a different environment. Most likely these differences represent various conformations of the protein.
   smFRET: Donor quenching, as in the case of eTCSPC, shifts the average donor lifetime towards shorter lifetimes. FRET lines are corrected for the multi-exponential properties of the donor decay.
   fFCS: We use the multi-exponential time-resolved fluorescence decay information to generate different filters to calculate the species cross-correlation. Although, protein dynamics can be extracted from single label variants, the structural information is lost. This is only possible from the FRET labeled samples.
3) Triplet-state of Donor:
   eTCSPC: Triplet state are long lived compared to the fluorescence lifetime. Therefore, on ensemble time-resolved fluorescence decays this effect is not visible.
   smFRET: Triplet or dark states kinetics are short-lived compared to the burst duration.
   fFCS: In a classical FCS experiment triplet or dark states appear as a "bunching" term in the correlation function. In fFCS we do not correlate fluctuations on signal, but rather we correlate fluctuations of species. In our case, they correspond to different conformations of T4L. We assume that triplet/dark states are not coupled to the conformations or the selected pseudo species. In other words, the photo-physics of the dye is independent of the conformation in which the molecule is. With this in mind, the *sCCF* will have positive and negative contributions from each species resulting in the fact that the "bunching" term is not present. We know that increasing the power can increase the triplet amplitude. To test this, we measured the sCCF of T4L-(DA) at different powers at objective and we did not observe any major differences in the relaxation times $t_{R1}$ and $t_{R2}$ (Figure S10) or shape of the *sCCF*. We also tested the addition of triplet quenchers Cycloocta-1,3,5,7-tetraenecarboxylic acid (COTc) but did not observe major deviations (Figure S10A-D).
4) Acceptor cis-trans isomerization:
   eTCSPC: If FRET to *cis* and *trans* is different the donor decay would reflect the *cis-trans* population. We assume that this effect is small therefore not visible.
   smFRET: This effect can be observed as acceptor quenching. The reason is that the *cis* state is dark. Spending more time in the *cis* state will reduce the overall counts observed from the acceptor. This effect can be seen in the two dimensional histograms as a vertical shift of the islands position on $F_D/F_A$ vs. lifetime $\langle \tau_{D(A)} \rangle_f$ representation.

fFCS: For fFCS we correlate only photons emitted by the donor fluorophore. Changes in the brightness of the acceptor are not correlated. However, something that can happen is that the absorption of the energy transferred from the donor can be different for *cis* and *trans* states. This is something that was not tested. But as in the case of the donor triplet we assume that, even in the case in which this occurs, the photophysics dynamics of the acceptor dye is decoupled from the conformational dynamics of the molecule.

5) Dye mobility:

eTCSPC: Dye mobility occurs at slower timescales than the time-resolved fluorescence decay of the fluorophore. For this reason, it is better to consider FRET due to all configurations of fluorophore positions during time-resolved fluorescence decays. We take this into consideration by having a distribution of distances instead of single lifetimes to identify each conformational state. These are included in the treatment of the FRET lines. In order to do so, ensemble time resolved anisotropy decays were measured. We assumed that fluorophore mobility follows the "wobble in a cone" model [26]. Table S5A-C summarizes the residual anisotropies ($r_\infty$) of D - donor, A - acceptor and A(D) - the sensitized by FRET emission of acceptor that were used to calculate dye order parameters and $\kappa^2$ distributions (Table S5D) according to refs. [2] [26] (Eq. 9 and 10 in Sindbert *et al.*). The assumption is that fluorophores move according to the "wobble in a cone" model. According to all distributions the assumption of $\kappa^2 = 2/3$ is very well justified.

smFRET: In smFRET one can inspect the anisotropy $r_{sc}$ vs. lifetime $\langle \tau_{D(A)} \rangle_f$ histograms. If anisotropy is too high then one would expect that the dye can have restricted mobility.

fFCS: The mobility of the dye alone is better resolved using a complete FCS technique [27].

6) FPS provides also a consistent view of the conformational states of T4L. Each distinct set of conformer specific FRET restrains are within the expected uncertainty of our tools. In addition, the kinetics found in all our variants are consistent with two global relaxation time ($t_{R1}$ = ~4 µs, $t_{R2}$ = ~ 230 µs) and the expected three conformational states.

7) Thermodynamic stability and proper folding of our mutants were verified by chemical denaturation using urea.

8) Fluorescence intensity decay were fit with various models and gave a consistent view of three FRET induced donor lifetimes or two FRET induced donor lifetimes where only one would be expected if the conformer $C_3$ did not exist.

**Table S5A.** Analysis of time-resolved fluorescence anisotropies $r(t)$ for donor only labeled samples [a] obtained by ensemble time-resolved fluorescence decays as described in [2].

| Samples | $r_{1,D}$ | $\rho_{1.D}$ [ns] | $r_{2,D}$ | $\rho_{2.D}$ [ns] | $r_{3,D}$ | $\rho_{3.D}$ [ns] |
|---|---|---|---|---|---|---|
| 5/44-(D(0)) | 0.049 | 0.18 | 0.080 | 1.92 | 0.246 | 12.30 |
| 5/132-D(0)) | 0.036 | 0.15 | 0.041 | 1.12 | 0.298 | 8.58 |
| 8/69-(D(0)) | 0.082 | 0.17 | 0.063 | 1.47 | 0.230 | 9.76 |
| 8/86-(D(0)) | 0.078 | 0.18 | 0.066 | 1.24 | 0.231 | 9.09 |
| 8/119-(D(0)) | 0.04 | 0.11 | 0.042 | 1.08 | 0.293 | 8.15 |
| 8/132-(D(0)) | 0.049 | 0.10 | 0.047 | 0.97 | 0.279 | 7.85 |
| 19/69-(D(0)) | 0.122 | 0.26 | 0.116 | 1.84 | 0.137 | 8.47 |
| 19/86-(D(0)) | 0.122 | 0.26 | 0.116 | 1.84 | 0.137 | 8.47 |
| 19/119-(D(0)) | 0.122 | 0.26 | 0.116 | 1.84 | 0.137 | 8.47 |
| 19/132-(D(0)) | 0.082 | 0.16 | 0.057 | 1.15 | 0.237 | 10.39 |
| 22/127-(D(0)) | 0.054 | 0.14 | 0.120 | 1.78 | 0.201 | 16.32 |
| 36/86-(D(0)) | 0.095 | 0.17 | 0.089 | 1.44 | 0.192 | 12.00 |
| 36/132-(D(0)) | 0.078 | 0.13 | 0.074 | 0.91 | 0.223 | 8.12 |
| 44/69-(D(0)) | 0.093 | 0.15 | 0.089 | 1.17 | 0.193 | 9.32 |
| 44/86-(D(0)) | 0.113 | 0.19 | 0.067 | 1.26 | 0.195 | 8.60 |
| 44/119-(D(0)) | 0.113 | 0.19 | 0.067 | 1.26 | 0.195 | 8.60 |
| 44-127-(D(0)) | 0.093 | 0.15 | 0.089 | 1.17 | 0.193 | 9.32 |
| 44/132-(D(0)) | 0.113 | 0.19 | 0.067 | 1.26 | 0.195 | 8.60 |
| 44/150-(D(0)) | 0.113 | 0.19 | 0.067 | 1.26 | 0.195 | 8.60 |
| 55/69-(D(0)) | 0.268 | 0.10 | 0.107 | 0.73 | 0.071 | 7.61 |
| 55/119-(D(0)) | 0.134 | 0.17 | 0.084 | 1.27 | 0.157 | 10.89 |
| 55/132-(D(0)) | 0.245 | 0.05 | 0.130 | 0.58 | 0.120 | 6.93 |
| 55/150-(D(0)) | 0.144 | 0.20 | 0.082 | 1.22 | 0.150 | 9.52 |
| 60/86-(D(0)) | 0.096 | 0.18 | 0.081 | 1.24 | 0.198 | 8.24 |
| 60/119-(D(0)) | 0.077 | 0.16 | 0.067 | 1.17 | 0.231 | 9.00 |
| 60/132-(D(0)) | 0.096 | 0.18 | 0.081 | 1.24 | 0.198 | 8.24 |
| 60/150-(D(0)) | 0.096 | 0.18 | 0.081 | 1.24 | 0.198 | 8.24 |
| 69/86-(D(0)) | 0.115 | 0.18 | 0.092 | 1.07 | 0.167 | 8.26 |
| 69/119-(D(0)) | 0.115 | 0.18 | 0.092 | 1.07 | 0.167 | 8.26 |
| 69/132-(D(0)) | 0.115 | 0.18 | 0.092 | 1.07 | 0.167 | 8.26 |
| 69/150-(D(0)) | 0.115 | 0.18 | 0.092 | 1.07 | 0.167 | 8.26 |
| 70/119-(D(0)) | 0.108 | 0.23 | 0.094 | 1.38 | 0.173 | 8.42 |
| 70/132-(D(0)) | 0.108 | 0.23 | 0.094 | 1.38 | 0.173 | 8.42 |

[a]: The fluorescence anisotropy decay $r(t)$ can be described as a sum of three exponentials:
$r(t) = r_1 \exp(-t/\rho_1) + r_2 \exp(-t/\rho_2) + r_3 \exp(-t/\rho_3)$ with $r_1 + r_2 + r_3 \leq r_0$
For Alexa488-hydroxylamine the fundamental anisotropy $r_{0,D}$ is 0.375, for Alexa647-maleimide $r_{0,A}$ is 0.39 and for FRET-sensitized anisotropy decay $r_{0,A(D)}$ is 0.38.

**Table S5B.** Analysis of time-resolved fluorescence anisotropies $r(t)$ for direct acceptor excitation of double labeled samples obtained by ensemble time-resolved fluorescence decays as described in [2]

| Samples | $r_{1,D}$ | $\rho_{1.D}$ [ns] | $r_{2,D}$ | $\rho_{2.D}$ [ns] | $r_{3,D}$ | $\rho_{3.D}$ [ns] |
|---|---|---|---|---|---|---|
| 5/44-(DA) | 0.081 | 0.03 | 0.144 | 0.70 | 0.165 | 10.68 |
| 5/132-DA) | 0.030 | 0.14 | 0.119 | 0.84 | 0.241 | 12.11 |
| 8/69-(DA) | 0.060 | 0.08 | 0.116 | 0.57 | 0.214 | 13.96 |
| 8/86-(DA) | 0.051 | 0.05 | 0.108 | 0.73 | 0.231 | 12.53 |
| 8/119-(DA) | 0.035 | 0.08 | 0.134 | 0.66 | 0.221 | 13.12 |
| 8/132-(DA) | 0.059 | 0.10 | 0.152 | 0.79 | 0.178 | 11.26 |
| 19/69-(DA) | 0.066 | 0.04 | 0.155 | 0.71 | 0.169 | 12.58 |
| 19/86-(DA) | 0.036 | 0.17 | 0.141 | 0.87 | 0.214 | 11.38 |
| 19/119-(DA) | 0.042 | 0.09 | 0.142 | 0.79 | 0.206 | 10.73 |
| 19/132-(DA) | 0.061 | 0.09 | 0.173 | 0.78 | 0.156 | 19.93 |
| 22/127-(DA) | 0.062 | 0.05 | 0.151 | 0.91 | 0.177 | 14.09 |
| 36/86-(DA) | 0.042 | 0.12 | 0.131 | 0.80 | 0.217 | 11.18 |
| 36/132-(DA) | 0.030 | 0.03 | 0.148 | 0.80 | 0.212 | 17.19 |
| 44/69-(DA) | 0.094 | 0.02 | 0.151 | 0.99 | 0.145 | 14.80 |
| 44/86-(DA) | 0.039 | 0.16 | 0.142 | 0.90 | 0.209 | 14.11 |
| 44/119-(DA) | 0.030 | 0.17 | 0.133 | 0.82 | 0.227 | 11.47 |
| 44-127-(DA) | 0.105 | 0.35 | 0.099 | 1.72 | 0.186 | 19.97 |
| 44/132-(DA) | 0.046 | 0.06 | 0.073 | 0.73 | 0.271 | 10.13 |
| 44/150-(DA) | 0.031 | 0.14 | 0.096 | 0.82 | 0.263 | 10.36 |
| 55/69-(DA) | 0.036 | 0.07 | 0.096 | 0.71 | 0.257 | 11.24 |
| 55/119-(DA) | 0.030 | 0.09 | 0.072 | 0.74 | 0.288 | 10.84 |
| 55/132-(DA) | 0.031 | 0.07 | 0.107 | 0.63 | 0.252 | 11.14 |
| 55/150-(DA) | 0.037 | 0.16 | 0.078 | 0.96 | 0.275 | 10.37 |
| 60/86-(DA) | 0.026 | 0.09 | 0.074 | 0.76 | 0.290 | 10.10 |
| 60/119-(DA) | 0.067 | 0.12 | 0.056 | 0.67 | 0.267 | 10.45 |
| 60/132-(DA) | 0.100 | 0.12 | 0.075 | 0.78 | 0.215 | 11.38 |
| 60/150-(DA) | 0.045 | 0.02 | 0.067 | 0.53 | 0.278 | 8.61 |
| 69/86-(DA) | 0.110 | 0.23 | 0.063 | 2.86 | 0.216 | 11.06 |
| 69/119-(DA) | 0.054 | 0.07 | 0.069 | 0.75 | 0.267 | 9.80 |
| 69/132-(DA) | 0.064 | 0.21 | 0.147 | 1.05 | 0.179 | 13.87 |
| 69/150-(DA) | 0.157 | 0.70 | 0.233 | 24.87 | | |
| 70/119-(DA) | 0.039 | 0.10 | 0.087 | 0.70 | 0.264 | 11.08 |
| 70/132-(DA) | 0.054 | 0.14 | 0.080 | 0.69 | 0.256 | 8.66 |

**Table S5C** Analysis of time-resolved fluorescence anisotropies *r*(*t*) for FRET-sensitized emission of acceptor of double labeled samples obtained by ensemble time-resolved fluorescence decays as described in [2] except for [1] and [2].

| Samples | $r_{1,D}$ | $\rho_{1.D}$ [ns] | $r_{2,D}$ | $\rho_{2.D}$ [ns] | $r_{3,D}$ |
|---|---|---|---|---|---|
| 5/44-(DA) | 0.052 | 0.192 | 0.010 | 8.126 | 0.062 |
| 5/132-DA) | 0.087 | 0.320 | 0.026 | 10.008 | 0.113 |
| 8/69-(DA) | 0.025 | 0.509 | 0.041 | ∞ | 0.066 |
| 8/86-(DA) | 0.032 | 0.438 | 0.049 | 380 | 0.081 |
| 8/119-(DA) | 0.061 | 0.127 | 0.012 | ∞ | 0.073 |
| 8/132-(DA) | 0.091 | 0.280 | 0.020 | 4.163 | 0.111 |
| 19/69-(DA) | 0.081 | 0.398 | 0.105 | 48.063 | 0.186 |
| 19/86-(DA)[1] | 0.209 | 0.756 | 0.0561 | 19.901 | 0.2651 |
| 19/119-(DA) | 0.041 | 0.512 | 0.091 | 202 | 0.132 |
| 19/132-(DA) | 0.1 | 0.373 | 0.112 | 88.561 | 0.212 |
| 22/127-(DA) | 0.044 | 0.702 | 0.018 | ∞ | 0.062 |
| 36/86-(DA) | 0.087 | 0.243 | 0.007 | 9.393 | 0.094 |
| 36/132-(DA) | 0.086 | 0.241 | 0.020 | 2.095 | 0.106 |
| 44/69-(DA) | 0.033 | 0.282 | 0.019 | 7.958 | 0.052 |
| 44/86-(DA)[2] | <0.06 | | | | |
| 44/119-(DA)[2] | <0.09 | | | | |
| 44-127-(DA) | 0.179 | 0.326 | 0.017 | 8.815 | 0.196 |
| 44/132-(DA) | 0.054 | 0.246 | 0.115 | 23.934 | 0.169 |
| 44/150-(DA) | 0.087 | 0.563 | 0.048 | 101.937 | 0.135 |
| 55/69-(DA) | 0.036 | 0.405 | 0.069 | 63.43 | 0.105 |
| 55/119-(DA) | 0.067 | 1.31 | 0.089 | 136.651 | 0.156 |
| 55/132-(DA) | 0.064 | 1.039 | 0.016 | 14.346 | 0.08 |
| 55/150-(DA) | 0.065 | 0.512 | 0.061 | 150.739 | 0.126 |
| 60/86-(DA) | 0.103 | 0.483 | 0.104 | 127.327 | 0.207 |
| 60/119-(DA) | 0.079 | 0.501 | 0.086 | 114.851 | 0.165 |
| 60/132-(DA) | 0.054 | 1.035 | 0.058 | 74.739 | 0.112 |
| 60/150-(DA) | 0.038 | 1.102 | 0.067 | 77.378 | 0.105 |
| 69/86-(DA) | 0.038 | 0.604 | 0.073 | ∞ | 0.111 |
| 69/119-(DA) | 0.045 | 0.603 | 0.059 | 84.864 | 0.104 |
| 69/132-(DA) | 0.039 | 0.294 | 0.049 | 72.456 | 0.088 |
| 69/150-(DA) | 0.049 | 0.595 | 0.048 | 210.295 | 0.097 |
| 70/119-(DA)[2] | <0.04 | 0.2416 | | | |
| 70/132-(DA)[2] | <0.04 | 0.2471 | | | |

[1] eTCSPC data not available. Fluorescence anisotropy decay was fitted from sub-ensemble single-molecule MFD data of the FRET population.
[2] eTCSPC data not available. Considering variants with a very high FRET efficiency, no satisfactory anisotropy decays from sub-ensemble single-molecule MFD data were obtainable due the short donor fluorescence lifetime. Here, steady-values anisotropies were taken as upper limit from single-molecule MFD measurements.

**Table S5D $\kappa^2$ distributions for the 24 DA samples.** Donor positions are labeled on green and acceptor positions on red. The mean $\kappa^2$ ($\langle\kappa^2\rangle$) is shown as a solid bar in blue, and $\kappa^2 = 2/3$ is shown in red. Therefore, the assumption of $\kappa^2 = 2/3$ is justified. Nevertheless, the $\kappa^2$ distribution adds to the uncertainty on our distances, which is considered as previously described in [2].

| E5pAcF S44C | E5pAcF N132C | R8pAcF N55C | R8pAcF Q69C | R8pAcF P86C |
|---|---|---|---|---|
| 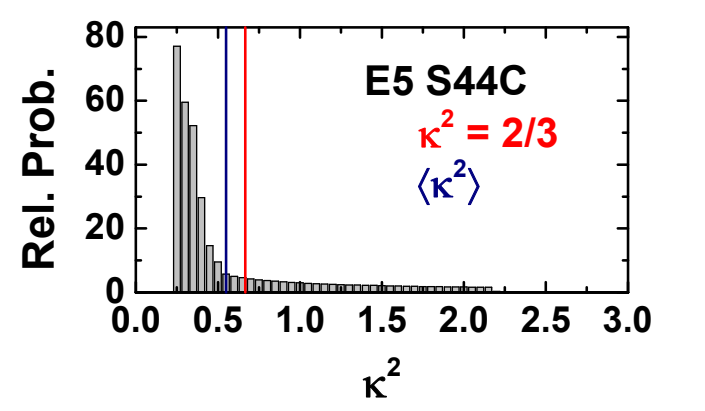 | 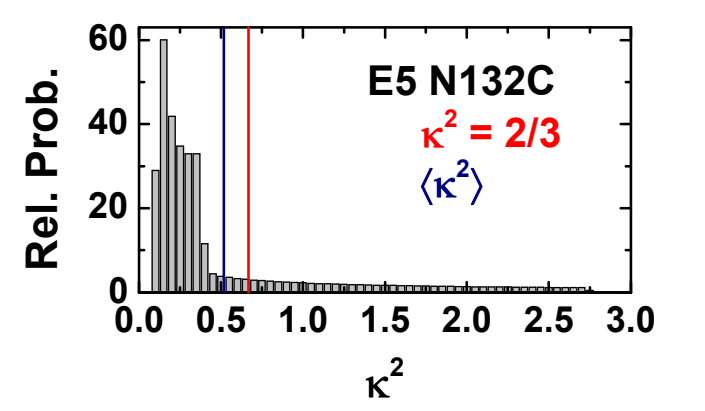 | 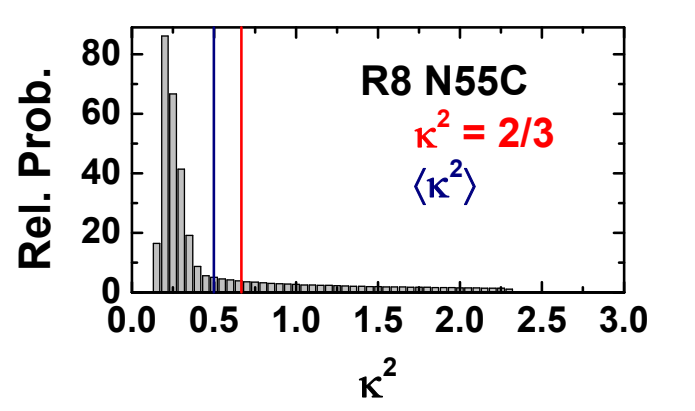 | 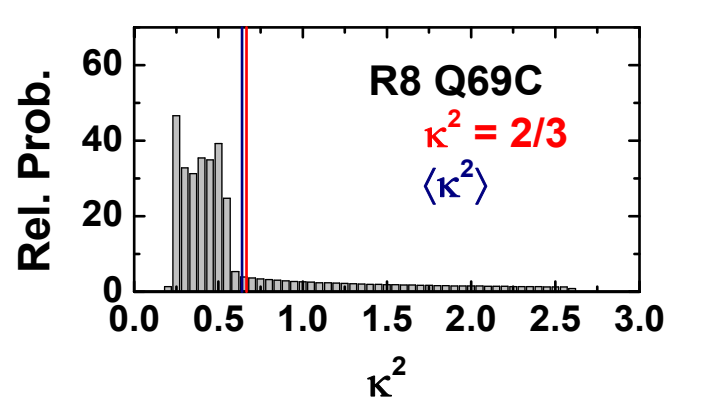 | 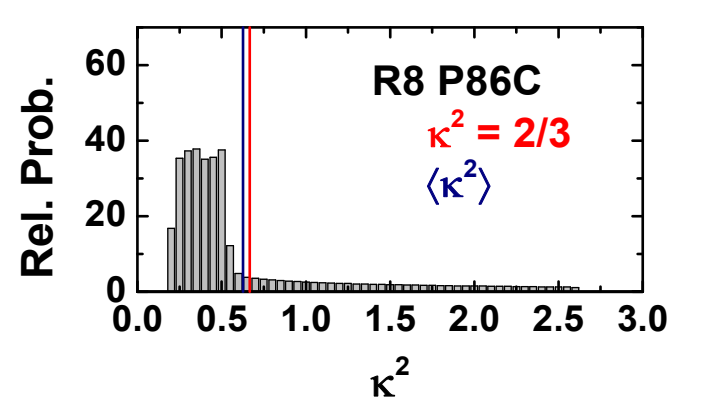 |
| **R8pAcF R119C** | **R8pAcF N132C** | **K19pAcF Q69C** | **K19pAcF P86C** | **K19pAcF R119C** |
| 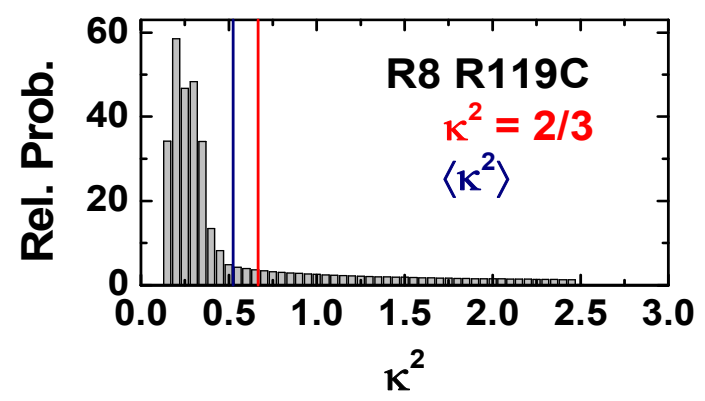 | 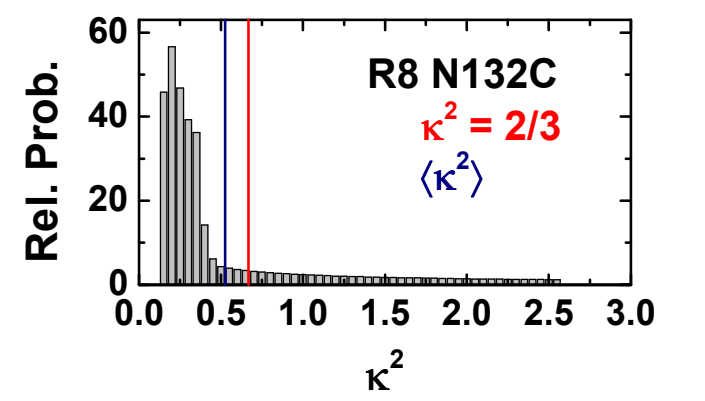 | 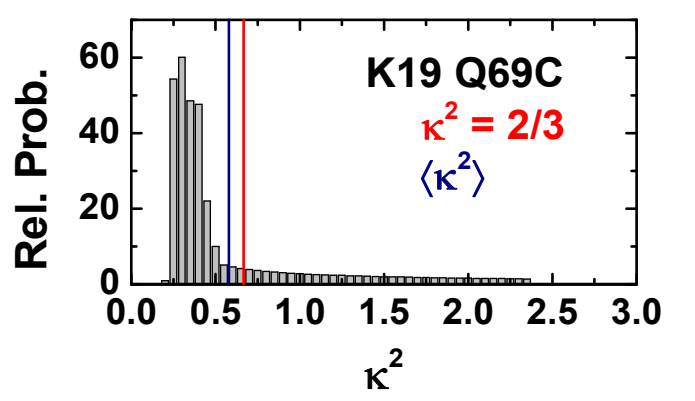 | 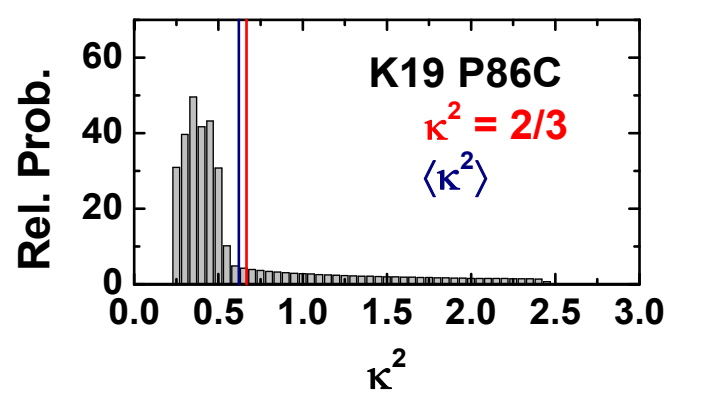 | 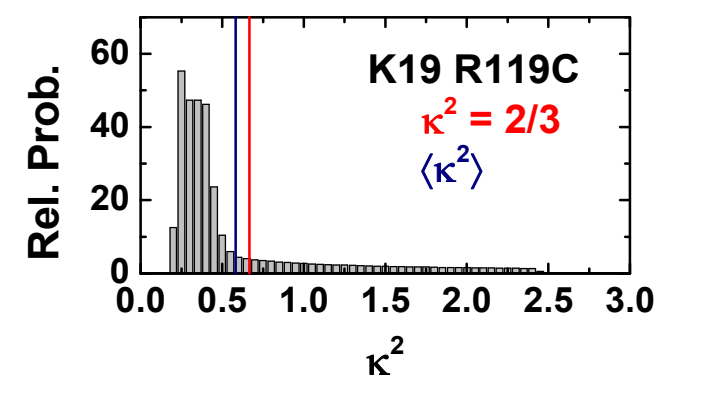 |
| **K19pAcF N132C** | **E22pAcF P86C** | **E22pAcF D127C** | **S36pAcF P86C** | **S36pAcF N132C** |
| 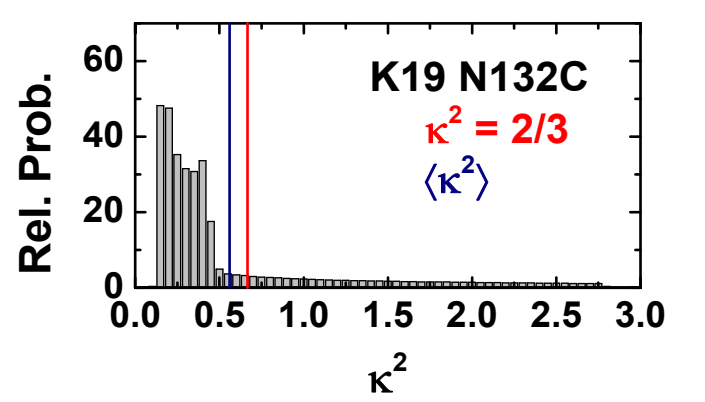 | 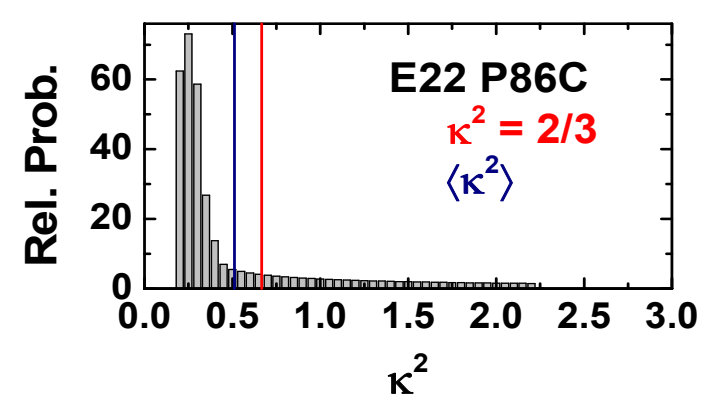 | 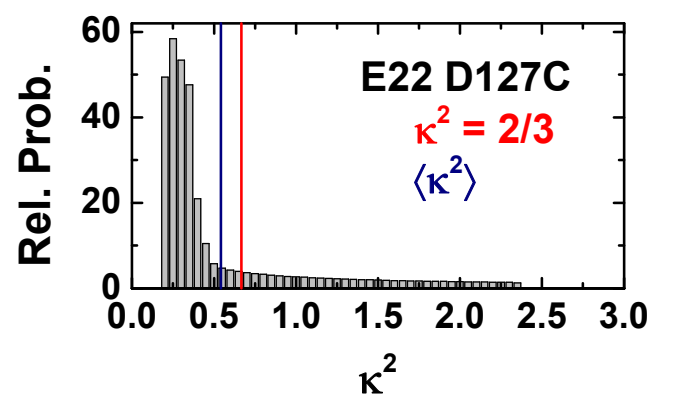 | 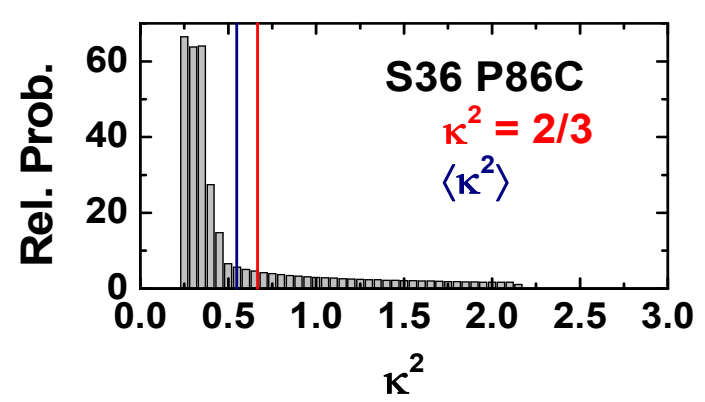 | 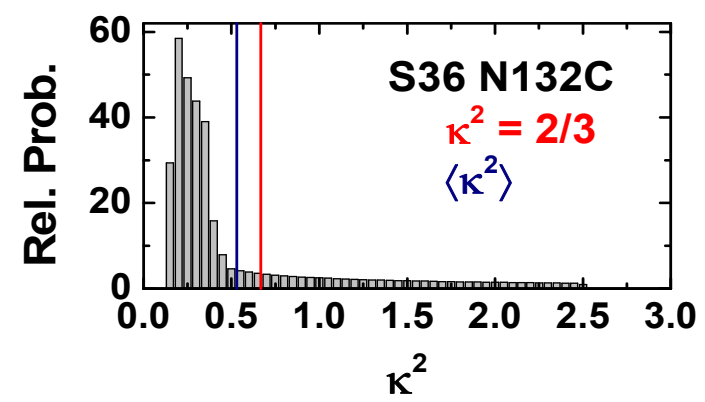 |
| **S44pAcF Q69C** | **S44pAcF P86C** | **S44pAcF R119C** | **S44pAcF D127C** | **S44pAcF N132C** |
| 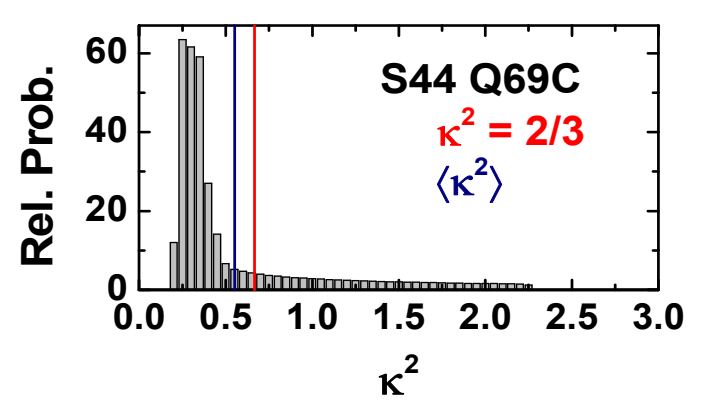 | 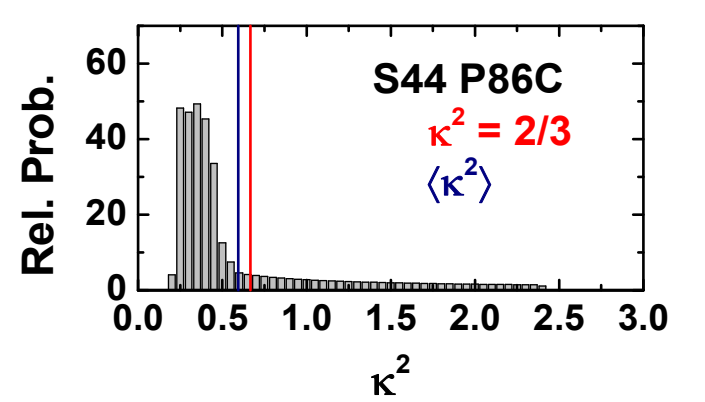 | 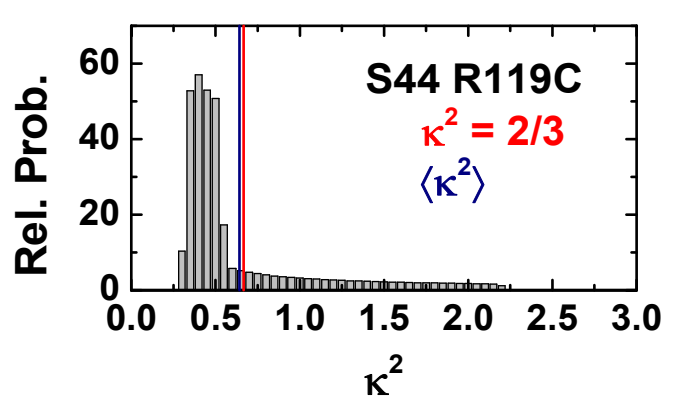 | 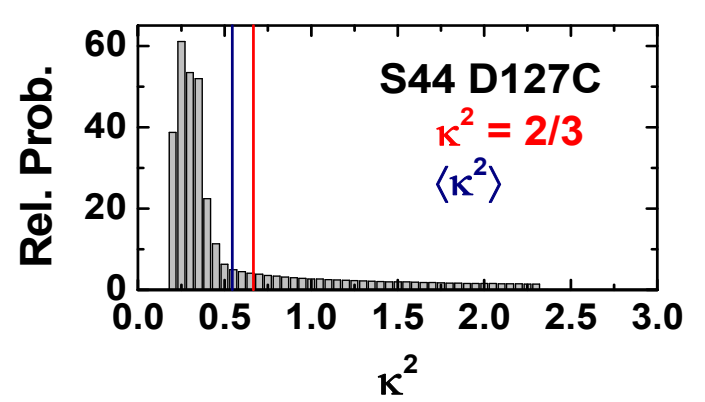 | 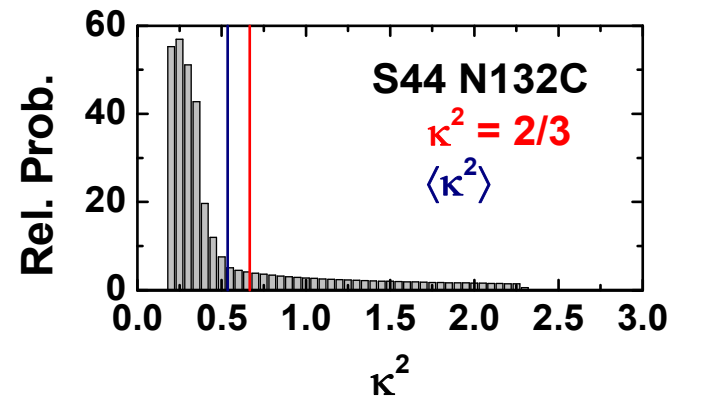 |

**S44pAcF I150C**

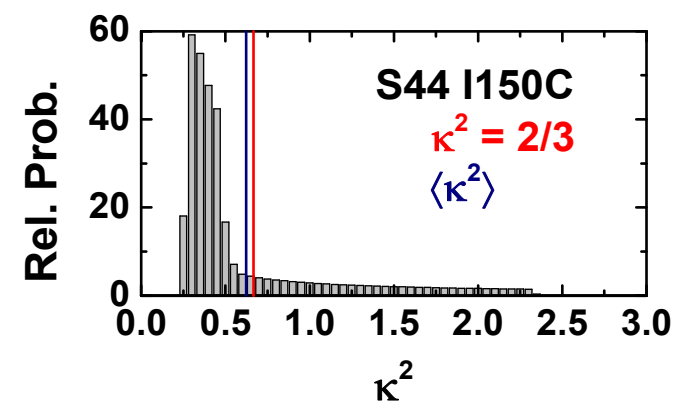

**N55pAcF Q69C**

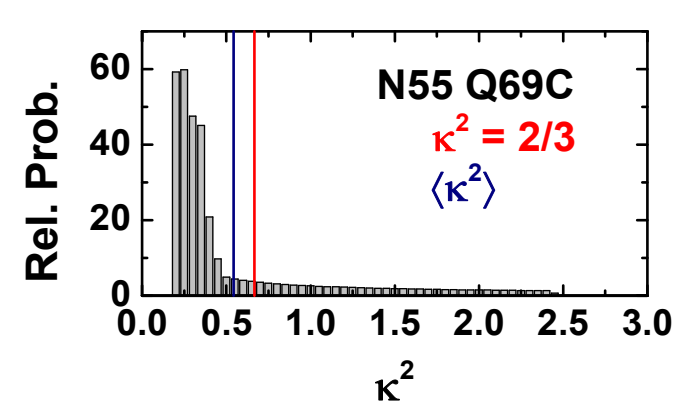

**N55pAcF R119C**

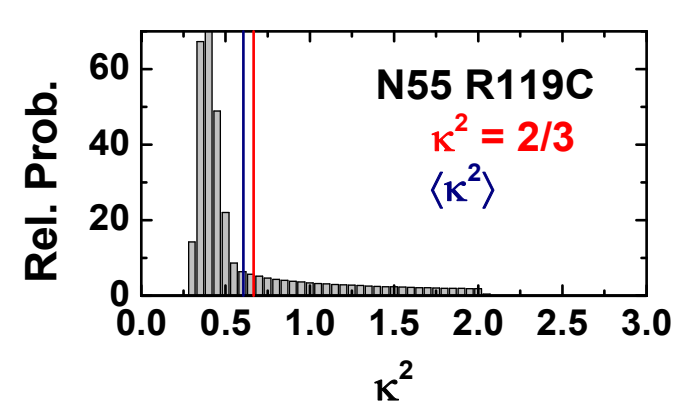

**N55pAcF N132C**

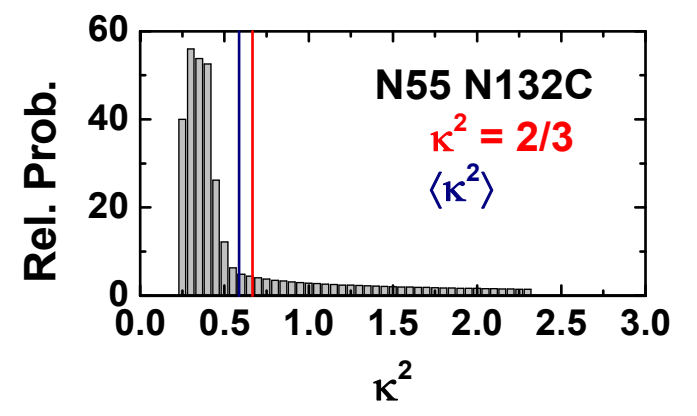

**N55pAcF I150C**

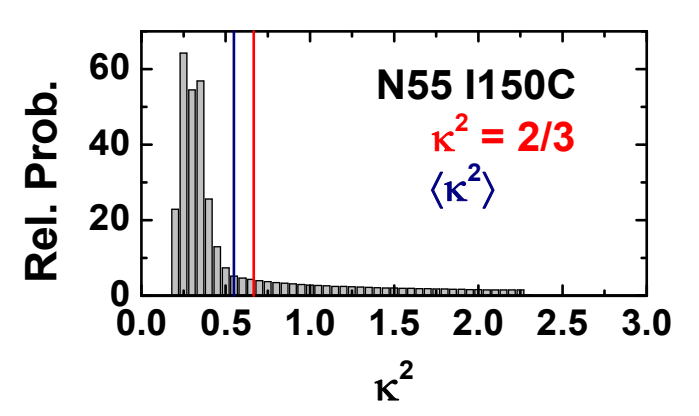

**K60pAcF P86C**

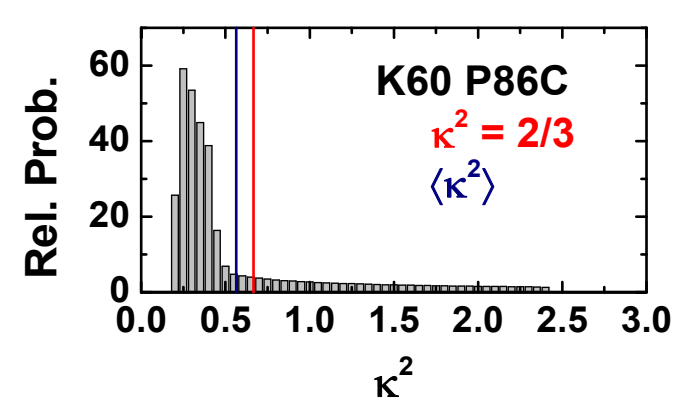

**K60pAcF R119C**

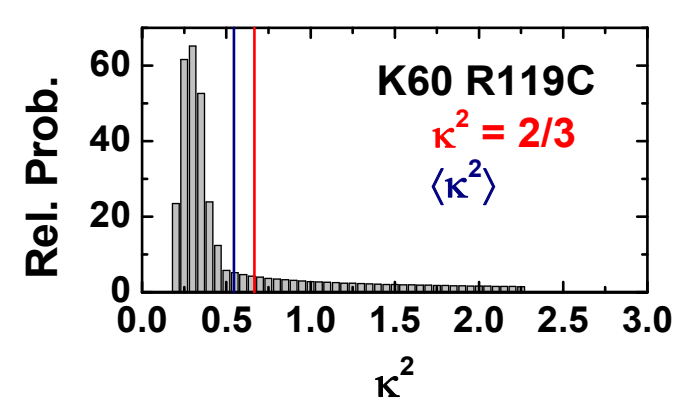

**K60pAcF N132C**

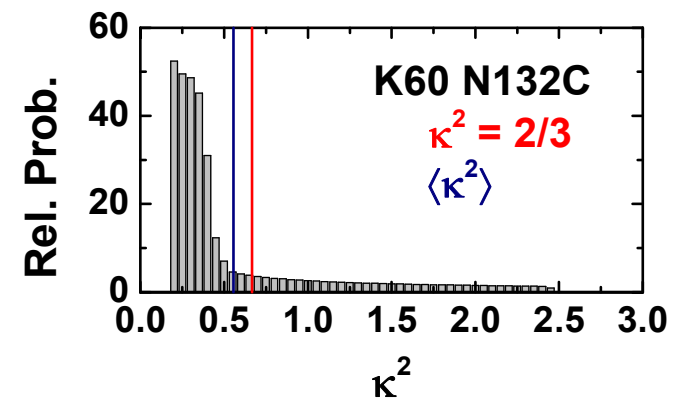

**K60pAcF I150C**

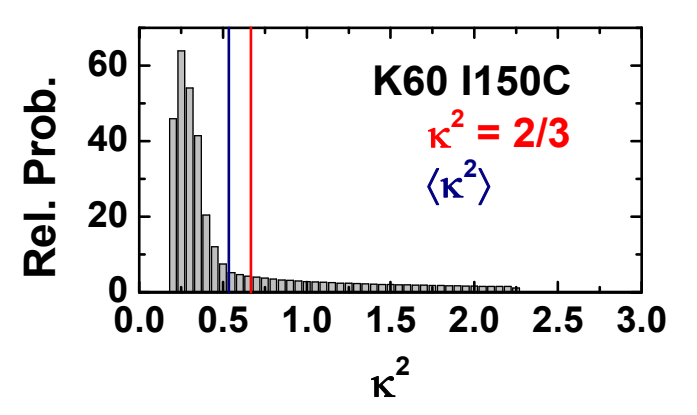

**Q69pAcF P86C**

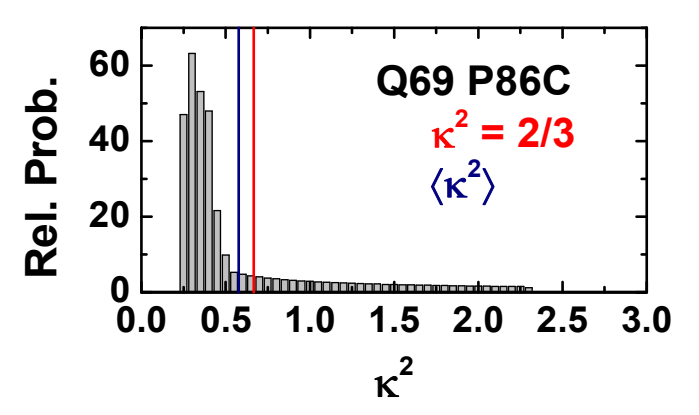

**Q69pAcF R119C**

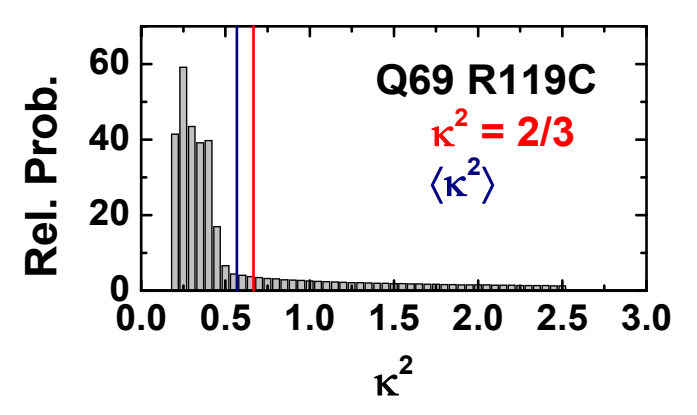

**Q69pAcF N132C**

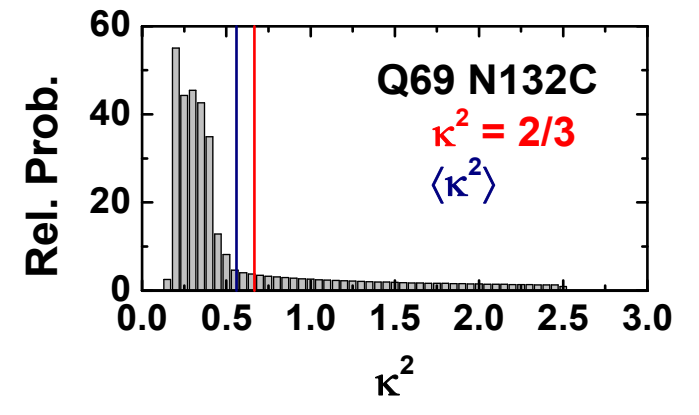

**Q69pAcF I150C**

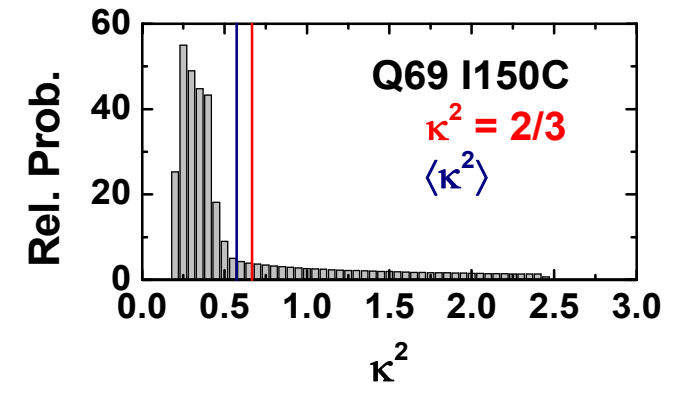

**D70pAcF R119C**

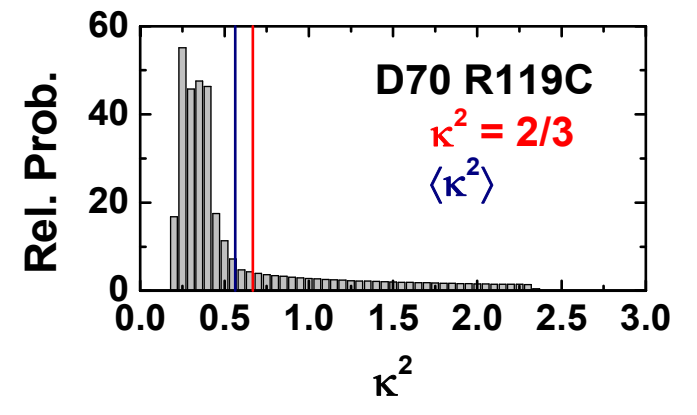

**D70pAcF N132C**

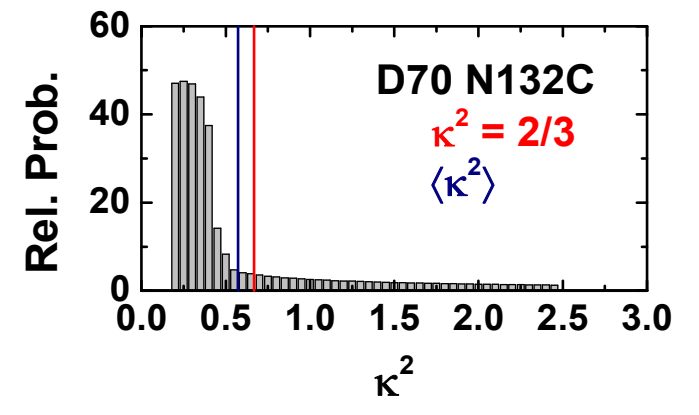

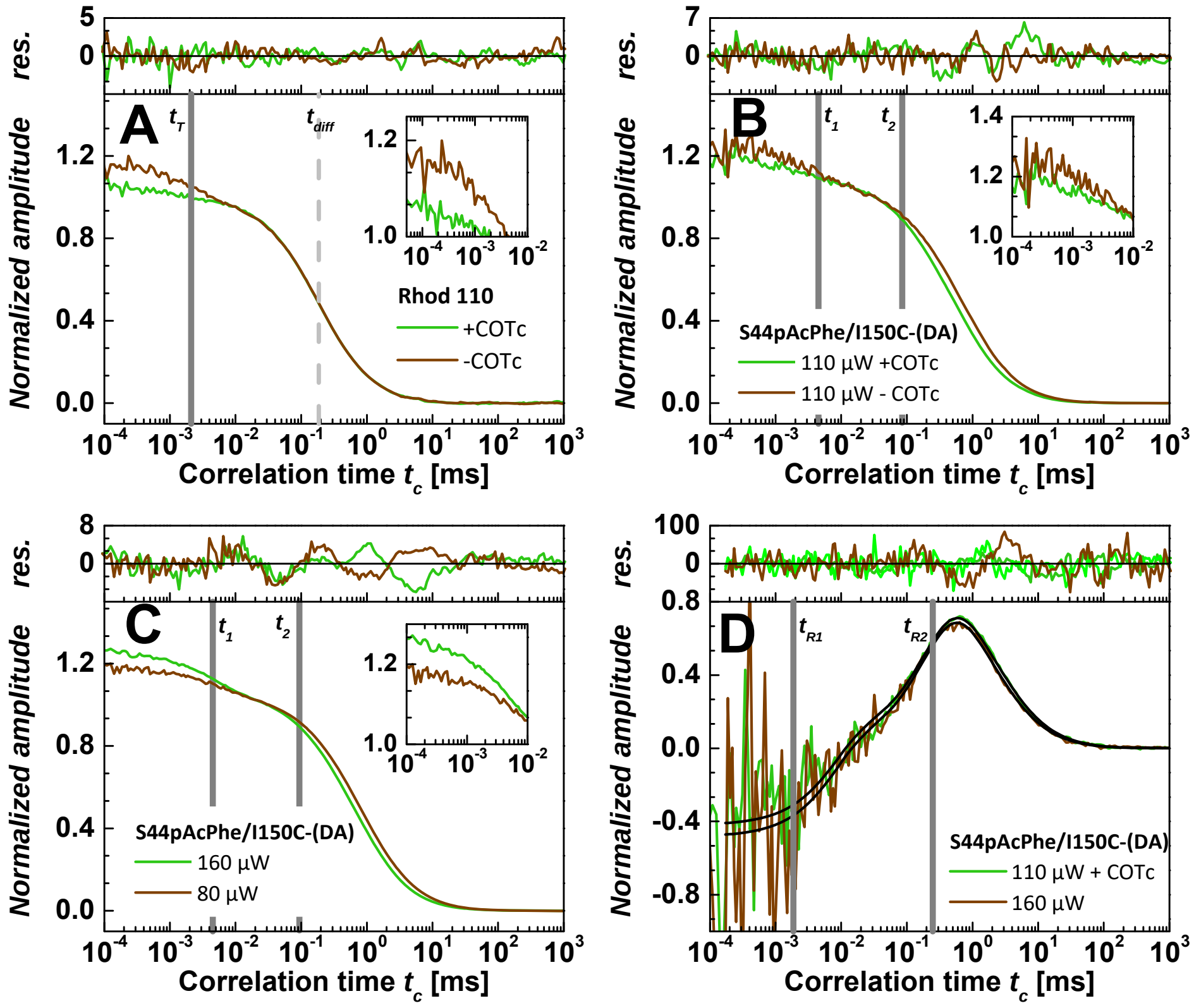

**Figure S10 Triplet or dark states do not influence the *sCCF* on the variant S44pAcF/I150C-(DA). (A)** The addition of the triplet quencher COTc into Rhod110 solution significantly reduces triplet fraction (see in inset). (**B**) Overlay of the standard auto/cross-correlation curves from signals in the green channels for the variant S44pAcF/I150C-(DA) without (-COTc) and with (+COTc) triplet quencher COTc in solution. Inset shows the regime where triple kinetics is observed. (**C**) Overlay of standard auto/cross-correlation of the green signals at 80 µW and at 160 µW power at objective. Two bunching terms are needed to fit the data ($t_T$ = 4.5 µs, and $t_b$ = 60 µs). The triplet fraction changes from 10 % at 80 µW to 15 % at 160 µW power at objective. Also changes in diffusion times are observed from 0.8 ms at 80 µW to 0.6 ms at 160 µW power at objective. Photobleaching can account for this change. Inset shows the reduction of the triplet fraction by COTc quencher. (**D**) *sCCF* of the variant S44pAcF/I150C-(DA) between pseudo-species $C_1$/$C_2$ and $C_3$ at different power at 80 µW and at 160 µW power at the objective. The relaxation times fitted globally are $t_{R1}$ = 6 µs and $t_{R2}$ = 240 µs, that are within the errors presented on Table S4C. Note that the amplitudes do not change as in the case of the standard auto-correlation.

**References (Supporting Information):**